\documentclass[a4paper]{jpconf}
\usepackage{graphicx}
\usepackage{amsmath}
\usepackage{amssymb}
\usepackage{epsfig}

\usepackage{enumitem}

\def\be{\begin{equation}}
\def\ee{\end{equation}}
\def\ba{\begin{eqnarray}}
\def\ea{\end{eqnarray}}

\def\bi{\begin{itemize}}
\def\ei{\end{itemize}}

\begin{document}
%
%\linenumbers
%
\title{Detection of Cosmic Rays from ground: an Introduction}

\author{Giuseppe Di Sciascio}

\address{INFN - Roma Tor Vergata}

\ead{giuseppe.disciascio@roma2.infn.it}

\begin{abstract}

Cosmic rays are the most outstanding example of accelerated particles. 
They are about 1\% of the total mass of the Universe, so that cosmic rays would represent by far the most important energy transformation process of the Universe.
Despite large progresses in building new detectors and in the analysis techniques, the key questions concerning origin, acceleration and propagation of the radiation are still open.
One of the reasons is that there are significant discrepancies among the different results obtained by experiments located at ground probably due to unknown systematic errors affecting the measurements.

In this note we will focus on detection of Galactic CRs from ground with EAS arrays.
This is not a place for a complete review of CR physics (for which we recommend, for instance \cite{spurio,gaisser,grieder,longair,kampert,blasi,kachelriess}) but only to provide elements useful to understand the basic techniques used in reconstructing primary particle characteristics (energy, mass and arrival direction) from ground, and to show why indirect measurements are difficult and results still conflicting.
\end{abstract}
\section{Introduction}
\label{intro}

Cosmic rays (CRs) are the most outstanding example of accelerated particles. 
Understanding their origin and propagation through the InterStellar Medium (ISM) is a fundamental problem which has a major impact on models of the structure and nature of the Universe.

Charged cosmic rays, gammas and neutrinos are strongly correlated in CR sources where hadronic accelerators are at work. 
Their integrated study is one of the most important and exciting fields in the \emph{'multi-messenger astronomy'}, the exploration of the Universe through combining information from different cosmic messengers: electromagnetic radiation, gravitational waves, neutrinos and cosmic rays.

The study of CRs from ground is based on two complementary approaches:
\begin{enumerate}
\item[(1)] Measurement of energy spectrum, elemental composition and anisotropy in the CR arrival direction distribution, the three basic observables crucial for understanding origin, acceleration and propagation of the radiation.
\item[(2)] Search of their sources through the observation of neutral radiation (photons and neutrinos), which points back to the emitting sources not being affected by the magnetic fields.
\end{enumerate}
Despite large progresses in building new detectors and in the analysis techniques, the key questions concerning origin, acceleration and propagation of CRs are still open.

\begin{figure}[ht]
\centering
\includegraphics[scale=0.90]{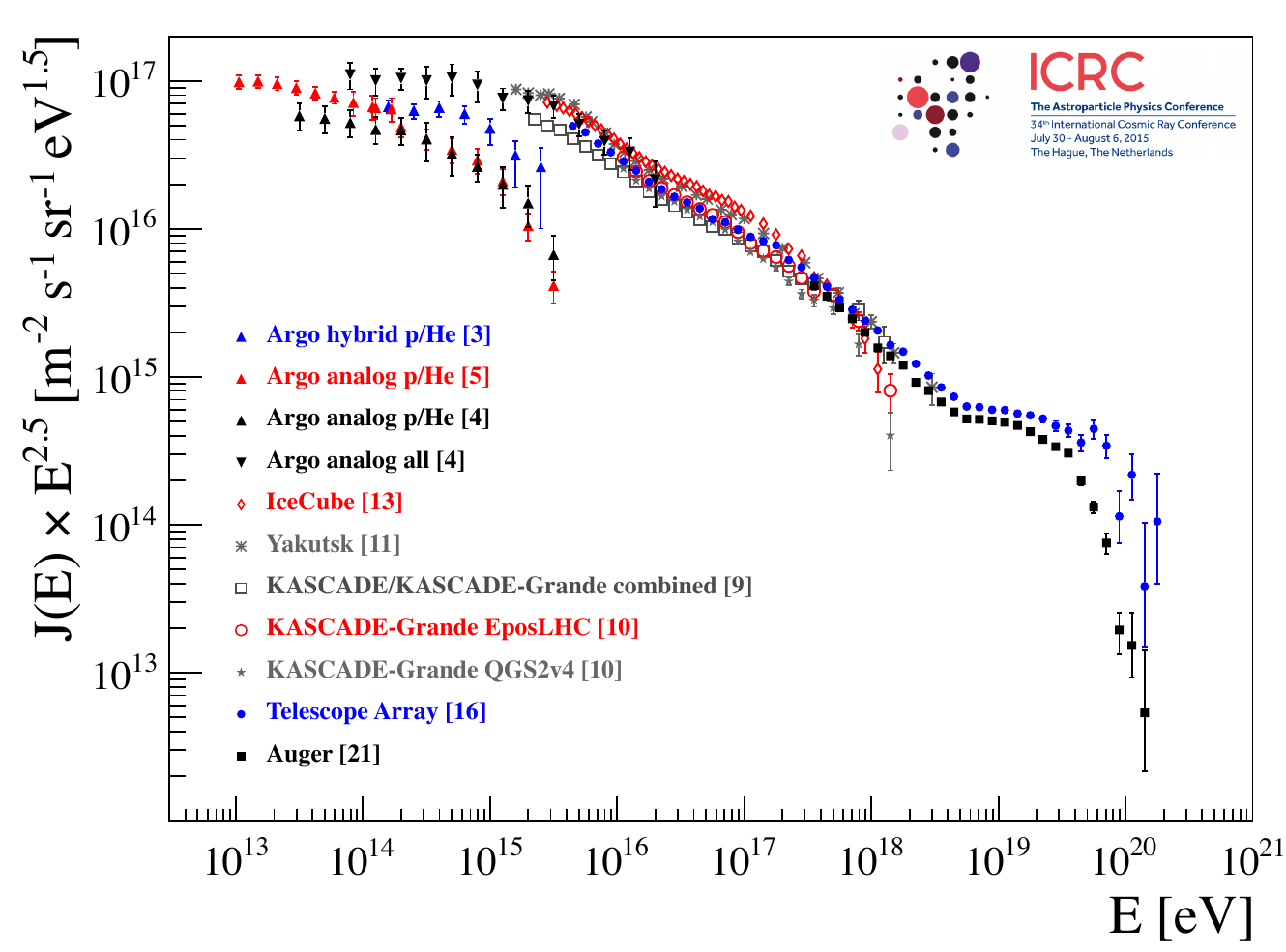}
\caption{All-particle energy spectrum of primary cosmic rays. The light component (p+He) measured by the ARGO-YBJ experiment is also shown. }
\label{fig:allpart-enespt}       % Give a unique label
\end{figure}

It is widely believed that the bulk of CRs up to about 10$^{17}$ eV are Galactic, produced and accelerated by the shock waves of SuperNova Remnants (SNR) expanding shells \cite{drury12}, and that the transition to extra-galactic CRs occurs somewhere between 10$^{17}$-10$^{19}$ eV.
The primary CR all-particle energy spectrum (namely the number of nuclei as a function of total energy) exceeds 10$^{20}$ eV showing a few basic characteristics (see Fig. \ref{fig:allpart-enespt}): 
%
%%\begin{itemize}
\begin{enumerate}[label=(\roman*)]
\item[(a)] a power-law behaviour $\sim$ E$^{-2.7}$ until the so-called \emph{``knee''}, a small downwards bend around few PeV; 
\item[(b)] a power-law behaviour $\sim$ E$^{-3.1}$ beyond the knee, with a slight dip near 10$^{17}$ eV, sometimes referred to as the \emph{``second knee''}; 
\item[(c)] a transition back to a power-law $\sim$ E$^{-2.7}$ (the so-called \emph{``ankle''}) around $10^{18}$~eV; 
\item[(d)] a cutoff probably due to extra-galactic CR interactions with the Cosmic Microwave Background (CMB) around 10$^{20}$ eV (the Greisen-Zatsepin-Kuzmin effect).
\end{enumerate}
%
%%%%%%%%%%%%%%%%%%%%%%%%%%%%%%%%%%%%
%%%%%%%%%%%%%%%%%%%%%%%%%%%%%%%%%%%%
%
All these features are believed to carry fundamental information that sheds light on the key question of the CR origin. 
In particular, understanding the origin of the \emph{"knee"} is the key for a comprehensive theory of the origin of CRs up to the highest observed energies.
In fact, the knee is clearly connected with the issue of the end of the Galactic CR spectrum and the transition from Galactic to extra-galactic CRs. 

If the knee is a source property we should see a corresponding spectral feature in the gamma-ray spectra of the CR sources. If, on the contrary, this feature is the result of propagation, we should observe a knee that is potentially dependent on location, because the propagation properties depend, in principle, on the position in the Galaxy.

To understand the origin of the knee we need to deepen our understanding of acceleration, escape and propagation of the relativistic particles, the main pillars that constitute the SuperNova paradigm for the origin of the radiation (see \cite{morlino17} and references therein).
We need to identify the sources and the mechanisms able to accelerate particles beyond PeV energies (the so-called \emph{"PeVatrons"}). We need to understand how particles escape from the sources and are released into the ISM. Finally, we need to understand how particles propagate through the Galaxy before reaching the Earth.

As we will discuss in the following sections, the SNR paradigm has two bases: firstly, the energy released in SN explosions can explain the CR energy density considering an overall efficiency of conversion of explosion energy into CR particles of the order of 10\% . 
Secondly, the diffusive shock acceleration operating in SNR can provide the necessary power-law spectral shape of accelerated particles with spectral index $-2.0$ that subsequently steepen to $-2.7$, as observed, due to the energy-dependent diffusive propagation effect (see \cite{drury17} and references therein).

SuperNovae are believed to be almost the only available power source. 
However, recent claims by H.E.S.S. of a possible detection of a PeVatrons in the Galactic Center, most likely related to a supermassive black hole \cite{hess-pev}, open new perspectives showing that galactic PeVatrons other than SNRs may exist.

Recently AGILE and Fermi observed GeV photons from two young SNRs (W44 and IC443) showing the typical spectrum feature around 1 GeV (the so-called \emph{'$\pi^0$ bump'}, due to the decay of $\pi^0\to\gamma\gamma$) related to hadronic interactions \cite{pizero-a,pizero-f}. 
This important measurement, however, does not demonstrate the capability of SNRs to produce the power needed to maintain the galactic CR population and to accelerate CRs up to the knee, at least. 
In fact, unlike neutrinos that are produced only in hadronic interactions, the question whether $\gamma$-rays are produced by the decay of $\pi^0$ from protons or nuclei interactions (\emph{'hadronic'} mechanism), or by a population of relativistic electrons via Inverse Compton scattering or bremsstrahlung (\emph{'leptonic'} mechanism), still needs a conclusive answer.

One of the main open problems in the SNR origin model is the maximum energy that can be attained by a CR particle in SNR. 
To accelerate protons up to the PeV energy domain a significant amplification of the magnetic field at the shock is required but this process is problematic \cite{gabici16}. 
However, if the knee is a propagation effect, the Galaxy could contain "super-PeVatrons", sources capable to accelerate particles well beyond the PeV. The study of these objects requires to observe the $\gamma$-ray sky at 100 TeV energies and beyond.

No direct observational evidence for the acceleration of PeV protons in SNRs has been reported yet, probably due to the fact that higher energy ($>$200 TeV) particles are believed to be accelerated in the early phases of the SuperNova explosion (i.e. in young SNRs). Therefore, we expect that very few SNRs are currently accelerating particles up to PeV energies.
In addition, the absorption of $\gamma$-rays may prevent observations of PeVatrons \cite{vernetto17}.
Finally, the sensitivity of current gamma-ray detectors in the 100 TeV range is very poor. 

As CRs are mostly charged nuclei, their paths are deflected and highly isotropized by the action of galactic magnetic field (GMF) they propagate through before reaching the Earth atmosphere. The GMF is the superposition of regular field lines and chaotic contributions. Although the strength of the non-regular component is still under debate, the local total intensity is supposed to be $B=2\div 4\textrm{ $\mu$G}$ \cite{beck01}. In such a field, the gyro-radius of CRs is given by ${r}_{a.u.}\approx 100\, R_{\textrm{\scriptsize{TV}}}$, where $r _{a.u.}$ is in astronomic units and R$_{\textrm{\scriptsize{TV}}}$ is the rigidity in TeraVolt. 
Clearly, there is very little chance of observing a point-like signal from any radiation source below $10^{17}{\rm eV}$, as they are known to be at least several hundreds parsecs away.

In the standard picture, mainly based on the results of the KASCADE esperiment, the knee is attributed to the steepening of the $p$ and \emph{He} spectra \cite{kascade}. 
According to a rigidity-dependent structure (Peters cycle), the sum of the fluxes of all elements, with their individual knees at energies proportional to the nuclear charge, makes up the CR all-particle spectrum \cite{peters}.
With increasing energies not only the spectrum becomes steeper, due to such cutoffs, but also heavier.

However, a number of results (in particular those obtained by experiments located at high altitudes) seem to indicate that the bending of the light component (p+He) is well below the PeV and the knee of the all-particle spectrum is due to heavier nuclei \cite{tibet,casamia,basje-mas}. 
Recent results obtained by the ARGO-YBJ experiment (located at 4300 m asl) clearly show, with different analyses, that the knee of the light component starts at $\sim$700 TeV \cite{hybrid15}, well below the knee of the all-particle spectrum that is confirmed by ARGO-YBJ at $\sim$4$\cdot$10$^{15}$ eV \cite{disciascio-rev}. 
%
%%%%%%%%%%%%%%%%%%%%%%%%%%%%%%%%%%%%%%%%%%%%%%%%%%%%%%%%%%
\begin{figure}
\centerline{\includegraphics[width=\textwidth,clip]{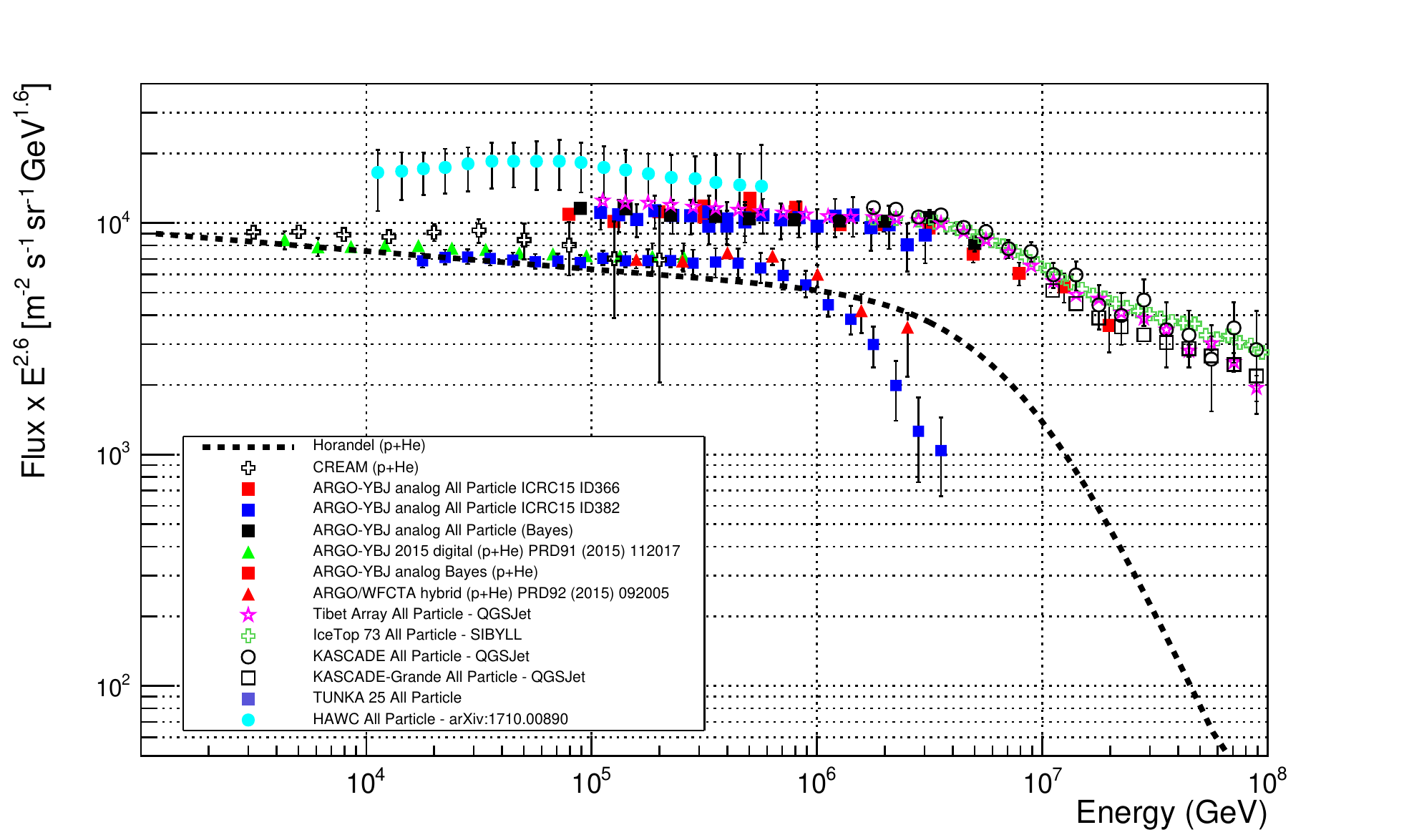} }
\caption{Cosmic ray all--particle and (p+He) energy spectra as measured by ARGO--YBJ compared with other experimental results \cite{cream11,tibetIII,icetop73,kascade,kascade-grande,hawc} and with the parametrization given in \cite{horandel}.}
\label{fig:phe-allp}       % Give a unique label
\end{figure}
%%%%%%%%%%%%%%%%%%%%%%%%%%%%%%%%%%%%%%%%%%%%%%%%%%%%%%%%%%
%
In Fig. \ref{fig:phe-allp} the CR all-particle and (p+He) energy spectra as measured by ARGO--YBJ compared with other experimental results are shown. A comparison with the parametrization of the light component by Horandel is also shown.

After more than half a century from the discovery of the knee experimental results are still conflicting with uncertainties on its origin.
This is not surprising for a lot of reasons.
The reconstruction of the CR elemental composition is often carried out by means of complex unfolding techniques based on the measurements of electronic and muonic sizes, procedures that heavily depend on the hadronic interaction models. 
The muonic size is much smaller than the electronic one with a wider lateral distribution, but the total sensitive area of muon detectors is typically only few hundred square meters and, due to the poor sampling, large instrumental fluctuations can be added to the stochastic ones associated to the shower development. In addition, the \emph{'punch-through effect'}, due to high energy secondary electromagnetic particles, could heavily affect the measurements.
Finally, some arrays have been operated close to the sea level and not in the shower maximum region where fluctuations are smaller and all nuclei produce the same electromagnetic size implying that the trigger efficiency is equal for all primary particles.
This imply that the sensitivity to the elemental composition of the classical $N_e/N_{\mu}$ unfolding technique is reduced at extreme altitudes and new mass-sensitive parameters must be used. In fact experiments operated at high altitude exploited, as an example, characteristics of the lateral distributions of secondary particles in the shower core region to select samples of showers induced by different nuclei.

At higher energies, KASCADE-Grande, IceTop and Tunka experiments observed a hardening slightly above 10$^{16}$ eV and a steepening at log10(E/eV) = 16.92$\pm$0.10 in the CR all-particle spectrum.
A steepening at log10(E/eV) = 16.92$\pm$0.04 in the spectrum of the electron poor event sample (heavy primaries) and a hardening at log10(E/eV) = 17.08$\pm$0.08 in the electron rich (light primaries) one were observed by KASCADE-Grande even if with modest statistical significance \cite{kascadeg-chiavassa}. The absolute fluxes of CRs with different masses measured by KASCADE-Grande are however strongly dependent on the adopted hadronic interaction models \cite{kascadeg-hadron}, thus requiring new high resolution data to clarify the observations.

As mentioned, the position of the knee of the proton spectrum is still controversial with important implications on the model of Galactic CRs (see Fig. \ref{fig:phe-allp}).
A general consequence of the SNR paradigm described above is, in fact, that the flux of galactic CRs should end with an iron dominated composition at energies $\sim$26 times larger than the knee in the proton spectrum. If such knee is indeed at PeV energies, as suggested by KASCADE analysis, then galactic CRs should end at about 10$^{17}$ eV, well below the ankle. 
To avoid an early appearance of the extragalactic CR component, in 2005 Hillas \cite{hillas2005} proposed in addition to the standard SNR component, a \emph{"component B"} of CRs of (probably) Galactic origin.
As a result, the transition occurs at the ankle and for the entire energy range from 10$^{15}$ eV to 10$^{18}$ eV a mixed elemental composition is expected. In this scenario, the second knee would be a feature of the component B.

If the proton spectrum ends below the PeV, roughly in agreement with the energy where SNR become inefficient accelerating particles, without invoking magnetic field amplification mechanisms \cite{lagage}, and with a number of gamma-ray astronomy observations \cite{hess2014}, this scenario should be reconsidered. 

Understanding the CR origin and propagation at high energy is made difficult by the poor knowledge of the elemental composition of the radiation as a function of the energy.
An integrated and statistically significant measurement of the energy spectrum, elemental composition and anisotropy in the PeV energy region can be carried out only by ground-based EAS arrays.
In fact, since the CR flux rapidly decreases with increasing energy and the size of detectors is constrained by the weight that can be placed on satellites/balloons, their collecting area is small and determines a maximum energy (of the order of a few hundred TeV/nucleon) related to a statistically significant detection. In addition, the limited volume of the detectors makes difficult the containement of showers induced by high energy nuclei, thus limiting the energy resolution of instruments in direct measurements.
Solving experimental conflicting results is essential for a comprehensive description of the CR energy spectrum up to the highest observed energies.

In this contribution we will focus on detection of Galactic CRs from ground with EAS array.
This is not a place for a complete review of CR physics (for which we recommend, for instance \cite{spurio,gaisser,grieder,longair,kampert,blasi,kachelriess}) but only to provide elements useful to understand the basic techniques used in reconstructing primary particle characteristics from ground, and to show why indirect measurements are difficult and results still conflicting.

The paper is organized as follows:

The basic facts about nature and propagation of Galactic CRs are summarized in Section 2. The sources and the mechanisms to accelerate the radiation are discussed in Section 3. Main experimental results are presented in Sections 4 and 5. Characteristics of four air shower arrays are summarized in Section 6. In Section 7 the main EAS observables are introduced and in Sections 8 and 9 the analysis techniques to reconstruct primary energy, elemental composition and anisotropy are discussed. As an example of shower analysis, the recent measurement of light (p+He) and all-particle energy spectra by the ARGO-YBJ experiment is presented in Section 10.
Lastly, the main characteristics of the new generation shower array LHAASO are introduced in the Section 11.

\section{The Nature and Propagation of Galactic Cosmic Rays}
\label{sec-2}

The study of the elemental composition of CRs is an important tool to investigate their origin and the acceleration and propagation mechanisms. 
For energies up to about 100 TeV/nucleon, the flux of different elements can be measured by \emph{"direct measurements"} carried out by experiments operated on balloons and satellites.

\begin{figure}[t]
\vfill  \begin{minipage}[t]{.48\linewidth}
  \begin{center}
    \mbox{\epsfig{file=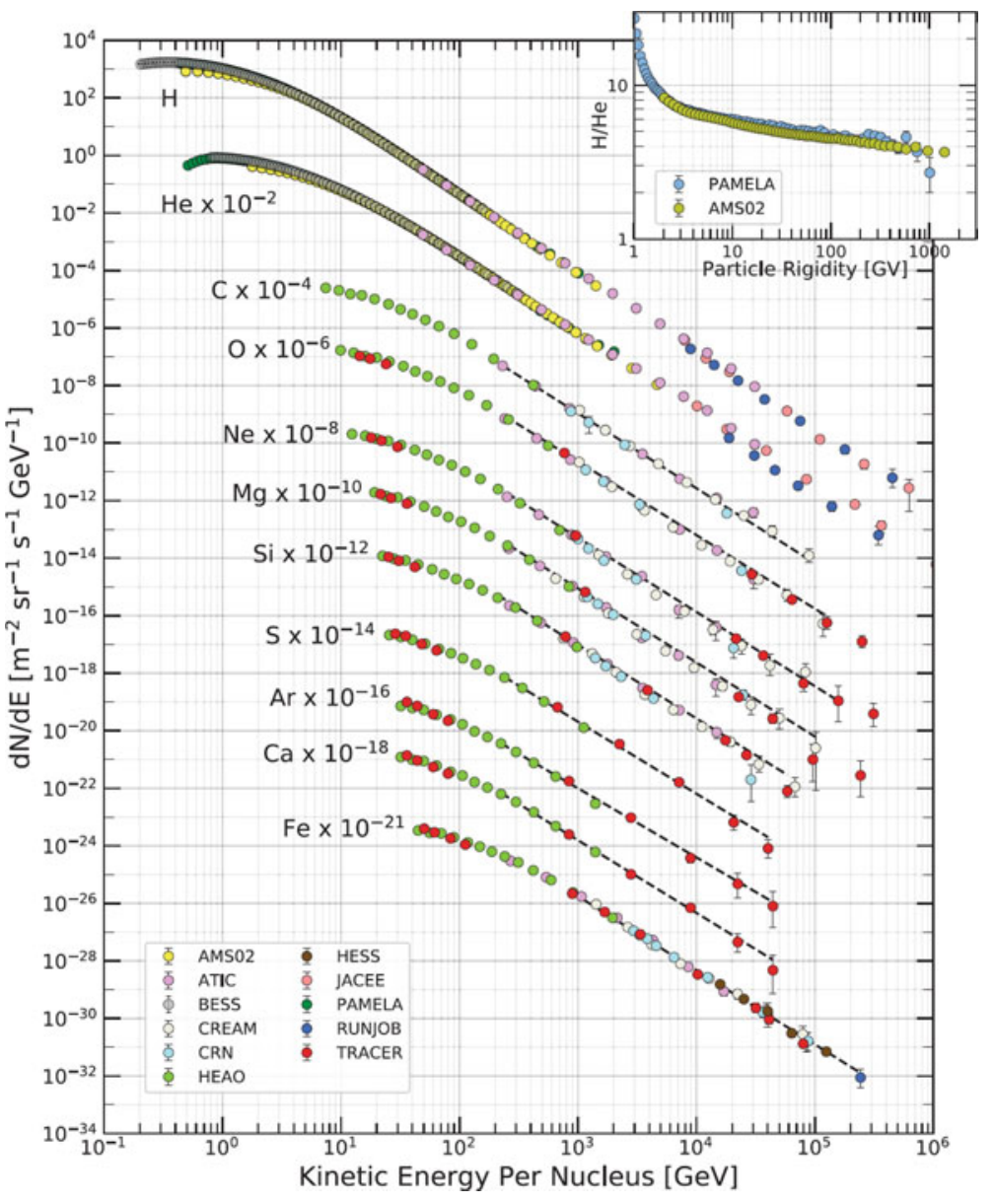,width=6cm}}
%%%    et-esiz.eps,width=6.8cm}}
  \vspace{-0.5pc}
\caption{\label{fig:cr-composition-enspectra}
Energy spectra of several components of the cosmic rays flux \cite{spurio}.}
  \end{center}
%%%\end{figure}
\end{minipage}\hfill
\hspace{-0.5cm}
\begin{minipage}[t]{.47\linewidth}
%%%%\begin{figure}[t]
%  
  \begin{center}
    \mbox{\epsfig{file=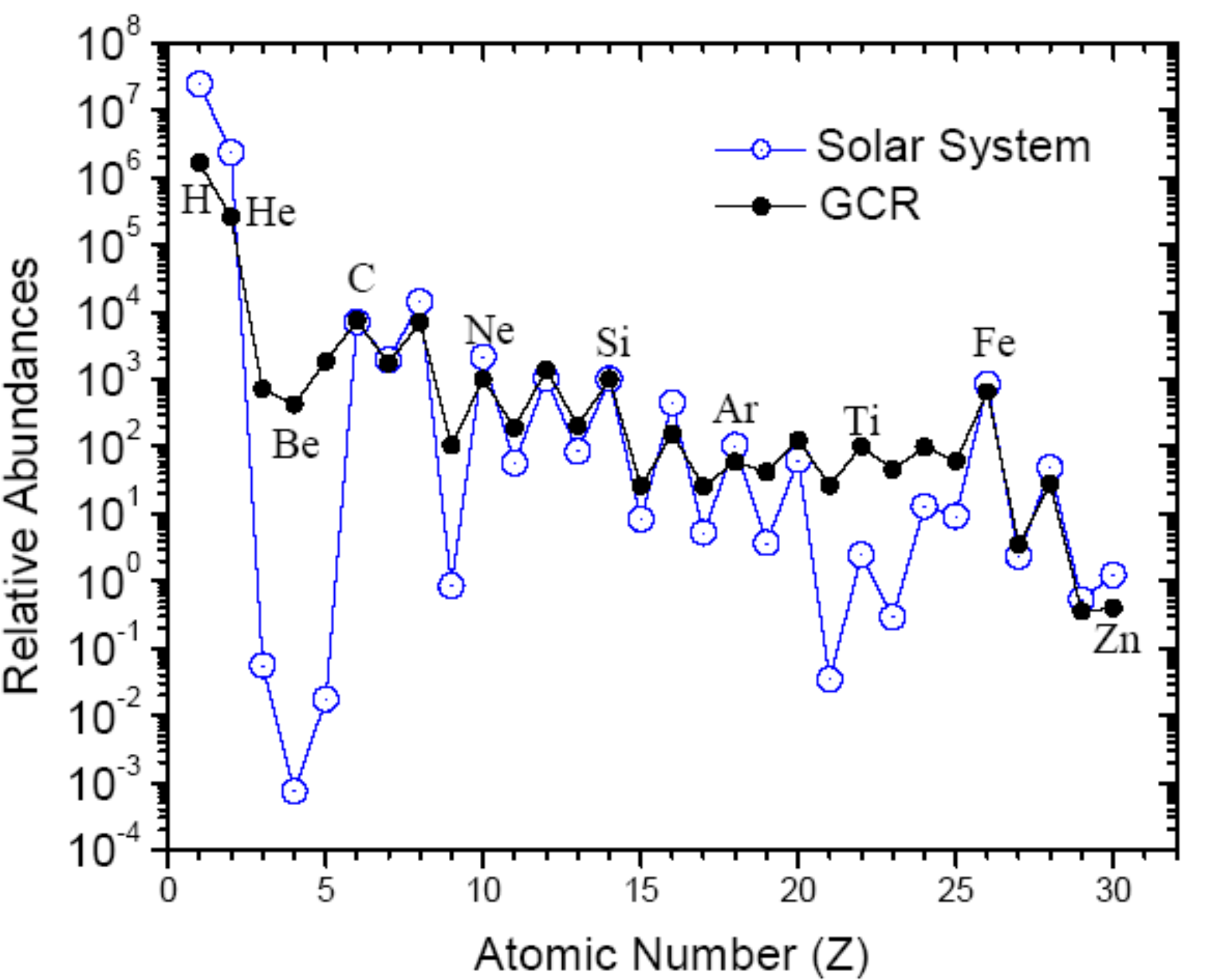,width=6cm}}
%%%%    et-musiz.eps,width=6.8cm}}
  \vspace{-0.5pc}
\caption{\label{fig:relabund-solar}
Abundance of elements measured in cosmic rays compared to the solar system abundance, from Ref. \cite{aband}.}
  \end{center}
\end{minipage}\hfill
\end{figure}

The energy spectra of several components of the CRs flux is shown in Fig. \ref{fig:cr-composition-enspectra} \cite{spurio}.

First, as it can be seen from the figure, the main component of CRs are protons, with additionally around 10\% of helium and a smaller admixture of heavier elements.

Second, the spectra shown in Fig. \ref{fig:cr-composition-enspectra} are above a few GeV power-laws, practically without any spectral features. The total CR spectrum is 
\be
 I(E) \sim 1.8\cdot E^{-\alpha} \, \frac{\rm particles}{\rm cm^2\, s\, sr\: GeV}
\ee
in the energy range from a few GeV to 100 TeV with $\alpha\approx 2.7$. 
The power-law form of the CR spectrum suggests that they are produced via \emph{non-thermal processes}, in contrast to other radiation sources.

Third, small differences in the exponent $\alpha$ of the power-law for different elements are visible. The relative contribution of heavy elements increases with energy.

Knowing the flux, we can define an energy density of CRs, assuming that these are uniformly and isotropically distributed in our Galaxy.
\be 
\rho_{CR}=\frac{4\pi\, N(\geqslant E)}{\beta c}\approx 1 \>\> {\rm eV/cm^3}
\ee
Considering that the energy density of star light is about 0.6 eV/cm$^3$, and that of the galactic magnetic field (whose average value is about 3 $\mu$Gauss) is 0.26  eV/cm$^3$, we understand how the CR share great part of the total energy available around us.
If we extrapolate such density homogeneously to the rest of the Universe, we obtain that CRs are about 1\% of the total mass of the Universe, so that they would represent by far the most important energy transformation process of the Universe! \cite{battistoni}

The relative abundance of elements measured in CRs (dark, filled circles) is compared to the one in Solar System (blue, open circles) in Fig. \ref{fig:relabund-solar}. 
%
%%%%%%%%%%%%%%%%%%%%%%%%%%%%%%%%%%%%%%%%%%%%%%%%%%%%%%%%%%%%%%%%%%%%
%%%%%%%%%%%%%%%%%%%%%%%%%%%%%%%%%%%%%%%%%%%%%%%%%%%%%%%%%%%%%%%%%%%%
%

The abundances in the solar system and in the cosmic radiation are in good agreement, which suggests, that, like in the Sun, the elements are produced by nucleosynthesis. But there are also important disagreements. 
The main difference of the two curves is that the Li-Be-B group (Z = 3-5) and the Sc-Ti-V-Cr-Mn (Z = 21-25) group are much more abundant in CRs than in the solar system. 
We explain this as a propagation effect, namely an effect due to nuclear \emph{spallation processes} occurring in the interactions of heavier nuclei with the protons of ISM between the sources of CRs and their arrival at the Earth. 
Indeed, in nucleosynthesis, these elements are only produced to a small amount. 
We can describe this process as:
\be
A + p \to A^*
\ee
with $A^*$ fragmenting in lighter nuclei. Thus, nuclei of the CNO group are likely to produce secondary Li, Be and B, while Fe can originate Sc, Ti, V, Cr or Mn. 

The question is: what is \emph{the amount of material} that CRs must cross in order to produce the observed abundances of secondary nuclei, without depleting too much the primary abundance itself? 
If the spallation cross sections are known, a measure of secondary/primary abundances can give an indication about the quantity of matter crossed from the production to the time of measure.

The evolution of abundances can be described by transport equations. 
In an approximative calculation in which only two species exist, primaries ($N_p$) and secondaries ($N_s$), and assuming that the only way the particle number changes is through spallation processes, we can write the following system \cite{battistoni}:
\begin{equation}
%%%%\begin{align*}
\label{eq:spallation}
\begin{split}
\frac{dN_p}{dX} &= - \frac{N_p}{\lambda_p} \\
 \frac{dN_s}{dX} &= - \frac{N_s}{\lambda_s} + \frac{N_p P_{sp}}{\lambda_p}
\end{split}
%%%\end{align*}
 \end{equation}
where $X$ is the amount of crossed material in g/cm$^2$, $\lambda_i$ is the interaction length for the nucleus $i$, and $P_{sp}$ is the probability of producing a given secondary from the spallation of a primary nucleus, which is in practice the ratio $\sigma_{sp}/\sigma_{tot}$.
The values of these last two parameters have to be deduced by experimental data from accelerators. 
The system Eqs. (\ref{eq:spallation}) can be solved as a function of the amount of crossed material $X$ with the initial condition $N_s(0)=0$ 
\be
 \frac{N_s}{N_p} = \frac{P_{\rm sp}\lambda_s}{\lambda_s-\lambda_p } 
 \left[ \exp\left( \frac{X_{\rm sp}}{\lambda_p}-\frac{X_{\rm sp}}{\lambda_s} \right) 
        -1 \right] \,.
\ee 
A simple calculation of the time spent in the galaxy by the CRs can be done measuring the secondary/primary ratio in the CRs.
With $\lambda_{\rm CNO}\approx 6.7\,$g/cm$^2$, $\lambda_{\rm LiBeB}\approx 10\,$g/cm$^2$, and $P_{\rm sp}\approx 0.35$ measured at accelerators, the observed ratio 0.25 is reproduced for $X_{\rm sp}\approx 4.3\,$g/cm$^2$.

If we approximate our Galaxy as a uniform thin disk of radius R = 15 kpc and thickness $h=300\:{\rm pc}\approx 10^{21}\:$cm, making the hypothesis that the matter density in the ISM around us is $\rho_{\rm ISM}\approx 1/$cm$^3$, a CR following a straight line perpendicular the disc crosses only $X=m_H\, n_H\,  h\approx 10^{-3}$g/cm$^2$. 

The value obtained for $X_{\rm sp}$ allows us to evaluate the total time (space) length that CRs spend (travel) between the source and their arrival on earth, assuming that they are mostly confined inside our galaxy. 
CRs propagate over distances of order 
\be
l_{\rm CR}=\frac{X_{\rm sp}}{m_p\cdot \rho_{\rm ISM}}\sim 3\cdot 10^{24}\>\> {\rm cm \sim1 Mpc}.
\ee
It can be easily seen that the residence time of CRs in the galaxy follows as $t={l_{\rm CR}}/{c}\approx 10^{14}$ sec $\approx 3\cdot 10^{6}$ years, which is much longer than the time to cross straightly the disk thickness. 
This result can only be explained if the propagation of CRs resembles a \emph{random-walk}.
Moreover, it suggests that acceleration and propagation can be treated separately. 

This is a very simplified picture and more general systems of coupled equations must be written to take also into account time-dependent evolution and the presence of many different species. 
However, even in this approximation, one is able to arrive at quite remarkable results. 

Summarizing, we are brought to consider the following scenario:
\begin{itemize}
\item some particles (nuclei) are produced and accelerated somewhere.
\item They leave sources and propagate in the ISM crossing $\approx$ 5 g/cm$^2$ of material.
\item During the propagation phase, these particles produce the observed abundances of light elements, through the interactions with the ISM.
\item There exists some mechanism that confines such accelerated particles in a confinement volume, possibly identified with our Galaxy, over about 10$^7$ years.
\item They lose energy in the ISM by electromagnetic processes (bremsstrahlung, Inverse Compton, synchrotron).
\item When they reach the Galaxy border they have some probability to escape the Galaxy.
\end{itemize}

In order to describe propagation of nuclei in the ISM we assume that there are sources distributed throughout the disk of the Galaxy that accelerate and inject particles of type $i$ at a rate $Q_i(E, x, t)$ per GeV per second per cm$^3$. In general the source term $Q_i$ depends on position and time, which means the energy spectrum injected by a particular source may evolve with time.

A general form of the diffusion-loss equation is the following \cite{battistoni}:

%%%\begin{equation}
\begin{align*}
\frac{dN_i(E,X,t)}{dt} = &\> Q_i(E,X,t) & \mathrm{(source\> term)} \\
                                    & + \vec{\nabla}\cdot \bigg(D_i \vec{\nabla}N_i(E)\bigg) & \mathrm{(diffusion)}\\
                                    & - \frac{\partial}{\partial E}\bigg[\frac{dE}{dt}N_i(E)\bigg] & \mathrm{(energy\> variation)}\\
                                    & - \vec{\nabla}\cdot \vec{u}N_i(E) & \mathrm{(convenction)}\\
                                    & - \bigg(\frac{v_i\rho}{\lambda_i} + \frac{1}{\gamma_i\tau_i}\bigg) N_i(E) & \mathrm{(loss\> by\> interaction\> and\> decay)}\\
                                    & + \frac{v_i\rho}{m_n} \sum_{k\ge i}\int\frac{d\sigma_{ik}(E,E')}{dE}N_k(E')dE' & \mathrm{(prod.\> of\> (i,E)\> by\> (k,E'))}
\end{align*}
%%%\end{equation}

The first line contains the injection (source) term $Q_i(E,X,t)$ of CR particles of $i$-type, while the second line describes diffusion with a diffusion coefficient $D_i$.
The third line describes continuous energy losses $\beta=d E/d t$ of a particle $i$: an important example is the synchrotron radiation. 
A convention term is contained in the forth line and the loss of particles of type $i$ by interactions or decays with $\lambda_d=\gamma_i\tau_i$ is described in the fifth line. 
The last line is the cascade term describing both the nucleonic cascade and the fragmentation processes where $\sigma_{ik}$ is the spallation cross section for nucleus $k\to i$.

\subsection{Leaky Box Model}

To describe the propagation of CRs, many models that differ mainly in the assumptions made about the source distribution and for the treatment of diffusion and convection have been elaborated.
The simplest phenomenological model is the so-called \emph{"Leaky Box Model"} (see, for example, \cite{spurio,gaisser,longair,battistoni}). 
The name derives from the fact that its main assumption is that particles diffuse freely in a confinement volume (the disc) with a small probability of escape each time they reach the boundary of the propagation region. This probability is independent on time ($\tau_{\rm esc}\gg c/h$), but is (possibly) dependent on energy. 

Diffusion and convection are replaced in the diffusion-loss equation by a characteristic escape time $N_i/\tau_{\rm esc}$. 
The approximation makes sense only if $c\tau_{\rm esc}\gg h$, so that the propagation time of a typical particle in the Galaxy is much greater than the half-thickness of the disk. 

If we consider only the diffusion term, neglecting all other effects, we obtain 
 \begin{equation}
 \frac{d N_i}{d t} =
 - \frac{N_i}{\tau_{\rm esc}} = D_i N_i \, .
  \end{equation}
This brings to an exponential distribution of the path lengths
\begin{equation}
N_i=N_0(t)\exp(-t/\tau_{\rm esc}) =N_0(x)\exp(-x/\lambda_{\rm esc}). 
\end{equation}
If we consider primary stable nuclei in a steady-state (like protons $N_p$, $dN_p/dt=0$), neglecting the fragmentation and the energy loss processes, the diffusion-loss equation becomes
\begin{equation}
\frac{N_p}{\tau_{\rm esc}}=Q_p - \bigg(\frac{\beta_p c \rho}{\lambda_p}\bigg) N_p. 
\end{equation}
Introducing $\lambda_{\rm esc}=\beta c\rho \times \tau_{\rm esc}$ as the amount of matter traversed by a particle with velocity $\beta c$ before escaping, we obtain
\begin{equation}
\label{eq:np}
N_p = \frac{Q_p\tau_{\rm esc}}{1+\lambda_{\rm esc}/\lambda_p}  \,.
\end{equation}
The escape time $\tau_{\rm esc}$ in the leaky-box model should be, similar to $D$ in the diffusion model, energy dependent. For the
simplest hypothesis that $\tau_{\rm esc}(E)$ of different elements depends only on the distance to the disc, $Z$, one obtains from a fit to data
\begin{equation}
  \lambda_{\rm esc}\approx 11 \frac{\rm g}{\rm cm^2} \left(\frac{4\,
      {\rm GV}}{R}\right)^\delta \qquad{\rm for}\quad p\geq 4\, Z\, {\rm GeV}
\end{equation}
with $\delta\approx 0.6$ and $\lambda_{\rm esc}$ = const. at lower energies.
The fit is given in terms of rigidity $R$ rather than kinetic energy per nucleon so that it can be compared to other nuclei, keeping in mind that propagation in magnetic fields is the same for different particles in terms of rigidity, but not in terms of energy per nucleon.

For protons, for which the interaction length $\lambda_p=55\,$ g/cm$^2\gg\lambda_{\rm esc}$ for all energies, only the numerator of eq. (\ref{eq:np}) is important and thus
\be
N_p = Q_p\tau_{\rm esc}\propto Q_p E^{-\delta} \,.
\ee
If the observed spectrum is $N_p \propto E^{-2.7}$ at high energy, the generation spectrum of protons should be steeper than the one observed, $Q_p\propto E^{-2.7+\delta} =E^{-2.1}$. 

For the other extreme case, the iron, the interaction length is $\lambda_{\rm Fe}=2.6\,$g/cm$^2$. 
Hence at low energies, iron nuclei are  destroyed by interactions before they escape, $\lambda_{\rm Fe}\ll\lambda_{\rm esc}$, and therefore the iron spectrum reflects the generation spectrum, $n_{\rm Fe}\propto Q_{\rm Fe}$. Starting from the energy where $\lambda_{\rm Fe}\sim\lambda_{\rm esc}$, the iron spectrum should become steeper. 
The observed iron spectrum is indeed flatter at low energies and steepens in the TeV range.

Some of the elements created in spallation processes are radioactive and hence, if the production rates of the different isotopes of a given element are known, information can be obtained about the time spent by these particles in our galaxy to reach the Earth from their sources \emph{(the mean age of the CRs}), and about the density of gas.
The most famous of these \emph{"cosmic ray clocks"} is the isotope $^{10}$Be which has a radioactive half-life of 1.5$\times$10$^6$ years, similar to the escape time found above and so is a very useful discriminant for determining the typical lifetime of the spallation products in the vicinity of the Earth.

$^{10}$Be is produced in significant quantities in the spallation of Carbon and Oxygen, the fraction of the total spallation cross-section for the production of $^{10}$Be being about 10\% of the total production cross-section of beryllium.
The $^{10}$Be nuclei undergo $\beta^-$ decays into $^{10}$B. Therefore, the relative abundances of the isotopes of $Be$ and $B$ provide a measure of whether or not all the $^{10}$Be has decayed and consequently an estimate of the mean age of the CRs observed in our vicinity.

The most precise estimate of the CRs' escape time using radioactive isotopes is due to the Cosmic Ray Isotope Spectrometer (CRIS) experiment, which was launched aboard NASA's Advanced Composition Explorer (ACE) satellite in 1997.
Averaged over the different isotopes, CRIS obtained a confinement $\tau_{CRIS}$ = 15.0 $\pm$ 1.6 My \cite{cris2001}.
From the CRs' escape time, CRIS also estimated the hydrogen number density. The average value corresponds to $n_{ISM}^{CRIS} = 0.34\pm 0.04$ H atom cm$^{-3}$.

The combination of the escape time and hydrogen number density measured by CRIS indicates an average escape length $\xi_{CRIS}\sim$ 7.6 g/cm$^{-2}$,to be compared with $X_{sp}\sim$ 5 g/cm$^{-2}$ obtained with simple estimate.

This value represents evidence that galactic CRs spend time in the galactic halo, where the matter density is lower than the canonical value assumed for the number density in the disk ($n_{ISM}\sim$ 1H atom cm$^{-3}$). A magnetic field confining CRs must therefore also be present in the galactic halo.

\section{Sources and acceleration of high energy cosmic rays}

As mentioned in the Introduction, it is widely believed that the bulk of CRs up to about 10$^{17}$ eV are Galactic, produced and accelerated by the shock waves of SNR expanding shells.

As suggested by Ginzburg and Sirovatsky \cite{ginzburg}, the SNR paradigm has two \emph{"order-of-magnitude"} arguments: firstly, the energy released in SN explosions can explain the CR energy density considering an overall efficiency of conversion of explosion energy into CR particles of the order of 10\% . 
Secondly, the diffusive shock acceleration operating in SNR can provide the necessary power-law spectral shape of accelerated particles with spectral index -2.0 that subsequently steepen to -2.7, as observed, due to the energy-dependent diffusive propagation effect.

In fact, let us define the Luminosity $L_{CR}$ of the Galaxy in terms of CRs, then:
\be
L_{CR}=\frac{\rho_{CR}\cdot V_{gal}}{\tau_{esc}}\sim\frac{1\, (eV/cm^3)\, 2\cdot 10^{66} (cm^3)}{3\cdot 10^{13} (s)}\sim 10^{41}\, \mathrm{(erg/s)}
\ee
The question then is \emph{what energy sources in the Galaxy are powerful enough to run an accelerator producing this output beam power?} The standard answer is that the only plausible energy source is the explosion of SuperNovae.

If we consider a typical SNR (the Crab Nebula), the radio observation allows to estimate the kinetic energy of accelerated electrons, which turns to be around 10$^{47}$ erg. We expect this to correspond to about 1\% of the kinetic energy of the protons, that is $\sim 10^{49}$ erg, if we assume that the value of e/p in the terrestrial environment is general. 
Therefore, if the Super Novae rate is $\sim (30 y)^{-1}$, as mainly deducted from the observation of distant galaxies, the corresponding luminosity $L_{SNR}$ would assume the value of $3\cdot 10^{40}$ erg s$^{-1}$, in reasonable agreement with $L_{CR}$ \cite{battistoni}.

The key point is that the energy has to be in a form that is capable of driving particle acceleration.

At present, the most successful description of the acceleration for the bulk of cosmic rays (up to about 10$^{14}$ eV), is the one related to shock waves from Super Novae. 
Qualitatively speaking, such an acceleration originates from the energy transfer of a moving macroscopic body (the shock wave from the Super Nova explosion) to the elementary particles or nuclei existing in the ISM, after many, small steps in which energy variation occurs. 

The acceleration mechanism is known as \emph{"First Order Fermi acceleration"}, who first considered the process of energy transfer from macroscopic regions of magnetized plasma to individual charged particles. The denomination "First Order" comes from the fact that the fractional energy increase $\eta\propto\beta$, as opposed to the ''Second Order", less efficient, acceleration in which $\eta\propto\beta^2$, as it would occur in the encounters with randomly moving magnetized clouds. 
This last scenario was the one originally considered by Fermi in his original paper, and the different result between First and Second order versions stems just from the different geometries and consequent angular averages.

The First order Fermi acceleration is able to reproduce appealing features, such as the power law spectrum of accelerated particles. We understand that once a particle crosses the shock has, after each collision, a certain probability $1 - P_{esc}$ to remain in the acceleration region and to be put back in the un-shocked region to restart an acceleration cycle having a characteristic duration time $T_{cycle}$.
It can be shown that
\be
P_{esc}=\frac{\rm rate\, into\, shock}{\rm rate\, out\, shock}\sim 4\frac{u_2}{c}
\ee
where $u_2$ is the speed with which the shocked material flows away from the shock, relatively to the shock rest frame.
After a certain number of collisions, if $P_{esc}$ and $\eta$ are energy-independent, then the \emph{integral} energy spectrum of the accelerated particles will be of the form:
\be
N(>E)\propto E^{-\alpha}
\ee
where $\alpha=P_{esc}/\eta$.
This is an important result, since, as discussed in the previous section, we expect that at source, the integral energy spectrum has to be $Q(> E)\propto E^{-\gamma+\delta+1}\sim E^{-1.1}$.

One of the most important open problems of this model concerns the difficulties in attaining particle energies up to PeV and beyond. 

\subsection{Main Characteristics of Supernova Explosion}

The average energy emitted as kinetic energy $E_c$ by a 10 $M_{\odot}$ (= 2$\times 10^{34}$ g) Supernova is roughly 1\% of the total binding energy, then for a \emph{Gravitational Energy}= 2$\times 10^{53}$ erg, we obtain $E_c$= 2$\times 10^{51}$ erg \cite{spurio}.
The velocity of the ejected mass (the shock wave) is of the order of:
\be
V_s\sim\sqrt{\frac{2\, E_c}{M}}=\sqrt{\frac{4\times 10^{51}}{2\times 10^{34}}}\sim 5\times 10^{8} \>\> {\rm cm/s}
 \to \frac{V_s}{c}\sim 2\times 10^{-2}
\ee
and corresponds to a non relativistic velocity but much larger than typical velocities of the interstellar medium.
More refined models (see \cite{hillas2005} for a recent review) assume that the velocity is higher for outer layers ($V_s/c\sim10^{-1}$), while the inner layers expand more slowly. The range of values:
\be
\frac{4}{3}\frac{V_s}{c}=\eta\sim 10^{-2} - 10^{-1}
\ee
correspond to the needed efficiency $\eta$ of the acceleration process required to explain the CRs acceleration by Supernovae explosions.

The shock front expands (we assume with constant velocity $V_s$ and with spherical symmetry) across the ISM, which has density $\rho_{ISM}\sim 1$ proton cm$^{-3}\sim 1.6\times 10^{-24}$ g cm$^{-3}$. During the expansion, the shock collects interstellar matter. When the swept-up mass is of the order of the mass of the ejected shells of the SuperNova we enter in the so-called \emph{Sedov phase}, when the shock has collected enough interstellar matter to greatly decreases its velocity. 

As the radius $R_{SN}$ of the shock front increases, the matter density $\rho_{ISM}\sim$mass/$R_{SN}^3$ inside the shock volume decreases. We assume that the shock becomes inefficient when $\rho_{SN}\sim \rho_{ISM}$. 
The radius within which the shock wave is able to accelerate particles can be derived using the condition
\be
\begin{split}
\rho_{SN} & = \frac{10 M_{\odot}}{4/3\, \pi\, R_{SN}^3} = \rho_{ISM}  \\
 & \to R_{SN} = \bigg(\frac{3\times 10 M_{\odot}}{4\pi\, \rho_{ISM}}\bigg)^{1/3} = \bigg(\frac{6\times 10^{34}}{4\pi\cdot 1.6\times 10^{-24}}\bigg)^{1/3} = 1.4\times 10^{19}\>\> {\rm cm} = 5\>\> {\rm pc}.
 \end{split}
\ee
The corresponding time interval $T_{SN}$ during which particles are accelerated is:
\be
T_{SN} = \frac{R_{SN}}{V_s} = \frac{1.4\times 10^{19}\>\> {\rm cm}}{3\times10^8\>\> {\rm cm/s}} \sim 3\times 10^{10} {\rm \>\>s}\sim \emph{O}(1000)\>\> {\rm y}
\ee
A large number of SuperNova explosions ((O(10$^4$)), with a short acceleration time duration (O(1000) y) with respect to the CR escape time (O(10$^7$)), contributes to fill the Galaxy with high energy particles.

\subsection{Maximum Energy Attainable in the Supernova Model}

There are two main parameters determining the maximum energy attainable in the SuperNova model: the finite age $T_{SN}$ and size $\lambda_{cycle}$ of the shock \cite{spurio}.

\begin{figure}
\centerline{\includegraphics[width=0.7\textwidth,clip]{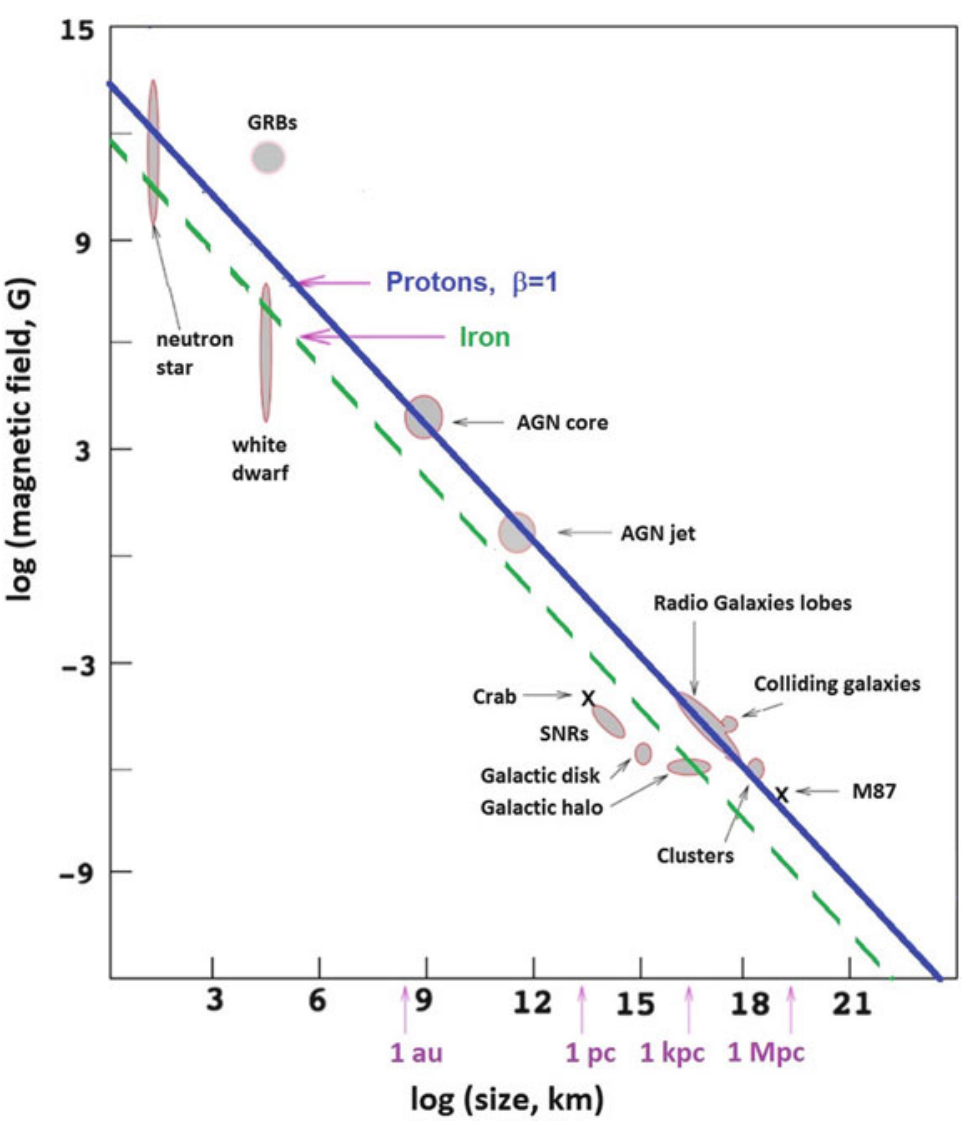} }
\caption{Hillas plot: acceleration of cosmic rays up to a given energy requires magnetic fields and sizes above the respective line. The full (dashed) line corresponds to the condition for B, L to accelerate protons (iron) at 10$^{20}$ eV. 
Some source candidates are still controversial. Plot taken from \cite{spurio}.}
\label{fig:hillas-plot}       % Give a unique label
\end{figure}

In any accelerator where the particles are magnetically confined while being accelerated the gyroradius of the particles has to be less that the size of the system. Thus for relativistic particles of momentum $p$ and energy $E = c\, p$, in any accelerator of size $R$ with magnetic fields of strength $B$ we have the following upper bound to the maximum energy attainable,
\be
r_{gyro}=\frac{p}{ZeB}=\frac{E}{ZeB}<R \to E<ZeBR.
\ee
 referred to as the \emph{Hillas limit} in reference to the well-know \emph{Hillas plot} where various astrophysical systems are plotted on a $B$, $R$ plane \cite{hillas2005} (see Fig. \ref{fig:hillas-plot}).
 %
%%%%%%%%%%%%%%%%%%%%%%%%%%%%%%%%%%%%%%%%%%%%%%%%%%%%%%%%%%
%%%%%%%%%%%%%%%%%%%%%%%%%%%%%%%%%%%%%%%%%%%%%%%%%%%%%%%%%%
%
The time between two successive scattering encounters with the shock front moving at velocity $v_s$ depends on the typical extension of the confinement region, given by the Larmor radius $r_{gyro}$, then
\be
T_{cycle}=\frac{\lambda_{cycle}}{v_s}=\frac{E}{ZeB\, v_s}.
\ee
The cycle time length depends on diffusion on both the unshocked and shocked material, which on turn depends on the strength of the magnetic field irregularities trapped in the shock, ultimately responsible of the scattering process.
The rate of energy increase is given by 
\be
\frac{dE}{dt}\sim \eta\frac{E}{T_{cycle}}\sim \eta ZeB\, v_s
\ee
Thus, the rate of energy gain is independent of the particle energy E. This is relevant, because the model is not constrained from a particular mechanism of pre-acceleration of the charged particles.

The maximum energy that a charged particle could achieve is then simply the rate of energy gain, times the duration $T_{SN}$ of the shock
\be
%%\begin{split}
E_{max}\sim\frac{dE}{dt}\times T_{SN}   = \eta\, ZeB\, R_{SN}  \sim ZeBR_{SN}\cdot\bigg(\frac{v_s}{c}\bigg) 
%% \end{split}
\ee
if the magnetic field is sufficiently tangled on the relevant scales for the scattering mean free path of charged particles to be comparable to the gyroradius, then the maximum rigidity to which particles are accelerated is of order the length scale of the system times the velocity scale times the effective magnetic field strength.
If we take fairly standard values for a SNR shock
\be
\begin{split}
E_{max} & \sim Z\cdot (4.8\times 10^{10})\cdot (4\times 10^{-6})\cdot (1.4\times 10^{19})\cdot (2\times 10^{-2})\\
 & \sim 500\cdot Z\>\> {\rm erg}\sim 300\cdot Z\>\> {\rm TeV}
\end{split}
\ee
The diffusive shock acceleration mechanism based on supernova explosions explains the spectrum of cosmic-ray protons up to few hundreds of TeV, an energy decade below the knee energy in the all-particle energy spectrum.
An important consequence is that $E_{max}$ depends on the particle charge $Z$. 
It means that a nucleus of charge $Z$ could achieve much higher total energy with respect to a proton. Thus, in this model, the knee is explained as a structure due to the different maximum energy reached by nuclei with different charge $Z$ (see Fig. \ref{fig:rigidity}).

\emph{Do SNRs Operate as PeVatrons ?}

As mentioned, according to the CR standard model, mainly driven by KASCADE results, at the knee the elemental composition of CR flux is dominated by light nuclei. This imply that SNR are expected to be able to accelerate protons up to PeV energies, namely that they are \emph{"PeVatrons"}.
To do so, significant amplification of the magnetic field at the shock is required.
But, as discussed in \cite{gabici2016}, acceleration up to PeV energies is problematic under various assumptions about the field amplification at SNR shocks. This implies that either a different (more efficient) mechanism of field amplification operates at SNR shocks, or that the sources of galactic CRs in the PeV energy range should be searched somewhere else.

The fact that the spectral measurements down to 60 MeV have enabled the identification of the $\pi^0$ decay feature in the case of IC 443 and W44 mid-aged SNRs, provided the first evidence for the acceleration of protons in SNRs. However, these two objects are far from being able to accelerate CRs up to PeV energies, and the spectral index for the $\gamma$-ray energy spectrum is much greater than 2. The quest for PeVatron galactic accelerators is still open.

\section{Experimental Results: Direct Measurements}

We can roughly divide the experimental methods adopted to measure fluxes and elemental composition of CRs into two categories: \emph{"direct"} and \emph{"indirect"} measurements. 
Generally speaking, for all particle types
\begin{itemize}
\item the higher the energy, the lower the flux;
\item the lower the flux, the larger the required detector area.
\end{itemize}
The direct measurements in principle detect and identify directly the primary particles performing experiments outside the atmosphere (stratospheric balloons, satellites) since the atmosphere behaves as a shield. 
Since the CR flux rapidly decreases with increasing energy and the size of detectors is constrained by the weight that can be placed on satellites/balloons, their \emph{"aperture"} (defined as the acceptance measured in m$^2\cdot$sr) is small and determines a maximum energy (of the order of a few hundred TeV/nucleon) related to a statistically significant detection. 
In fact, the number of detected event $N_{evts}$ is given by 
\be
N_{evts} = Flux \times Area \times Time
\ee
where \emph{Flux} is the CR flux, \emph{Area} is the detection area and \emph{Time} is the total observation time.
As it can be seen, the detection area limits the smallest measurable flux.
In addition, the limited volume of the detectors makes difficult the containement of showers induced by high energy nuclei, thus limiting the energy resolution of the instruments in direct measurements.

At higher energies instead, the flux is so low (about 1 particle/m$^2$/year in the knee region) that the only chance is to have earth-based detectors of large area, operating for long times. In that case, the atmosphere is considered as a target, and one study the primary properties in an \emph{"indirect"} way, through the measurement of secondary particles produced in the atmosphere.

In this note we will focus on ground-based measurements of Galactic CRs with EAS arrays.
But firstly, let us review, very shortly, the main achievements from direct measurements. 
Characteristics of detectors used in direct measurements are discussed in books cited in the Introduction which can be consulted for further details. In the following we refer to the description given in \cite{spurio}.

Roughly speaking, there are three kind of detectors:
\begin{itemize}
\item Totally passive detectors (emulsions. track-etch plastics, etc.).
\item Totally active detectors (wire chambers, Cherenkov light detectors, semiconductor detectors, calorimeters with different technologies, etc.). 
\item Mixed (passive+active detectors) apparata.
\end{itemize}

In the low energy region, up to about 1 GeV/n, detectors on satellites can identify individual CRs. In some case different isotopes of the same element can be separated, fully characterized by simultaneous measurements of their energy, charge, and mass (\emph{E, Z, A}). The charge and the time of flight (ToF) can be measured with the so-called $dE/dx$ method. Usually the ToF system provides also the trigger for other sub-detectors.

Experimentally more challenging is the measurement of the energy, usually obtained with a homogeneous calorimeter, selecting non-interacting stopping particles. For this reason, this technique works up to energies of a few GeV only.

In the energy range from the GeV to about 1 TeV, the energy can be measured using magnetic spectrometers or Cherenkov detectors. Individual elements are identified, characterized by their charge Z through the $dE/dx$ method. At high energy, also Transition Radiation Detectors (TRDs) are used.

In calorimeters, the particles need to be (at least partly) absorbed. Calorimeters of limited dimension have been used because of weight and size constraints of balloon and space experiments. The weight of a detector with a thickness of one hadronic interaction length and area of 1 m$^2$ amounts to about 1 ton. In some cases, multiple energy measurements are needed in order to cover the largest possible energy range and to perform a cross-calibration of detectors with different systematic uncertainties.

\begin{figure}
\centerline{\includegraphics[width=0.7\textwidth,clip]{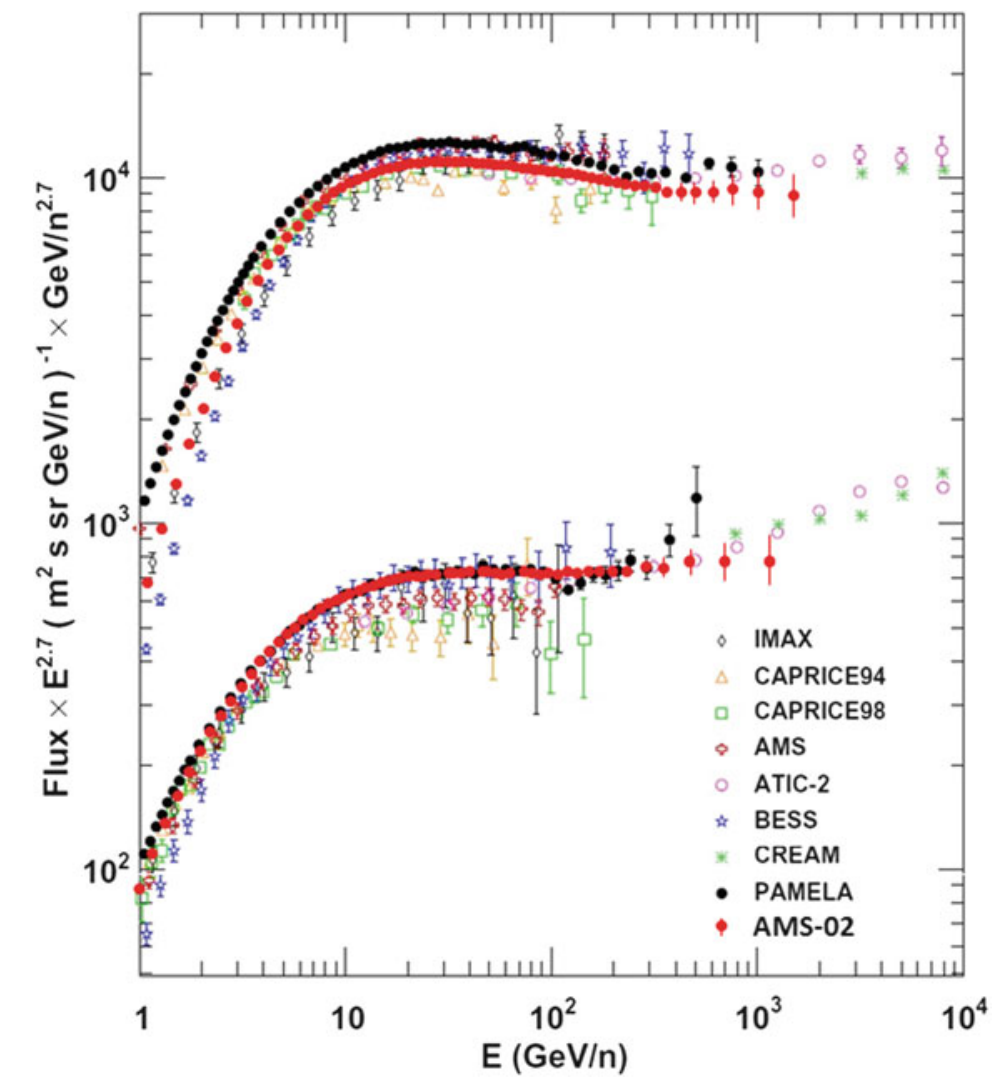} }
\caption{Proton (upper set) and helium (lower set) energy spectra above 1 GeV/n obtained by balloon-borne (CAPRICE, IMAX, BESS, ATIC-2, CREAM) and space-borne (AMS-01, AMS-02, PAMELA) experiments. Plot taken from \cite{spurio}.}
\label{fig:phe-spectra}       % Give a unique label
\end{figure}

Measurements of proton and helium primary spectra by balloon and space-borne experiments are shown in Fig. \ref{fig:phe-spectra} (for references see \cite{spurio}).
 %
%%%%%%%%%%%%%%%%%%%%%%%%%%%%%%%%%%%%%%%%%%%%%%%%%%%%%%%%%%
%%%%%%%%%%%%%%%%%%%%%%%%%%%%%%%%%%%%%%%%%%%%%%%%%%%%%%%%%%
%
Until recently the paradigm was that all the primary Galactic CRs, after correcting for solar modulation and spallation during propagation, were essentially just one feature-less power law between a few GeV per nucleon and the "knee" at around 3$\times$10$^{15}$ eV. 
Recent observations by Pamela \cite{pamelahe} and AMS02 \cite{ams02he}, confirming previuos evidences reported by ATIC \cite{atic} and CREAM \cite{creamhe}, clearly showed that this is not the case.
 These experiments have resulted in two important discoveries concerning the nuclear species:
 \begin{itemize}
 \item The Helium energy spectrum is harder than the proton one.
 \item Both spectra show a break and a spectral hardening at around a rigidity of 200 GV.
 \end{itemize}
In addition to these results, the measurements of the positron, electron and anti-proton components are also throwing up new ideas and challenges, see e.g. \cite{lipari-pantip} for further details.
The observation of a Helium energy spectrum harder than the proton one has the interesting consequence that the knee region could be dominated by Helium and CNO masses with a proton knee below the PeV, as suggested by the ARGO-YBJ results (Fig. \ref{fig:phe-allp}) \cite{hybrid15}.

\section{Experimental Results: Indirect Measurements}

Approaching the hundred TeV energy region, even in space-borne experiments, the energy assignment is indirect since it is generally based on the energy deposition of particles produced in the interaction of primaries in the detector itself. The reconstruction of the total energy is then obtained by comparison with some model prediction, and therefore, at least in that region, the boundary line between "direct" and "indirect" experiments is more uncertain.

At ground the study of cosmic rays is based on the reconstruction and interpretation of Extensive Air Showers (EAS) observables, mainly electromagnetic component, muon and hadron components, Cherenkov photons, nitrogen fluorescence, radio emission.
Therefore, different detectors must be used to detect different observables.

\begin{figure}
\centerline{\includegraphics[width=0.7\textwidth,clip]{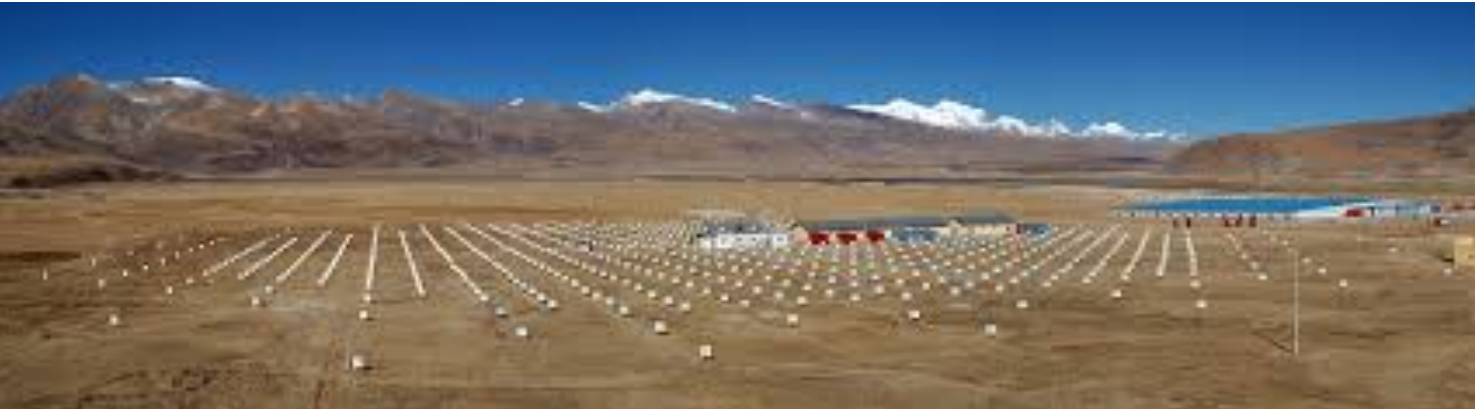} }
\caption{Example of a typical air shower array (Tibet AS$\gamma$ experiment located at the YangBaJing Cosmic Ray Observatory in Tibet (P.R. China) 4300 m asl.)}
\label{fig:array}       % Give a unique label
\end{figure}

Two different approaches are exploited:
\begin{itemize}
\item \emph{Arrays}, to sample the shower tail particles reaching the ground. In High Energy Particle language, a shower array is a \emph{"Tail Catcher Sampling Calorimeter"}. The atmosphere is the absorber and the detectors at ground are the device to measure a (poor) calorimetric signal. Arrays are wide field of view detectors able to observe all the overhead sky with a duty cycle of $\sim$100\%. Measurements are limited by large shower-to-shower fluctuations.
\item \emph{Telescopes}, to detect Cherenkov photons or nitrogen fluorescence and observe the EAS longitudinal profile. The atmosphere acts as a \emph{"Homogeneous Calorimeter"}. The duty cycle is low ($\sim$10-15\%) because telescopes can be operated only during clear moonless nights and the field of view small (a few degrees). On the contrary, pointing capability and energy resolution are excellent.
\end{itemize}
Shower arrays are made by a large number of detectors (scintillators or water Cherenkov tanks, for example) distributed over very large areas, of order of 10$^{5}$ m$^2$ (see Fig. \ref{fig:array}). The shower \emph{"size"}, the total number of charged particles, and the shower arrival direction are the two key parameters reconstructed by arrays.
%
%%%%%%%%%%%%%%%%%%%%%%%%%%%%%%%%%%%%%%%%%%%%%%%%%%%%%%%%%%
%%%%%%%%%%%%%%%%%%%%%%%%%%%%%%%%%%%%%%%%%%%%%%%%%%%%%%%%%%
%

One of the main characteristics of arrays is the \emph{coverage}, the ratio between the total sensitive area of the detectors and the instrumented area. In classical arrays this ratio is very small, of order 10$^{-2}$ - 10$^{-3}$, precluding the possibility to study details of the lateral distribution, as for example the shower core region. In addiction, the coverage is one of the parameters determining the energy threshold of an array. In fact, low energy primaries produce small showers that can be detected only by arrays with high coverage.
%energy threshold?????

\subsection{Extensive Air Showers: the Heitler-Matthews model}

A general idea of the main characteristics of Extensive Air Showers and of how different mass primaries produce showers with different properties can be obtained from some relatively simple arguments, as suggested by Heitler \cite{heitler} and Matthews \cite{matthews}.

\begin{figure}[h]
\vfill  \begin{minipage}[h]{.47\linewidth}
  \begin{center}
    \mbox{\epsfig{file=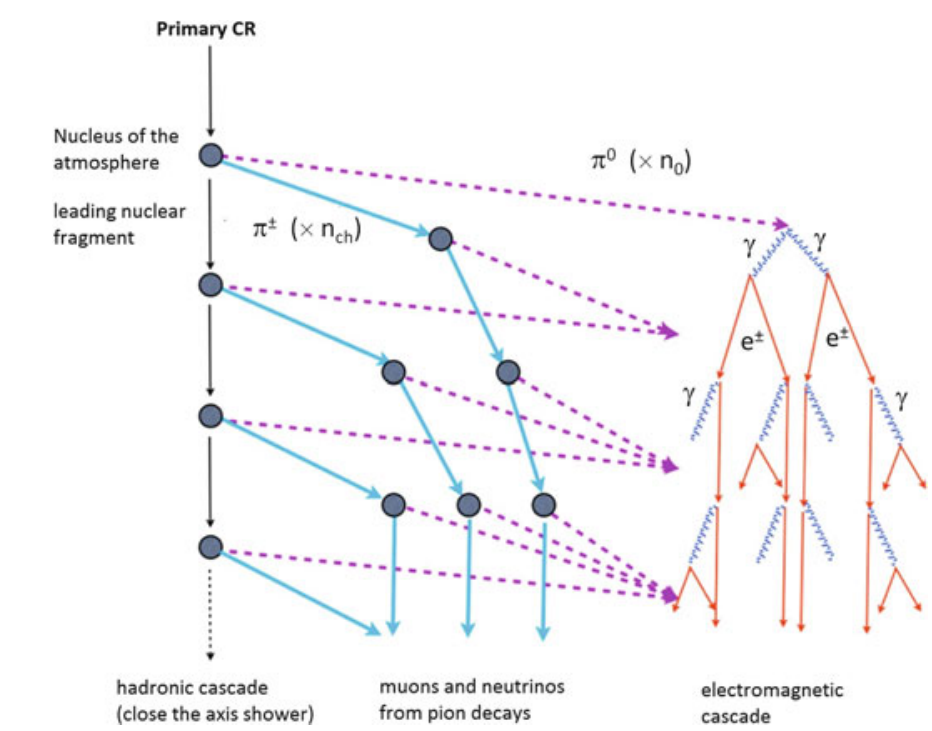,width=6.8cm}}
  \vspace{-0.5pc}
 \caption{Schematic evolution of cascades initiated by hadrons \cite{spurio}. At each step, roughly 1/3 of the energy is transferred from the hadronic cascade to the electromagnetic one.
    \label{fig:eas-schema}}
  \end{center}
%%%\end{figure}
\end{minipage}\hfill
\hspace{-1.cm}
\begin{minipage}[h]{.47\linewidth}
%%%%\begin{figure}[t]
  \begin{center}
    \mbox{\epsfig{file=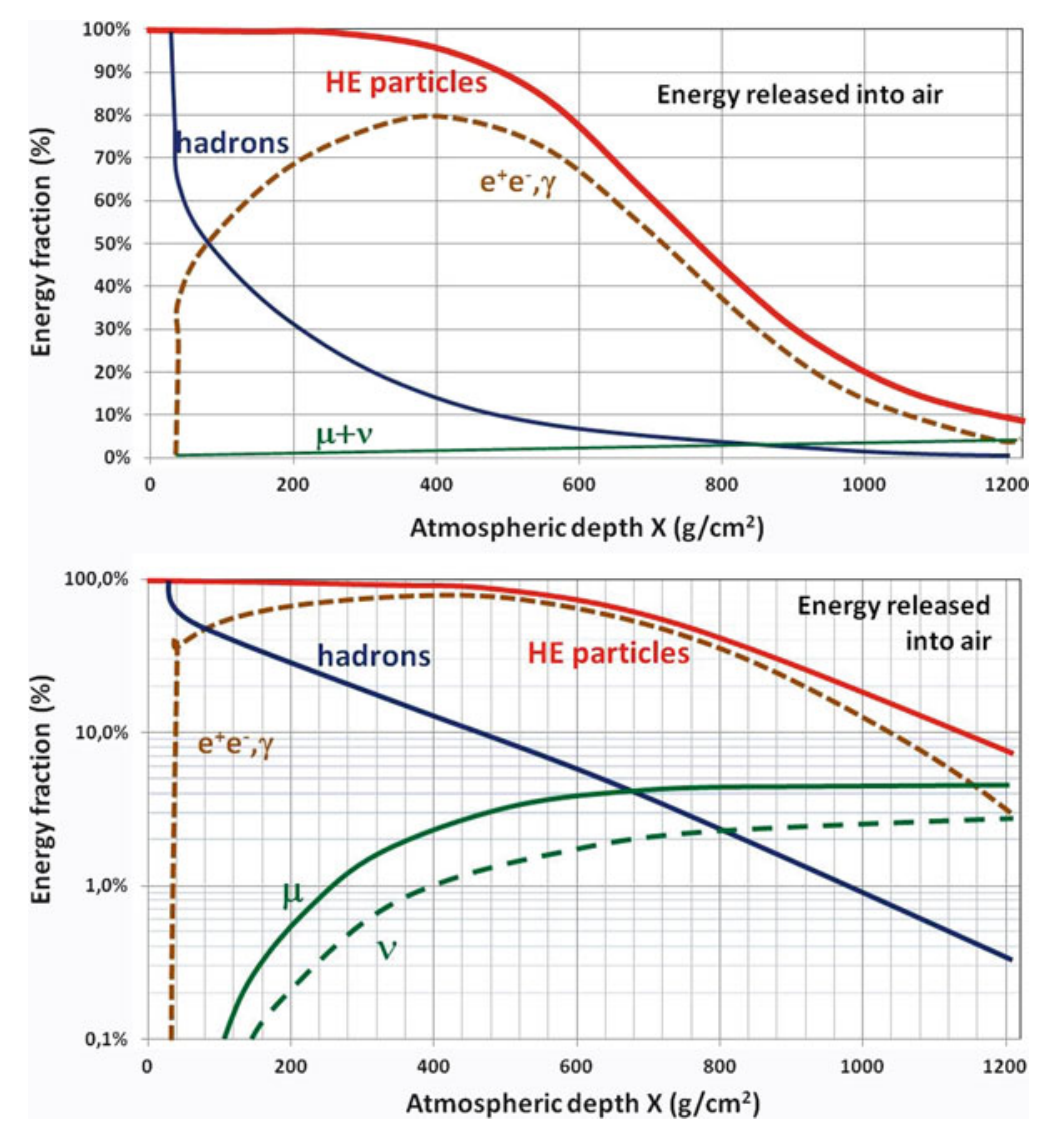,width=6.8cm}}
  \vspace{-0.5pc}
    \caption{Fraction of energy transferred to the different components of the cascade induced by a primary proton of 10$^{19}$ eV \cite{spurio}. Part of the energy is released into air by excitation/ionization processes. The top graph uses a linear scale for the energy fraction; the bottom uses a log scale for a better image of the "older part of the shower.
 \label{fig:eas-energy}}
  \end{center}
\end{minipage}\hfill
\end{figure}

The collision of a primary CR with a nucleus of the atmosphere produces one large nuclear fragment and many charged and neutral pions (with a smaller number of kaons) (Fig. \ref{fig:eas-schema}). A significant fraction of the total energy is carried away by a single \emph{"leading"} particle. This energy is unavailable immediately for new particle production. 
Roughly speaking, half of the energy of the primary particle is transferred to the nuclear fragment and the other half is taken by the pions (and kaons). The fraction of energy transferred to the new shower particles is referred as \emph{inelasticity}. Accurate description of the leading particles is crucial because these high-energy nucleons feed energy deeper into the EAS.

Approximately equal number of positive, negative and neutral pions are produced.
The neutral pions immediately (8.4$\times$10$^{-17}$ s) decay in a pair of photons. These photons, producing electron and positron pairs, induce different electromagnetic sub-showers, the most intense component of an EAS, through multiplicative processes (mainly pair production and bremsstrahlung). At each interaction before the charged pions decay, nearly a third of the energy of this hadronic component is released into the electromagnetic component. 
As the number of particles increase, the energy per particle decreases. They will also scatter, losing energy, and many will range-out. 
Thus, the number of particles (or, with less ambiguities in the definition, the quantity of energy transferred to secondaries and eventually released in the atmosphere) will reach a maximum at some depth $X_{max}$ which is a function of energy, of the nature of the primary particle and of the details of the interactions of the primaries and secondaries in the cascade. 
After that, the energy/particle is so degraded (will be below some \emph{"critical energy"}) that energy losses dominate over particle multiplication process, and the shower "size" will decrease as a function of depth: it grows "old". The critical energy is process dependent: for instance, for low energy electrons the relevant energy is that at which energy losses by ionization become important, while for e.g. underground muons is the energy necessary to produce a muon capable to penetrate the rock through the detector.

Figure \ref{fig:eas-energy} shows the energy fraction (both in linear and logarithmic scales) of the electromagnetic, hadronic, muonic and neutrino components as functions of the atmospheric depth, as obtained with the CORSIKA Monte Carlo simulation for a primary proton with E$_0$ = 10$^{19}$ eV. The energy released into air refers to the energy fraction transferred from high-energy particles to the excitation and ionization of the medium. As you can see, at sea level 90\% of the primary energy of the CR particle is dissipated in the atmosphere during the shower development! 

The charged pions will decay as well, with a longer lifetime (2.6$\times$10$^{-8}$ s) allowing many charged pions to interact with the atmosphere, perhaps multiple times, before decaying. At each interaction more pions are created, again in equal numbers of positive, negative and neutral.
Once the pions have reached low enough energy, they will decay into muons and neutrinos ($\pi^+\to\mu^+\nu_{\mu}$ or $\pi^-\to\mu^-\bar{\nu_{\mu}}$). This resulting muons propagate unimpeded to the ground. 
The muon cascade grows and maximizes, but the decay is slower as a consequence of the relative stability of the muon and small energy losses by ionization and pair production.

%
%%%%%%%%%%%%%%%%%%%%%%%%%%%%%%%%%%%%%%%%%%%%%%%%%%%%%%%%%%%%%%%%%%%%%
%%%%%%%%%%%%%%%%%%%%%%%%%%%%%%%%%%%%%%%%%%%%%%%%%%%%%%%%%%%%%%%%%%%%
%

These are the most common processes, but not at all the only ones. As an example, successive hadronic interactions of the primary cosmic ray, interactions/decays of kaons and muon decays must be also considered.
Therefore, detailed simulations must be used to describe all the characteristics of these random processes.

These atmospheric showers are known as Extensive Air Showers (EAS). Their longitudinal evolution is a function of the nature and energy of the primary particle. Their lateral extension depends on the average transverse momentum of the hadronic component, and, in the case of the electro-magnetic component (which, speaking in number of particles, is the most important one) it is strongly affected by the multiple Coulomb scattering. 

The main features of an  electromagnetic shower profiles can be described within the simple Heitler's toy model of particle cascades \cite{heitler}.

\begin{figure}[ht]
\centering
\includegraphics[scale=0.90]{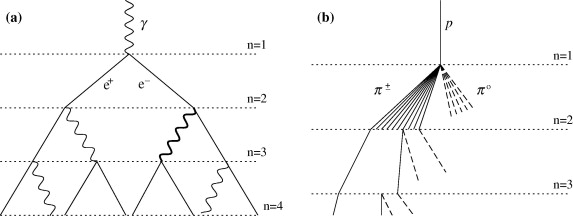}
\caption{Schematic view of an electromagnetic cascades (a), and of a hadronic shower (b). In the hadron shower, dashed lines show $\pi^0$ which do not re-interact but decay producing e.m. sub-showers. Figure taken from \cite{matthews}}.
\label{fig:eastoymodel}       % Give a unique label
\end{figure}

Let us suppose that a particle (electron, positron or photon) with energy $E_0$ splits its energy equally into two particles after traveling a radiation length $X_0$ in the air, and let this process be repeated by the secondaries (see Fig. \ref{fig:eastoymodel}).
%
%%%%%%%%%%%%%%%%%%%%%%%%%%%%%%%%%%%%
%%%%%%%%%%%%%%%%%%%%%%%%%%%%%%%%%%%%
%
Let $X$ describe the depth in the atmosphere and define the depth at which the average CR interacts with the atmsphere to be $X$ = 0 g/cm$^2$.
After $n$ radiation lengths we obtain a particle cascade which has evolved into $N$ = 2$^n$ particles of equal energy $E$ = E$_0/N$.
Multiplication stops when the energies of the particles are too low for pair production or bremssthralung. This energy is the critical energy $E_C^{em}$ in the air (= 80 MeV, below which the collisional energy losses are dominant).

The maximum number of particles $N_{max}$ is reached at this moment, when all particles have the same energy $E_C^{em}$, $E_0$ = $E_C^{em} \cdot N_{max}$.
The depth $X_{max}$ at which the shower reaches the maximum size is  $X_{max}$ = $n_{max} \cdot X_0$, where $n_{max}$ is the number of radiation lengths required for the primary energy to be reduced to $E_C^{em}$.

Since $N_{max}$ = $2^{n_{max}}$, we have
\begin{equation}
n_{max} = ln\, \bigg(\frac{E_0}{E_C^{em}}\bigg) \cdot \frac {1}{ln\, 2}
\end{equation}
so that
\begin{equation}
\label{eq:xmax}
X^{em}_{max} = \frac{X_0}{ln\, 2} \cdot ln\, \bigg(\frac{E_0}{E_C^{em}}\bigg).
\end{equation}
Finally, it is interesting to estimate the \emph{elongation rate} $\Lambda$, i.e. the rate of increase
of $X_{max}$ with the primary energy
\be
\Lambda^{em}=\frac{d\, X_{max}}{d\, log_{10}\, (E_0)}.
\ee
From the relation (\ref{eq:xmax}), we have $\Lambda^{em}$ = 2.3$\cdot X_0$ = 85 g/cm$^2$ per decade of energy.

This simple model predicts two basic features of e.m. shower development:
\begin{itemize}
\item $N_{max}$ increases proportional to the primary energy $E_0$
\item $X_{max}$ increases logarithmically with primary energy, at a rate of 85 g/cm$^2$ per decade of energy.
\end{itemize}

Air showers initiated by protons have been modeled by Matthews \cite{matthews} following an approach similar to the Heitler's one.
Protons travel one interaction length and interact producing $N_{ch}$ charged pions and $\frac{1}{2}\cdot N_{ch}$ neutral pions, which immediately decays into photons, initiating an e.m. shower.
As for the e.m. cascade we assume equal division of energy during particle production.

After $n$ interactions the $N_{\pi}$ = $(N_{ch})^n$ charged pions produced carry a total energy of $(\frac{2}{3})^n \cdot E_0$. 
The remainder of the primary energy goes into e.m. showers from $\pi^0$ decays. 
The energy per charged pions after $n$ interaction is then $E_{\pi}$ = $\frac{E_0}{(3/2 N_{ch})^n}$.

The process stop when the pions energy fall below the critical energy $E_C^{\pi}$ where they decay to muons. 
The number of muons is $N_{\mu}$ = $N_{\pi}$ = $(N_{ch})^{n_{max}}$, where $n_{max}$ is the number of interaction length required for the charged pion's interaction length to exceed its decay length $n_{max}=\frac{ln(E_0/E_C^{\pi})}{ln(3/2N_{ch})}$.
Thus the total energy is divided into two channels, hadronic and electromagnetic
\begin{equation}
E_0 = E_C^{e.m.}\cdot N_e + E_C^{\pi}\cdot N_{\mu}.
\end{equation}
This equation represents energy conservation. The relative magnitude of the contribution from N$_{\mu}$ and N$_e$ does not depend on the details of the model, but only on the respective critical energies, the energy scales at which e.m. and hadronic multiplication cease.
An important conclusion of the Matthews description of the hadronic cascades is that the energy is given by a linear combination of muon and electron sizes. This result is insensitive to fluctuations in the division of energy between the hadronic and electromagnetic channels and independent on the mass of the primary particle.

The muon size is given by
\begin{equation}
ln\, N_{\mu}=ln\, N_{\pi} = n_C ln\, N_{ch} = ln\, N_{ch}\cdot \frac{ln\, (E_0/E_C^{\pi})}{ln\, (3/2 N_{ch})}
= \beta \cdot ln\, \bigg(\frac{E_0}{E_C^{\pi}}\bigg).
\end{equation}
Following \cite{matthews} we can estimate $\beta = 0.85$ for $E_0$ in the range 10$^{14}$ - 10$^{17}$ eV.
As a consequence we obtain
\begin{equation}
N_{\mu} = \bigg( \frac{E_0}{E_C^{\pi}}\bigg)^{0.85} \sim 9900 \bigg(\frac{E_0}{10^{15} eV}\bigg)^{0.85}
\end{equation}
in good agreement with the results of detailed MC simulations.
Therefore, the muon size grows with primary energy more slowly than proportionally.

In the first hadronic interaction $\frac{1}{3} E_0$ goes into e.m. channel via $\pi^0$ decay. 
This interaction occurs at an atmospheric depth $X^* = \lambda_I \cdot ln\, 2$ = 59 g/cm$^2$, where in this case $\lambda_I$ is the interaction length of the primary proton, and produces $\frac{1}{2N_{ch}}\pi^0$ decaying into $N_{ch}$ $\gamma$'s, each with energy $E_0/3 N_{ch}$.
All these photons initiate parallel e.m. sub-showers. 

From equation (\ref{eq:xmax}) we have
\begin{align*}
X^p_{max} & = X^* + X_0 \cdot ln\, \bigg(\frac{E_0}{3 N_{ch}\cdot E_C^{e.m.}}\bigg) \\
 & = X^* + X_{max}^{e.m.} - X_0 \cdot ln\, (3\, N_{ch})\>\>\>\>\>\>{\rm g/cm^2}
\end{align*}
where $X_{max}^{e.m.}$ is the atmospheric depth of the maximum of $\gamma$-induced showers with $E_0$ primary energy and $N_{ch}$ is the multiplicity of charged pions in the first interaction.
The elongation rate for showers induced by protons is then 
\be
\Lambda^p = \Lambda^{\gamma} + \frac{d}{d\, log_{10} E_0}\bigg[ X^* - X_0\cdot ln\, (3\, N_{ch})\bigg] = 58 \>\>\>\>\>\>{\textrm{g/cm$^2$ per decade,}}
\ee
reduced from the elongation rate for purely electromagnetic showers. This estimation verifies Linsley's elongation rate theorem \cite{linsley77}, which point out that e.m. showers represent an upper limit to the elongation rate of the hadronic showers.

In the framework of the \emph{superposition model} each nucleus is taken to be equal to $A$ individual single nucleons, each with energy $E_0/A$ and each acting independently.
The shower resulting from the interaction of the primary nucleus $A$ can be treated as the sum of $A$ proton induced independent showers all starting at the same point.
Thus, while a proton creates one shower with energy $E_0$, an iron nucleus of the same total energy is expected to create the equivalent of 56 proton showers, each with energy $E_0/56$. The average properties of showers are well reproduced by this model, though the fluctuations are clearly underestimated. 
By substituting  the lower primary energy $({E_0}/A)$ into the previous expressions and summing $A$ such showers we obtain the following relations for a shower induced by a nucleus $A$
\begin{equation}
X_{max}^A = X_{max}^p - X_0\cdot ln\, A
\end{equation}
\begin{equation}
N_{\mu}^A = N_{\mu}^p\cdot A^{0.15}
\end{equation}
\begin{equation}
\label{eq:primener}
E_0= E_C^{e.m.}\cdot N_e + E_C^{\pi}\cdot N_{\mu}
\end{equation}
The toy model predictions can be summarized as follows:
\begin{itemize}
\item $X_{max}$ is smaller for heavier nuclei (logarithmic dependence on $A$)
\item $X_{max}$ is the same for same $E_0/A$ but different $E_0$. As a consequence, the proton-induced showers result, on average, in a larger number of particles at the observation level compared to iron-induced events. Thus, a measure of $X_{max}$ is an important feature for
mass discrimination. But the shower-to-shower fluctuations are as large as the shift of $X_{max}$ between proton and iron. This limits an event-by-event assignement of a primary mass.
\item Nuclear showers have more muons than proton showers, at the same total primary energy. In fact, due to the smaller energy per nucleon ($E_0/A$), the secondary pions are less energetic. This favours a pion decay as well as the fact that heavier nuclei interact higher in atmosphere, where the air density is smaller.
\item The energy relation (\ref{eq:primener}) is independent from $A$ because it intrinsically accounts for all of the primary energy being distributed into a hadronic channel and into e.m. showers.
\end{itemize}
Despite the simplicity and limitations of this model, the findings of detailed MC shower simulations are quite well reproduced.

\section{EAS Experiments}

Modern air shower arrays are designed to measure simultaneously different shower observables to study their correlations. 
In Tables~\ref{tab:one} and \ref{tab:one-prime} the characteristics of air shower arrays operated in the last two decades to study Galactic CR physics from ground are summarized.
%
%%%%%%%%% Tables 1-2 %%%%%%%%%%%%%%%%%%%%%%%%%%%%%%%%%%%%%%%%%%%%
\begin{table}[h]
\caption{\label{tab:one} Characteristics of different EAS-arrays}
\begin{center}
\begin{tabular}{ccccccc}
\br
  Experiment &  g/cm$^2$ & Detector & $\Delta$E & e.m. sens. & Instr. & Coverage \\
                     &                  &                &       (eV)       &         area (m$^2$)             & area (m$^2$) & \\
\mr
ARGO-YBJ  & 606            & RPC/hybrid    & $3\cdot 10^{11} - 10^{16}$ & 6700 & 11,000 & 0.93 \\
                    &                   &      &                                            &          &  & (c.c.)\\
BASJE-MAS & 550           & scint./muon & $6\cdot 10^{12} - 3.5\cdot 10^{16}$ &    & $10^4$ &  \\
 TIBET AS$\gamma$ & 606 &  scint./burst det.  & $5\cdot 10^{13} - 10^{17}$ & 380 & 3.7$\times$10$^4$ & 10$^{-2}$ \\
 CASA-MIA & 860 & scint./muon & 10$^{14} - 3.5\cdot 10^{16}$ & 1.6$\times$10$^3$ & 2.3$\times$10$^5$ & 7$\times$10$^{-3}$ \\
KASCADE & 1020 & scint./mu/had & $2 - 90\cdot 10^{15}$ & 5$\times$10$^2$ & 4$\times$10$^4$ & 1.2$\times$10$^{-2}$ \\
KASCADE-& 1020 & scint./mu/had  & $10^{16} - 10^{18}$ & 370 & 5$\times$10$^5$ & 7$\times$10$^{-4}$ \\
Grande &  &  &  &  &  &  \\
Tunka & 900 & open Ch. det. & 3$\cdot 10^{15} - 3\cdot 10^{18}$ & --- & 10$^6$ & --- \\
IceTop & 680 & ice Ch. det. & $10^{16} - 10^{18}$ & 4.2$\times$10$^2$ & 10$^6$ & 4$\times$10$^{-4}$ \\
 LHAASO & 600 & Water C& $10^{12} - 10^{17}$ & 5.2$\times$10$^3$ & 1.3$\times$10$^6$ & 4$\times$10$^{-3}$ \\
                    &        & scint./mu/had     &                                            &          &  & \\
                    &        & wide-FoV Ch. Tel.     &                                            &          &  & \\
\br
\end{tabular}
\end{center}
\end{table}

\begin{table}[h]
\caption{\label{tab:one-prime} Characteristics of different EAS-arrays (cont.)}
\begin{center}
\begin{tabular}{ccccc}
\br
 Experiment &Altitude & $\mu$ Sensitive Area & Instrumented Area & Coverage \\
  & (m) & (m$^2$) & (m$^2$) & \\
\mr
LHAASO & 4410 & 4.2$\times$10$^4$ & 10$^6$ & 4.4$\times$10$^{-2}$ \\
 TIBET AS$\gamma$ & 4300 & 4.5$\times$10$^3$ & 3.7$\times$10$^4$ & 1.2$\times$10$^{-1}$ \\
 KASCADE & 110 & 6$\times$10$^2$ & 4$\times$10$^4$ & 1.5$\times$10$^{-2}$ \\
 CASA-MIA & 1450 & 2.5$\times$10$^3$ & 2.3$\times$10$^5$ & 1.1$\times$10$^{-2}$ \\
\br
\end{tabular} 
\end{center}
\end{table}
%%%%%%%%%%%%%%%%%%%%%%%%%%%%%%%%%%%%%%%%%%%%%%%%%%%%%%%%%
%

In the following the main characteristics of EAS-TOP, KASCADE, HEGRA experiments, the first multi-component arrays, and of ARGO-YBJ, the first and so far only full-coverage apparatus, will be described.

\subsection{EAS-TOP experiment}

\begin{figure}[h]
\vfill  \begin{minipage}[h]{.47\linewidth}
  \begin{center}
    \mbox{\epsfig{file=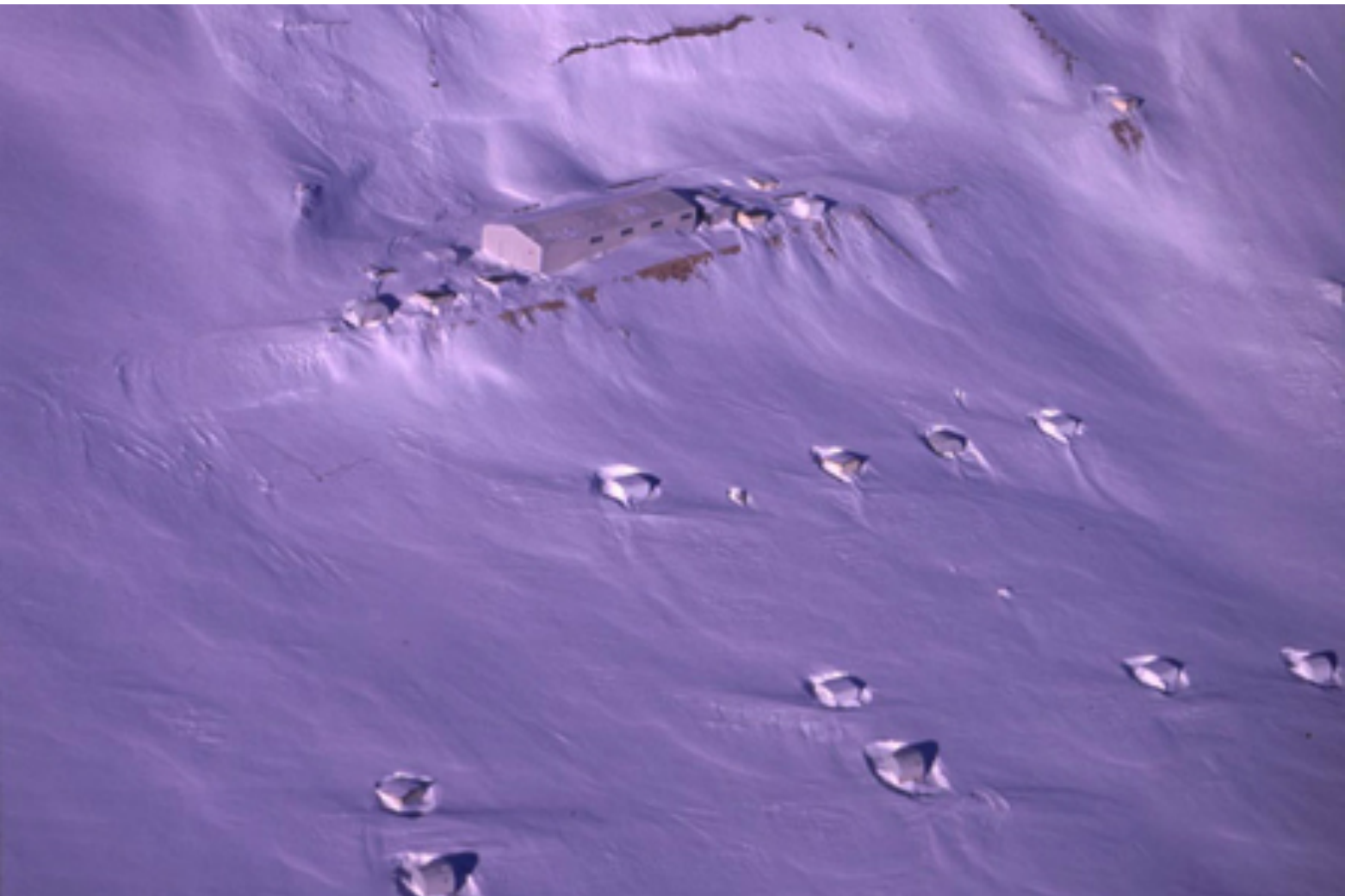,width=6.8cm}}
  \vspace{-0.5pc}
 \caption{Aereal view of the EAS-TOP array.
    \label{fig:eastop-array}}
  \end{center}
%%%\end{figure}
\end{minipage}\hfill
\hspace{-1.cm}
\begin{minipage}[h]{.47\linewidth}
%%%%\begin{figure}[t]
  \begin{center}
    \mbox{\epsfig{file=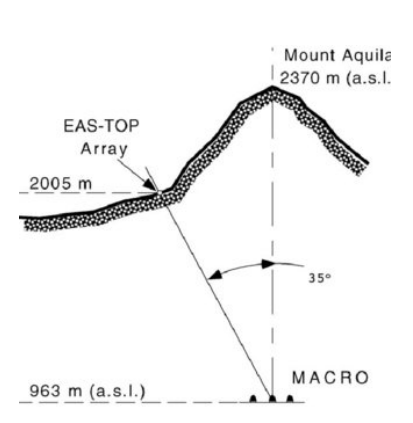,width=6.8cm}}
  \vspace{-0.5pc}
    \caption{Location of the EAS-TOP experiment.
 \label{fig:eastop-location}}
  \end{center}
\end{minipage}\hfill
\end{figure}

The EAS-TOP experiment has been in operation between January 1989 and May 2000 to study the CR physics in the energy range
10$^{12}$ -- 10$^{16}$ eV. The array was located at Campo Imperatore, INFN Gran Sasso National Laboratories (Italy), 2000 m
a.s.l., 810 g/cm$^2$ atmospheric depth (Fig. \ref{fig:eastop-array}).
It consists of detectors of the different components of EAS:
electromagnetic particles, muons, hadrons and atmospheric Cerenkov light \cite{eastop_gen}. 
Moreover, its location has been chosen to have the further possibility of running in coincidence with the muon detectors MACRO and LVD experiments operating inside the deep underground Gran Sasso Laboratories. The sites are separated by $\Delta h = 1000$ m in altitude
(corresponding to $\sim$ 3000 m w.e., and $E_{\mu}>1.4$ TeV). The relative zenith angle is 
$\langle\theta\rangle \sim 30^{\circ}$ (Fig. \ref{fig:eastop-location}).
%
%%%%%%%%%%%%%%%%%%%%%%%%%%%%%%%%%%%%%%%%%%%%%%%%%%%%%%%%%%%%%%%%%%%%%
%%%%%%%%%%%%%%%%%%%%%%%%%%%%%%%%%%%%%%%%%%%%%%%%%%%%%%%%%%%%%%%%%%%%
%

\begin{figure}[t]
\centering
\includegraphics[scale=0.90]{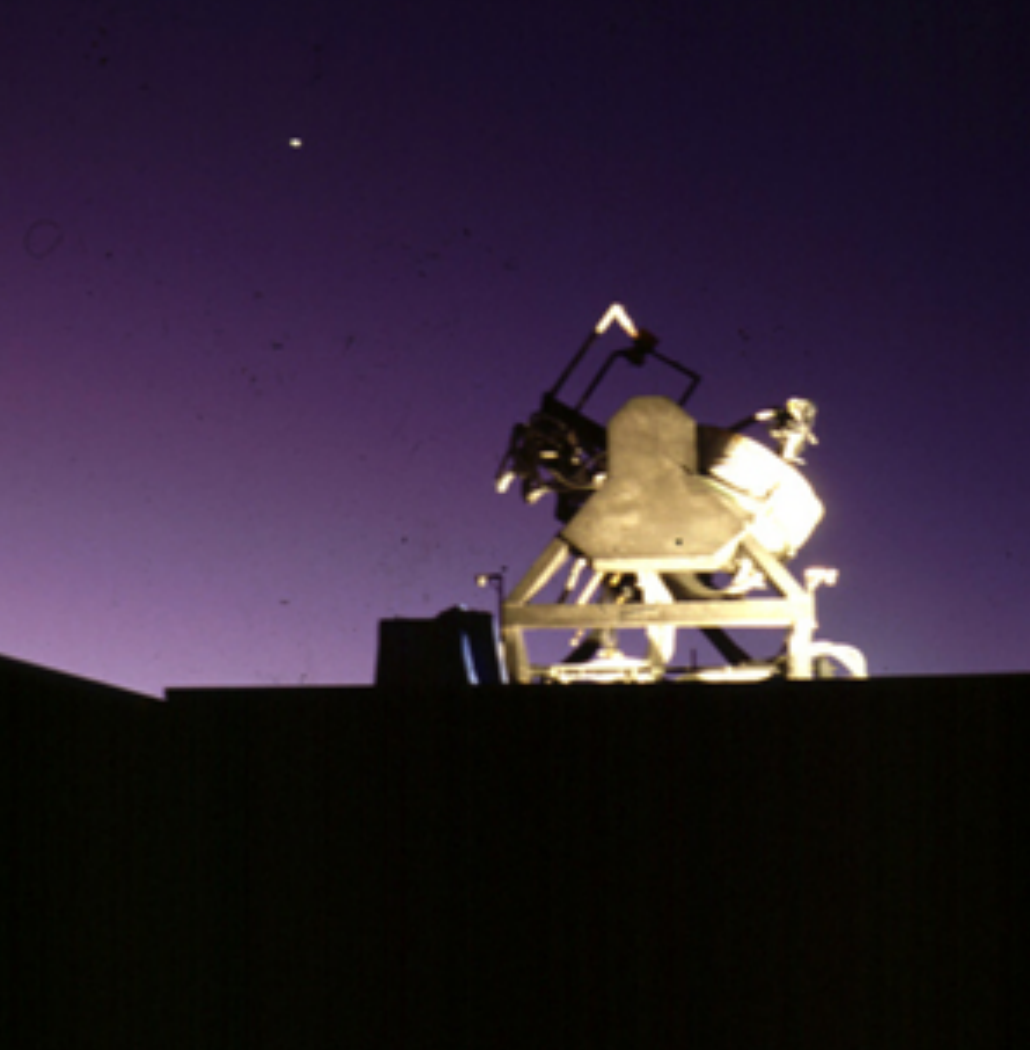}
\caption {Cherenkov telescope of the EAS-TOP experiment. }
\label{fig:eastop-chere}
\end{figure}

The different detectors are:
\begin{itemize}
\item {\em Array of scintillation detectors}, consisting of 35 scintillator modules (10 m$^2$ each) distributed over an area of $\sim 10^5$ m$^2$, for the measurement of the shower size, the core position and the arrival direction \cite{eastop_gen}. Each module is split into 16 individual scintillators (80$\times$80 cm$^2$) read out by one phototube each for timing and particle density measurements from $\sim$ 0.1 m$^{-2}$ to $\sim$ 40 m$^{-2}$. The four central scintillators are equipped with an additional similar phototube with a maximum linearity divider, for larger particle density measurements ($\approx$ 400 m$^{-2}$). The array is organized in ten subarray, that include a central module and five or six modules positioned on circles of radii 50 -- 80 m, interconnected with each other. Any fourfold coincidence ($\Delta$ = 350 ns) of the central module of a subarray, together with three consecutive modules on the circle, triggers the data acquisition of the array. The events can be divided into different classes \cite{eastop_gen}:
\begin{enumerate}
\item {\it Internal events}: at least a whole 6- or 7-fold subarray has triggered and the highest particle density has been recorded by a module not located at the edges of the array (frequency $\approx$ 1.5 Hz);
\item {\it External events}: events with less than a full subarray fired (4 -- 6 modules), or for which the largest number of particles has been recorded on the border of the array. The shower core positions therefore are expected at the edge or outside the array (frequency $\approx$ 20 Hz).
\end{enumerate}
The angular resolution is 0.85$^{\circ}$ for all internal triggers, and 0.5$^{\circ}$ for $N_e>$ 10$^5$.

Inside 13 e.m. stations additional 10 m$^2$ scintillator detectors are positioned below the e.m. modules, each shielded by 30 cm of iron, for the muons detection ($E_{\mu}>1$ GeV).

\item {\em Hadronic calorimeter}, consisting in a parallelepiped of (12$\times$12$\times$3) m$^3$ made of 9 identical planes. Each of them is formed by two streamer tube layers \cite{eastop_streamer} for muon tracking, one layer of \emph{"quasi proportional"} tubes for hadron calorimetry \cite{eastop_prop} and a 13 cm thick iron absorber, for a total depth of 818 g/cm$^2$, i.e., $\simeq$ 6.2 nuclear mean free paths \cite{eastop_calo}. The active layers of the upper plane are unshielded and operate as a fine grain detector of the e.m. component of EAS cores. The distance between two successive layers is $\sim$ 31 cm, expect for the 7th and 8th planes, between which $\simeq$ 24 cm are left to lodge an additional scintillator layer (constituted by six (80$\times$80 cm$^2$) scintillators) as a triggering and timing measurement facility at a depth of $\simeq$ 1.5 absorption lengths. Two vertical detector planes (3$\times$12) m$^2$ are made of one layer of streamer chambers to improve the triggering and tracking capabilities for very inclined muons \cite{eastop_calo}.

The streamer and \emph{"quasi proportional"} tubes consist of 8-cell tubes 12 m long, with (3$\times$3) cm$^2$ single tube cross section. The tubes are operated with an Argon/Isobutane 50/50 gas mixture. The streamer tubes use a 100$\mu$m wire and are supplied with HV = 4650 V. The tube walls are coated with graphite (R = 1 K$\Omega$/square). The two-dimensional read-out is performed using signals from the anode wires (X view) and orthogonal external pick-up strips (Y view).

The high particle densities (for hadron calorimetry and the study of EAS cores) are recorded with the chambers of the upper layer of each plane that use a reduced wire diameter of 50$\mu$m and operate at voltage HV = 2900 V. These detectors operate in saturated proportional mode with the gain reduced by a factor $\simeq$ 100. The signal charge is picked up by an external pad (40$\times$38) cm$^2$ matrix placed on top of the detectors, for a total area of 128 m$^2$. On average, the ADC dynamic is saturated at $\sim$1200 particles/pad. To achieve maximum transparency the wall resistivity has been increased to 200 -- 700 K$\Omega$/square. The particle density range where the signal charge is proportional to the number of incident particles is thus increased, and we refer to this detector as a "quasi proportional" one. The average induced charge/pad for a single particle is $\simeq$ 0.82 pC. In order to study the detector response to different particle densities and check the results of a simulation including the modelling of the tube response, a prototype apparatus has been tested at a 50 GeV positron beam at the
CERN-SPS accelerator at CERN.

A set of scintillators is included in the detector for different aims \cite{eastop_calo}:
\begin{itemize}
\item four 80$\times$80 cm$^2$ scintillators (P$_1$ -- P$_4$), identical to those of the e.m. detectors, are placed on the top of 9th plane. Other four identical (M$_1$ -- M$_4$) are placed outside the detector at the four corners of the 9th plane. They are used to select contained EAS cores on the calorimeter surface;
\item six identical scintillators (S$_1$ -- S$_4$) are placed on the 7th plane for hadronic triggering purposes and timing measurements;
\item eight 40$\times$38 cm$^2$ scintillators (T$_1$ -- T$_8$), placed on top of the 7th and 9th planes, are arranged to form four telescopes to select muons hitting the detector near the input and output of the gas distribution system for monitoring procedures.
\end{itemize}
As a consequence, the calorimeter can be considered as an ensemble of three different detectors (tracking detector, proportional
tubes and scintillator counters) which, because of the different time propagations of their signals, have to be independently triggered \cite{eastop_calo}.
The energy resolution of the calorimeter in the measurement of single hadron energies is $\simeq$ 15$\%$ at 1 TeV and $\simeq$25$\%$ at 5 TeV.
\item {\it atmospheric Cerenkov light}, by means of 8 telescopes positioned at $\sim$ 100 m one from the other (Fig. \ref{fig:eastop-chere}). Each of them houses three mirrors, 90 cm diameter and 67 cm focal length. Two of them where seen by arrays of 7 photomultipliers for a full field of view of $5\cdot 10^{-2}$ $sr$, for measurements of the total Cerenkov light signal, and a wide acceptance for operating in coincidence with the e.m. and muon detectors; 
\item {\it EAS radio emission}, via 3 antennas 15 m high, located on different sides of the array, at distances of 200 m, 400 m and 550 m from each other, operating in two wave bands: 350 -- 500 kHz and 1.8 -- 5 MHz.
\end{itemize}
%
%%%%%%%%%%%%%%%%%%%%%%%%%%%%%%%%%%%%%%%%%%%%%%%%%%%%%%%%%%%%%%%%%%%%
%%%%%%%%%%%%%%%%%%%%%%%%%%%%%%%%%%%%%%%%%%%%%%%%%%%%%%%%%%%%%%%%%%%%
%

\subsection{KASCADE experiment}

The KArlsruhe Shower Core and Array DEtector (KASCADE) experiment was located on the site of the Forschungszentrum Karlsruhe, Germany
($8^\circ$\ E, $49^\circ$\ N; 110 m a.s.l.). The array, distributed on a surface of about 200 $\times$ 200 m$^2$ (Fig.\ref{fig:kascade_layout}), measures the electromagnetic, muonic and hadronic components of extensive air showers by means of the following three major components \cite{kascade_gen}:
%
%%%%%%%%%%%%%%%%%%%%%%%%%%%%%%%%%%%%%%%%%%%%%%%%%%%%%%%%%%%%%%%%%%%%%
\begin{figure}[h]
\vfill  \begin{minipage}[h]{.47\linewidth}
  \begin{center}
    \mbox{\epsfig{file=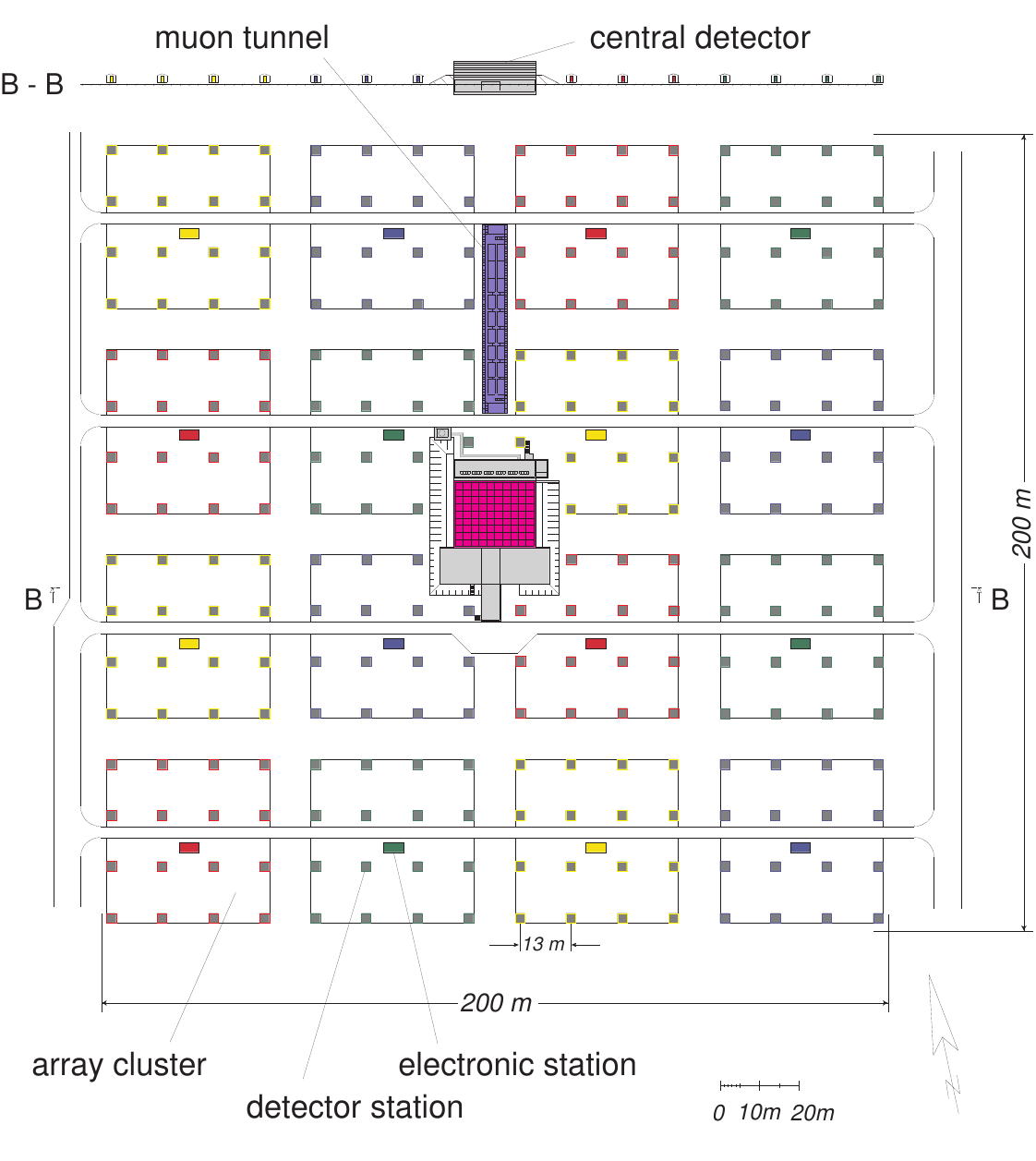,width=6.8cm}}
  \vspace{-0.5pc}
 \caption{Schematic layout of the KASCADE experiment.
    \label{fig:kascade_layout}}
  \end{center}
%%%\end{figure}
\end{minipage}\hfill
\hspace{-1.cm}
\begin{minipage}[h]{.47\linewidth}
%%%%\begin{figure}[t]
  \begin{center}
    \mbox{\epsfig{file=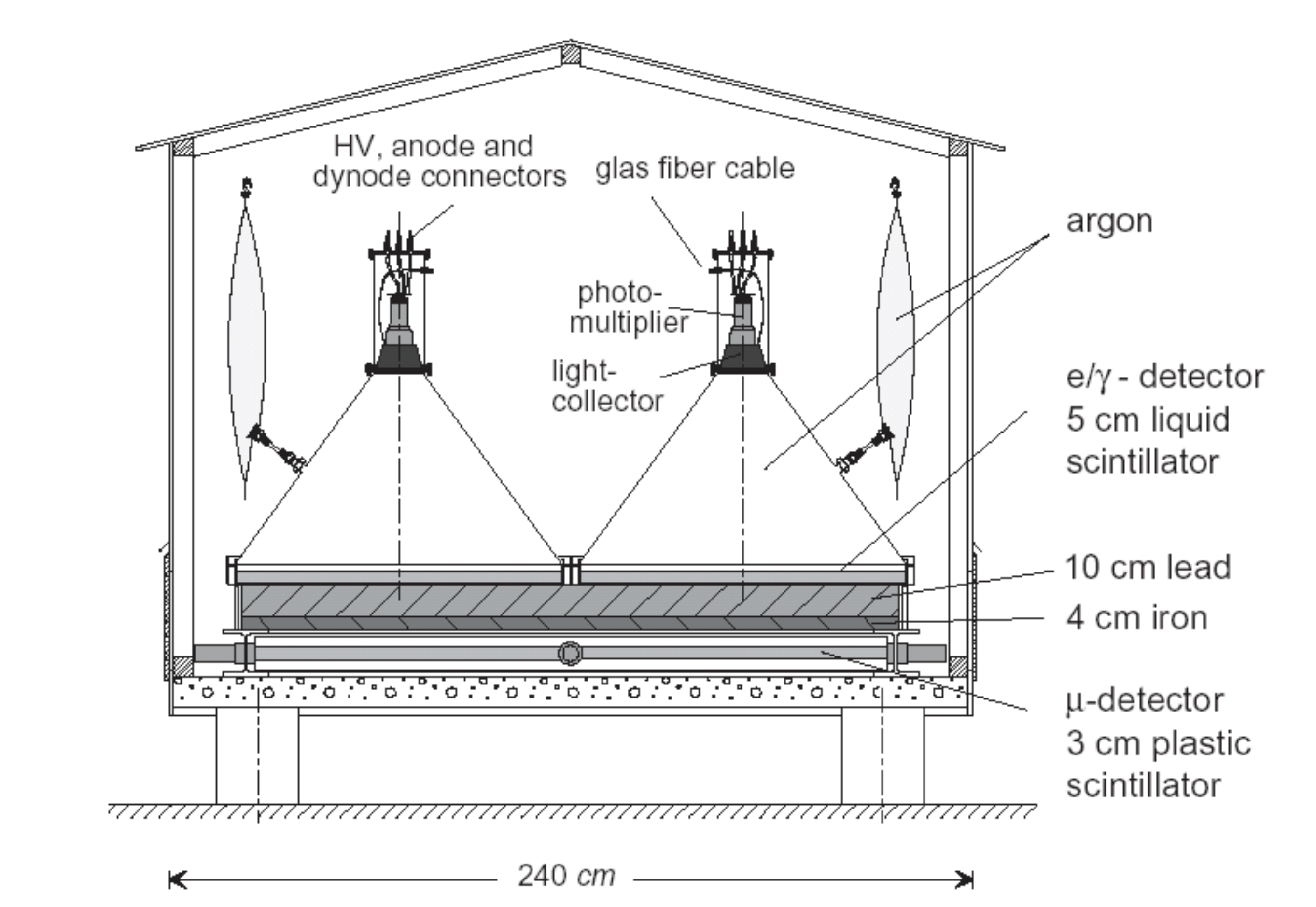,width=6.8cm}}
  \vspace{-0.5pc}
    \caption{Sketch of an array station for $e/\gamma$ and
    muon detection.
 \label{fig:kasc_scint}}
  \end{center}
\end{minipage}\hfill
\end{figure}
%%%%%%%%%%%%%%%%%%%%%%%%%%%%%%%%%%%%%%%%%%%%%%%%%%%%%%%%%%%%%%%%%%%%
%
%%%%%%%%%%%%%%%%%%%%%%%%%%%%%%%%%%%%%%%%%%%%%%%%%%%%%%%%%%%%%%%%%%%%
\begin{figure}[t]
\centerline{\includegraphics[width=0.8\textwidth]{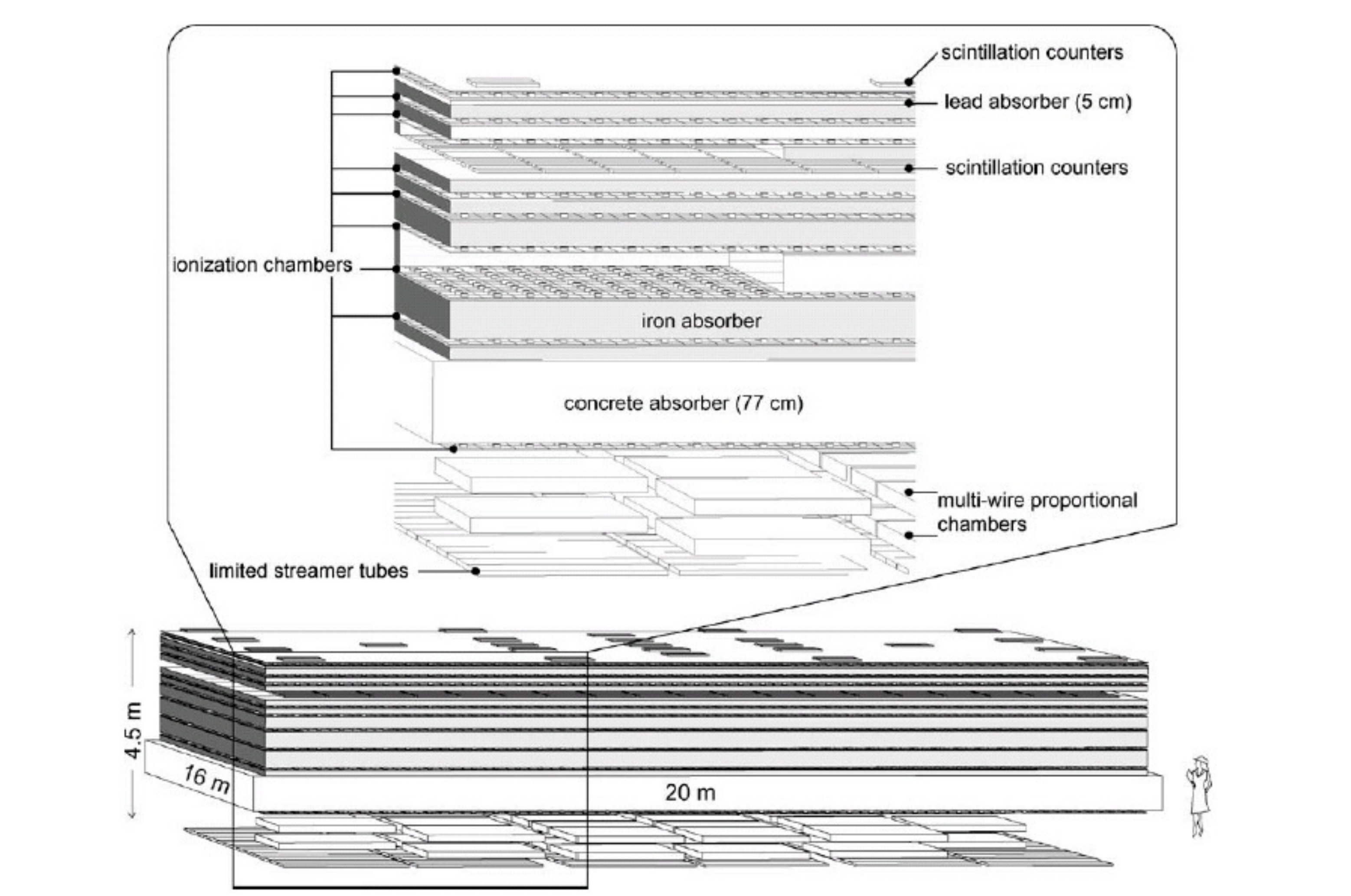}}
\caption {Schematic layout of the central detector. General view (lower picture) and partly exploded view (upper picture). }
\label{fig:kasc_centrdet}
\end{figure}
%%%%%%%%%%%%%%%%%%%%%%%%%%%%%%%%%%%%%%%%%%%%%%%%%%%%%%%%%%%%%%%%%%%%
%
\begin{itemize}
\item {\em Array of scintillation detectors}, consisting of 252 detector stations electronically organized in clusters of 16 stations, the inner clusters only with 15 stations, and placed on a square grid with 13 m spacing. A profile view of a detector station is shown in Fig.\ref{fig:kasc_scint}. The detector stations contain 4 liquid scintillation counters ($e/\gamma$ detectors) of 0.79 m$^2$ area each, positioned on a lead/iron shielding plate (10 cm Pb and 4 cm Fe) covering 4 plastic scintillators of 0.81 m$^2$ each ($\mu$ detectors, $E_\mu^{\rm thres}$= 230 MeV). As result, a detector coverage of 1.3 $\%$ for the e.m. and 1.5 $\%$ for the muonic component is obtained. The four inner clusters have no muon detectors installed since close to the shower core the punch-through of hard e.m. and hadronic component makes them redundant. The twelve outer clusters contain only two $e/\gamma$ detectors per station. The reconstruction of the EAS data measured with the array provides the basic
information about lateral distributions and total intensities of the electron-photon (shower size $N_e$) and muon components ($N_{\mu}^{tr}$), the location of the EAS core and the direction of incidence.
 \item {\em Central Detector system} (320 m$^{2}$) consisting of a highly-segmented hadronic calorimeter with eight tiers of iron absorber interspersed with 9 layers read out by 44,000 channels of warm liquid ionization chambers. A picture of the central detector is shown in Fig.\ref{fig:kasc_centrdet}. The calorimeter is of the sampling type, the energy being absorbed in the iron stacks and sampled by the ionization chambers. Its performance is described in detail in \cite{kascade_kalo}. The iron slabs are 12--36~cm thick, becoming thicker in deeper parts of the calorimeter. Therefore, the energy resolution does not scale as $1/\sqrt{E}$, but is rather constant varying slowly from $\sigma/E = 20\%$ at
100~GeV to 10$\%$ at 10~TeV. The concrete ceiling of the detector building is the last part of the absorber and the ionization
chamber layer below acts as tail catcher. In total, the calorimeter thickness corresponds to about 11.5 nuclear interaction lengths in vertical direction, so that hadrons up to $\sim$ 25 TeV are absorbed. At this energy, containment losses are at a level of 5$\%$. At 50 TeV signal losses of about 5$\%$ have to be taken into account.

The liquid ionization chambers use the room temperature liquids tetramethylsilane (TMS) and tetramethylpentane (TMP). One of the principal motivations to use liquid ionization chambers in a CR experiment is their long term stability and the high dynamic range. A detailed description of their performance can be found in \cite{kascade_liqscint}. Liquid ionization chambers exhibit a linear signal behaviour with a very large dynamic range. A stability of better than 2$\%$ over two years of operation has been attained.

From their signals the impact point, the direction and the energies of individual hadrons are reconstructed. In particular, the number of EAS hadrons with energies larger than 100 GeV, the energy of the most energetic hadron observed in the shower  ($E_{\rm h}^{\rm max}$) and the energy sum of all reconstructed hadrons ($\sum E_{\rm h}$) are deduced as shower observables.
 Typically, for a 1 TeV hadron an energy resolution with a rms--value of 20$\%$ and an angular resolution of 3$^{\circ}$ is achieved \cite{kascade_gen}. Two hadrons of nearly equal energy are resolved with 50$\%$ probability if their axes are separated by 40 cm \cite{kascade_hadcore}.

A layer of 456 scintillation detectors, each with a size of 0.45 m$^2$, is mounted in the gap below the third absorber plane, at a depth of 2.2 nuclear interaction lengths. It is used for triggering the central detector system, for muon detection (with a threshold of $E_\mu^{\rm thres}$= 490 MeV), and to determine arrival time distributions~\cite{kascade_time}. A description of the system can be found in \cite{kascade_triglay}.

On top of the central detector 25 scintillation counters are installed forming the top cluster. They cover 7.5$\%$ of the area and serve, among others, as trigger source for small EAS, i.e., for primary protons below 0.4 PeV. They also fill the gaps of the central four missing array stations. Directly below the top cluster, a layer of liquid ionization chambers are in operation. If the shower core hits the central detector, the core can be
determined precisely by the e.m. punch-through to the first active layer of ionization chambers in case of large EAS or by the top layer for small EAS. This allows to study EAS cores in great detail.

In the basement of the central building, below the iron stack and 77 cm of concrete, 2 layers of multi-wire proportional chambers (MWPCs) are arranged as a tracking hodoscope, covering an area of 122 m$^2$. A detailed description of the chamber system and operational tests can be found in \cite{kascade_mwpc}. The MWPCs register muons with an energy threshold $E_{\mu}^{\rm thres}$ = 2.4 GeV and provide the observable $N_{\mu}^{\star}$, i.e. number of reconstructed muons in the MWPCs. Due to the good position resolution, the MWPCs register also the spatial distribution of the high-energy muons together with traversing secondaries produced in the absorber by high-energetic hadrons, whose pattern has been shown to carry valuable information about the mass of the primary particle~\cite{kascade_haungs96}. To reduce the
ambiguities at higher densities, especially near the shower core, a third layer of chambers with pad read-out has been installed below the MWPCs. For this purpose limited streamer tubes have been chosen.

 \item {\em Muon Tracking Detector} (MTD). North of the central  detector the MTD represents a second device for muon measurements by tracking. In Fig.\ref{fig:kascade_mtd} is shown a side view of the installation. The total length of the detector extends to 32 m and provides an effective detection area of 120 m$^2$ for vertical particles. A large spacing of 82 cm between three horizontal planes of limited streamer tubes ensures a precise determination of the muon angle and the possibility to extrapolate the track back to find its production height by means
 of triangulation. The three horizontal layers are supplemented on the sides by vertical chambers in order to accept also inclined
 tracks, increasing the acceptance to $\sim$ 500 m$^2$ sr.
 A detailed description ot the muon tracking detector can be found in \cite{kascade_mtd1,kascade_mtd2}.
The shielding of concrete, iron and soil corresponds in vertical direction to 18 $X_0$ and entails a muon threshold of 0.8 GeV.
\end{itemize}
%
%%%%%%%%%%%%%%%%%%%%%%%%%%%%%%%%%%%%%%%%%%%%%%%%%%%%%%%%%%%%%%%%%%%%%
\begin{figure}[h]
\vfill  \begin{minipage}[h]{.47\linewidth}
  \begin{center}
    \mbox{\epsfig{file=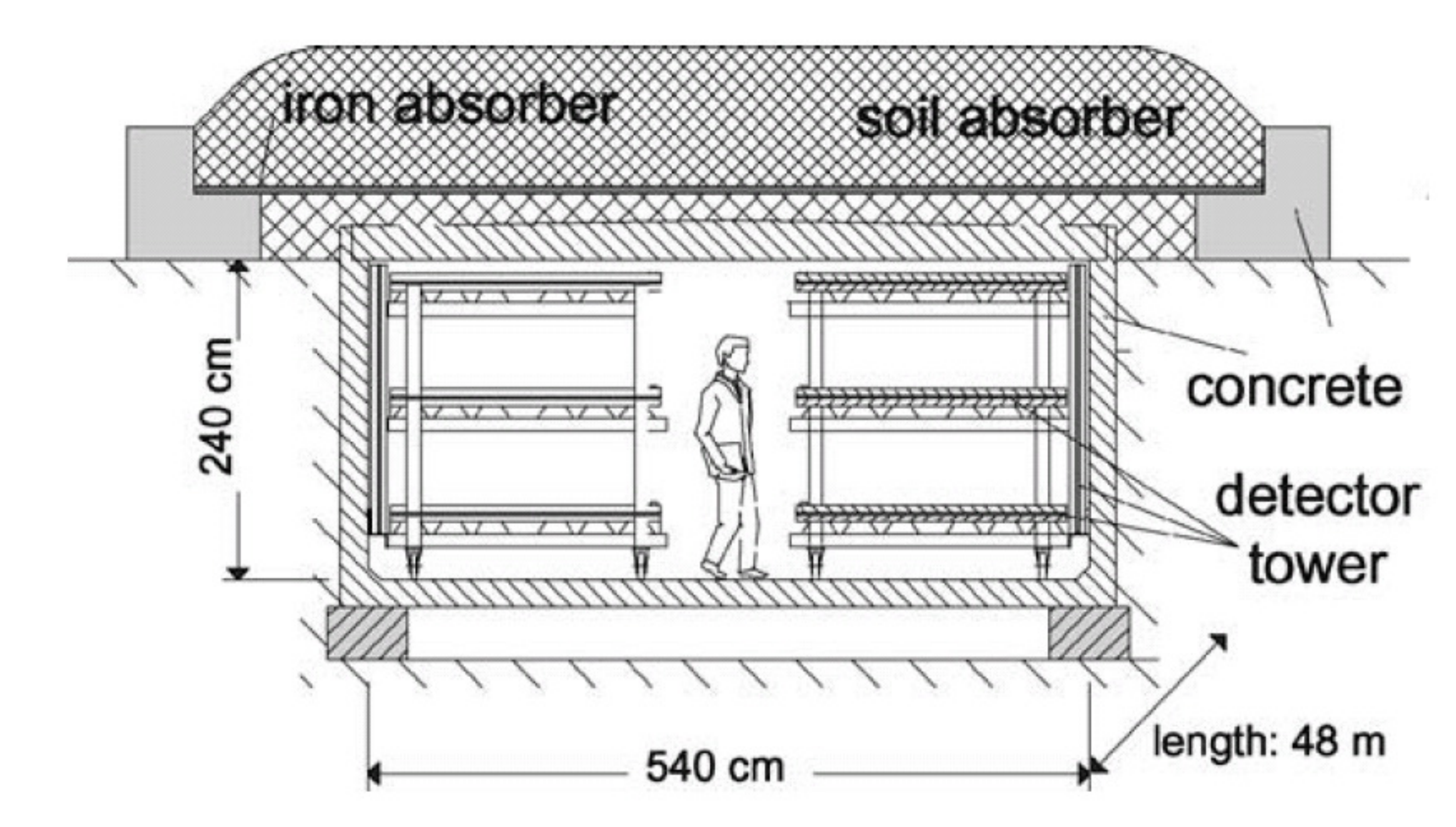,width=6.8cm}}
  \vspace{-0.5pc}
    \caption{Sketch of the muon tracking detector with 3 horizontal
    and one vertical layer of limited streamer tubes.
    \label{fig:kascade_mtd}}
  \end{center}
%%%\end{figure}
\end{minipage}\hfill
\hspace{-1.cm}
\begin{minipage}[h]{.47\linewidth}
%%%%\begin{figure}[t]
  \begin{center}
    \mbox{\epsfig{file=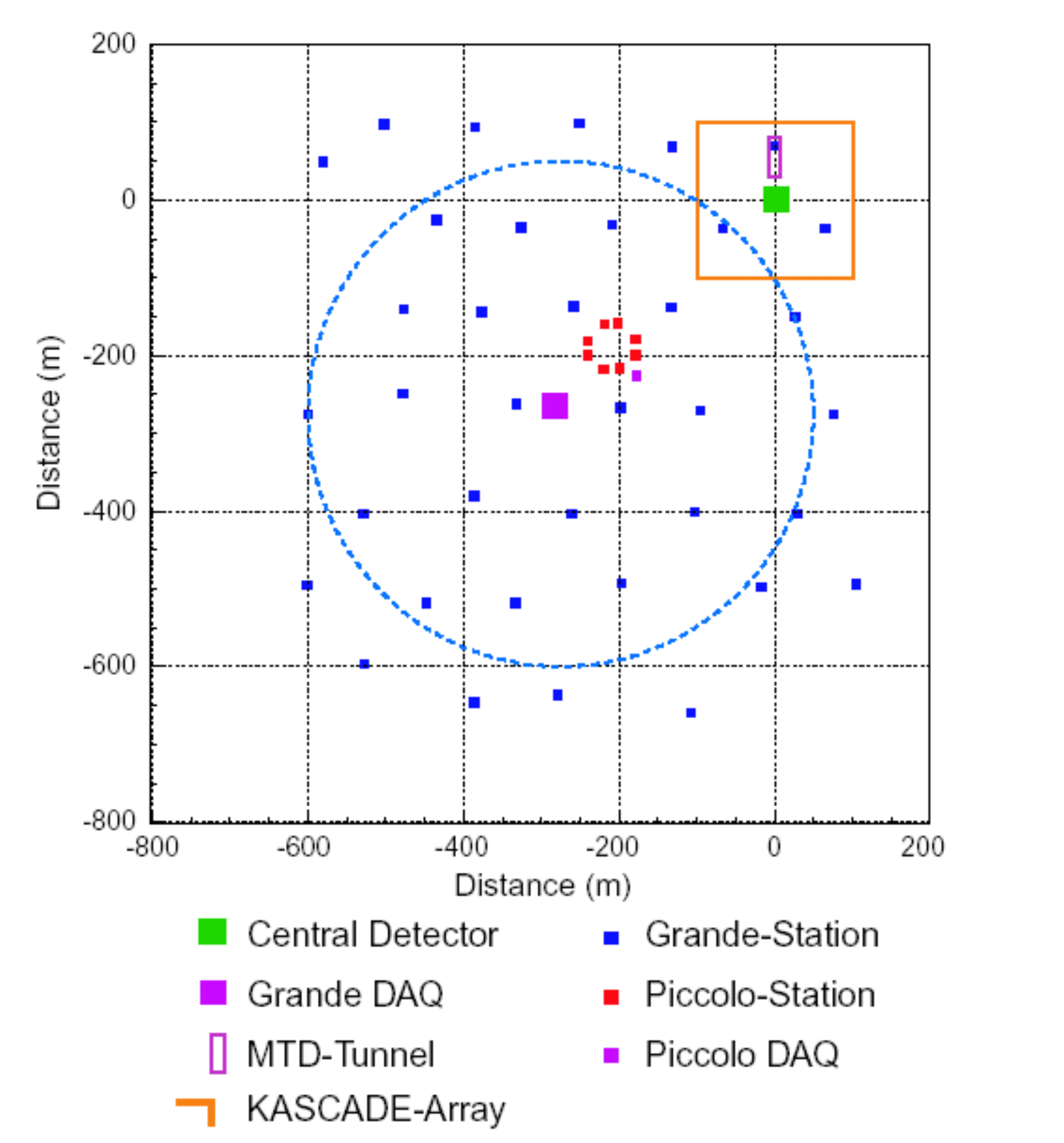,width=6.8cm}}
  \vspace{-0.5pc}
    \caption{Schematic layout of the KASCADE-Grande
    experiment.
 \label{fig:kascade_grande}}
  \end{center}
\end{minipage}\hfill
\end{figure}
%%%%%%%%%%%%%%%%%%%%%%%%%%%%%%%%%%%%%%%%%%%%%%%%%%%%%%%%%%%%%%%%%%%%
%
\begin{table}[t]
\begin{center}
\caption{\label{detector:tab} KASCADE detector components.} 
    \begin{tabular}[h]{llcr}
    \br
    Detector & Particles & Total Area (m$^2$) & Threshold \\
    \mr
    Array, liquid scintillators & $e/\gamma$ & 490 & 5 MeV \\
    Array, plastic scintillators& $\mu$      & 622 & 230 MeV \\
    MTD, streamer tubes  & $\mu$      & 128$\times$4 layers & 800 MeV\\
    Central Detector:    &            &     &  \\
    calorimeter, liquid ionization chamb. & h & 304$\times$8 layers  & 50 GeV \\
    trigger layer, plastic scintillators & $\mu$ & 208 & 490 MeV\\
    top cluster, plastic scintillators & $e/\gamma$ &23 & 5 MeV\\
    top layer, liquid ionization chamb. & $e/\gamma$ & 304 & 5 MeV\\
    MWPCs  & $\mu$ & 129$\times$4 layers & 2.4 GeV\\
    limited streamer tubes & $\mu$ & 250 & 2.4 GeV\\
    \br
    \end{tabular}
\end{center}
\end{table}
Information concerning the total detector area and the threshold energy for vertical particles of the different detector components are summarized in Table 1 \cite{kascade_gen}.

Different triggers in the experiment are used to study a broad variety of physics problems. The principal trigger of KASCADE for the study of the primary spectrum and the composition around the knee is a cluster detector multiplicity, fulfilled in at least one array cluster. The envisaged trigger threshold of about 0.5 PeV for iron initiated showers necessitates a minimum multiplicity of $n_i$ = 20 $e/\gamma$ array detectors have a signal over threshold out of 60 in a inner cluster, and $n_o$ = 10 out of 32 for the outer clusters, resulting in an overall rate of 3 Hz from the
whole array. The resulting energy threshold is a few times 10$^{14}$ eV, depending on zenith angle and primary mass.

Since 1996 the experiment has taken data continuously up to the end of 2002. In the following years the KASCADE-Grande array \cite{kascade_grande} has been operated as a joint application of the KASCADE experiment and the EAS-TOP array detectors to cover a surface of 0.5 km$^2$ (Fig.\ref{fig:kascade_grande}) and to allow measurements of the CR energy spectrum up to $\approx$10$^{18}$ eV.

\subsection{HEGRA experiment}

\begin{figure}[h]
\vfill  \begin{minipage}[h]{.47\linewidth}
  \begin{center}
    \mbox{\epsfig{file=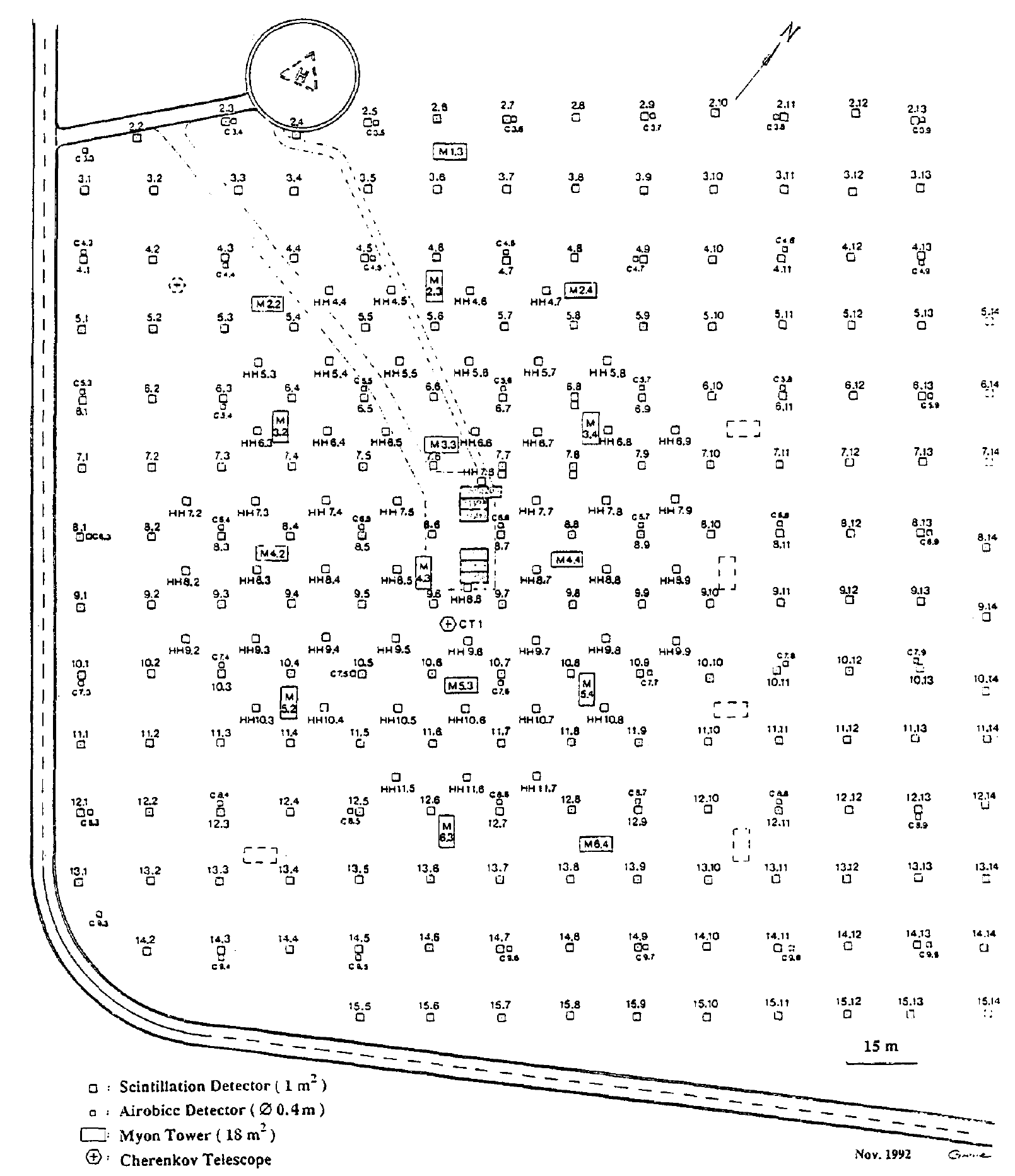,width=6.8cm}}
  \vspace{-0.5pc}
    \caption{Layout of the HEGRA experiment \cite{hegra}.
    \label{fig:hegra_layout}}
  \end{center}
%%%\end{figure}
\end{minipage}\hfill
\hspace{-1.cm}
\begin{minipage}[h]{.47\linewidth}
%%%%\begin{figure}[t]
  \begin{center}
    \mbox{\epsfig{file=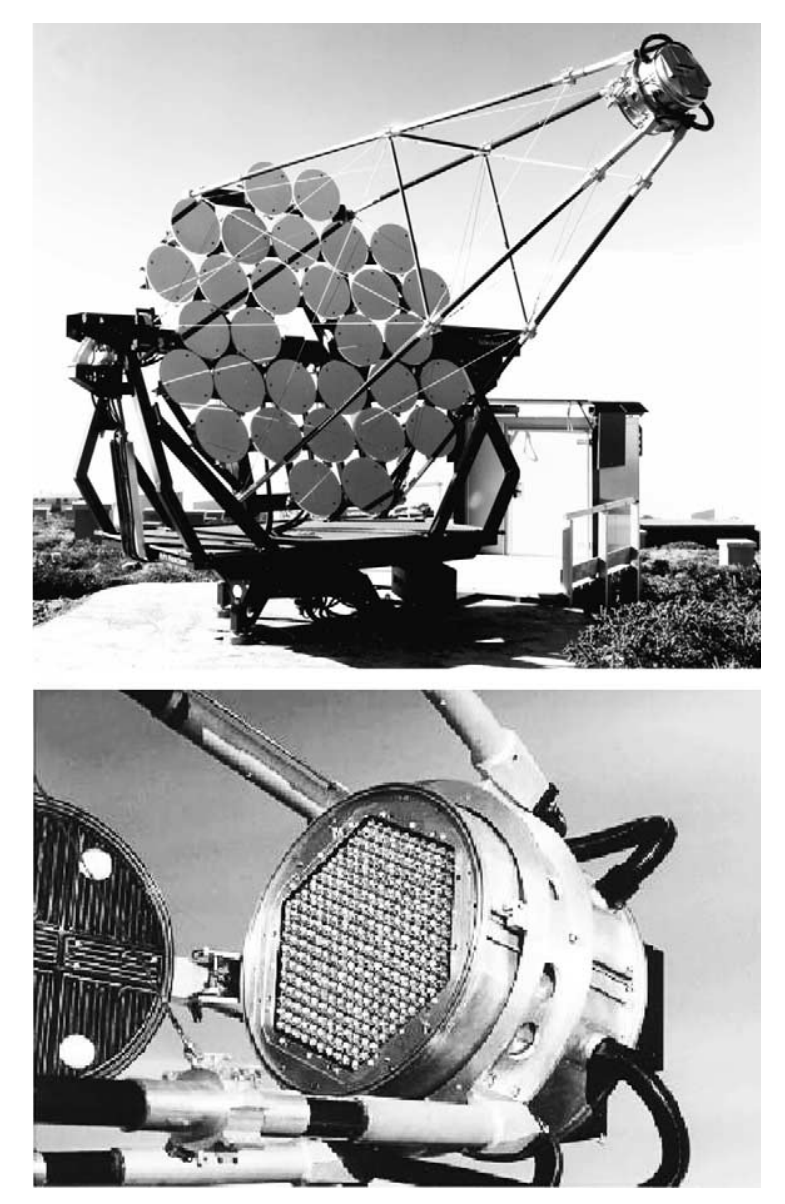,width=6.8cm}}
  \vspace{-0.5pc}
    \caption{Up: One of the system telescopes. Down: Front view of the camera \cite{hegra-iact}.
 \label{fig:hegra-iact}}
  \end{center}
\end{minipage}\hfill
\end{figure}

The High Energy Gamma Ray Astronomy (HEGRA) experiment, located at the site of the Observatorio del Roque de Los Muchachos ($28^{\circ}45'43''$ N, $17^{\circ}53'27''$ W, 2200 m a.s.l., 790 g/cm$^2$) on the Canary Island La Palma, Spain, have been operating in changing setups in the years 1989 -- 2000.
The HEGRA arrays is a hybrid installation characterized by different detectors installed on an area 200$\times$200 m$^2$ with a total coverage of about 3$\%$ (Fig.\ref{fig:hegra_layout}) \cite{hegra}.
%
%%%%%%%%%%%%%%%%%%%%%%%%%%%%%%%%%%%%%%%%%%%%%%%%%%%%%%%%%%%%%%%%%%%%%
%%%%%%%%%%%%%%%%%%%%%%%%%%%%%%%%%%%%%%%%%%%%%%%%%%%%%%%%%%%%%%%%%%%%
%
The basic elements are:
\begin{enumerate}
\item A scintillator array, consisting in 243 individual stations arranged on a square lattice with 15 m spacing \cite{hegra}. The inner part of the array is denser to increase the detection efficiency and to lower the energy threshold. Each counter consists of 1 m$^2$ lead covered (4.8 mm thickness) plastic scintillator. This array measures energies $\geq$ 30 TeV. The arrival time and amplitude sampling of the shower-front particles allows one to reconstruct the incoming direction of primary particles to an accuracy of $\sim 0.7^\circ$.
\item An array of 17 Geiger towers to identify and track muons and to measure particle density and energy distribution in different
distances to the shower core thus providing a calorimetric shower information \cite{hegra-tower}. Each tower consists of six planes of Geiger tubes placed in a support structure of gaseous concrete ($\rho$ = 0.8 g/cm$^3$) and has and active surface of $\sim$ 16 m$^2$. The upper two planes are each followed by a layer of 2.5 cm (4.5 radiation lengths) of lead. The distance between planes amounts to $\sim$ 20 cm. Each plane consists of 160 aluminium Geiger tubes with a length of 6 m and an inner quadratic profile of 1.5$\times$1.5 cm$^2$. The Geiger
towers may be divided into two different function zones. With the lead layers below the first and second Geiger plane, the upper planes work as a calorimeter, while the lower planes serve to reconstruct directed information such as tracks of muons of remaining hadrons.

A peculiar characteristics of the Geiger towers is the ability to trace the particles and to determine their identity. Obviously the track reconstruction efficiency depends on the complexity of the hit pattern. Due to the high particle densities near the shower core, the reconstruction efficiency is an increasing function of the core distance. The number of clusters defining a track vary from 4 to 6. The inset shows the fractions of different particle species for all reconstructed tracks as a function of the core distance. As expected, the muon purity is increasing with the core distance while the fraction of fake tracks dominates the core region.

Due to the rather small detector sampling density the number of detectable muons at HEGRA energies (3 -- 4 per hadronic shower) is
too small for a sufficient stand alone hadron shower suppression.
Nevertheless, high energy (E $>$ 250 MeV) electrons and converting gammas outside the shower core region ($r\geq$ 30 m), which both
are reconstructed as punch through tracks, are strong indicators of hadron induced showers. By means of all available information, the Geiger towers allow a $\gamma$-hadron separation which considerably goes beyond ordinary muon counting \cite{hegra-gamhad}.
\item An array of 6 imaging air Cerenkov telescopes, measuring in the energy range $5\cdot 10^{11} - 10^{14}$ eV (see Fig. \ref{fig:hegra-iact}) \cite{hegra-iact,hegra-iact2}. This array allows one to measure the arrival directions of primary particles to an accuracy of $\sim 0.1^\circ$ and their relative energy to an accuracy of $\sim 20 \%$.
\item The wide angle integrating air Cherenkov "AIROBICC" (AIr-shower Observatory By Angle Integrating Cerenkov Counters) array of 97 detectors measuring energies $\geq$ 10 TeV \cite{hegra-airobicc}. AIROBICC measures within $\pm 35^\circ$ around the zenith the integrated longitudinal development of EAS. By multiple sampling of the amplitude and relative arrival time of the shower front to an accuracy $\leq$ 500 ps, it allows one to determine the incoming direction and energy of primary particles to a precision of $\sim$0.1-0.2$^{\circ}$ and 15-20 \%, respectively. The single detector focus the light from a large part of the sky, $\sim$1 steradian, allowing to observe simultaneously a large number of sources.
\end{enumerate}

\subsection{ARGO-YBJ experiment}

ARGO-YBJ is a multipurpose experiment consisting in a dense sampling air shower array with 93\% sensitive area located at very high altitude and devoted to the integrated study of gamma rays and cosmic rays with an energy threshold of a few hundreds GeV (for a summary of the ARGO-YBJ results see \cite{discia-rev}).

\begin{figure}[!t]
\centerline{\includegraphics[width=0.8\textwidth]{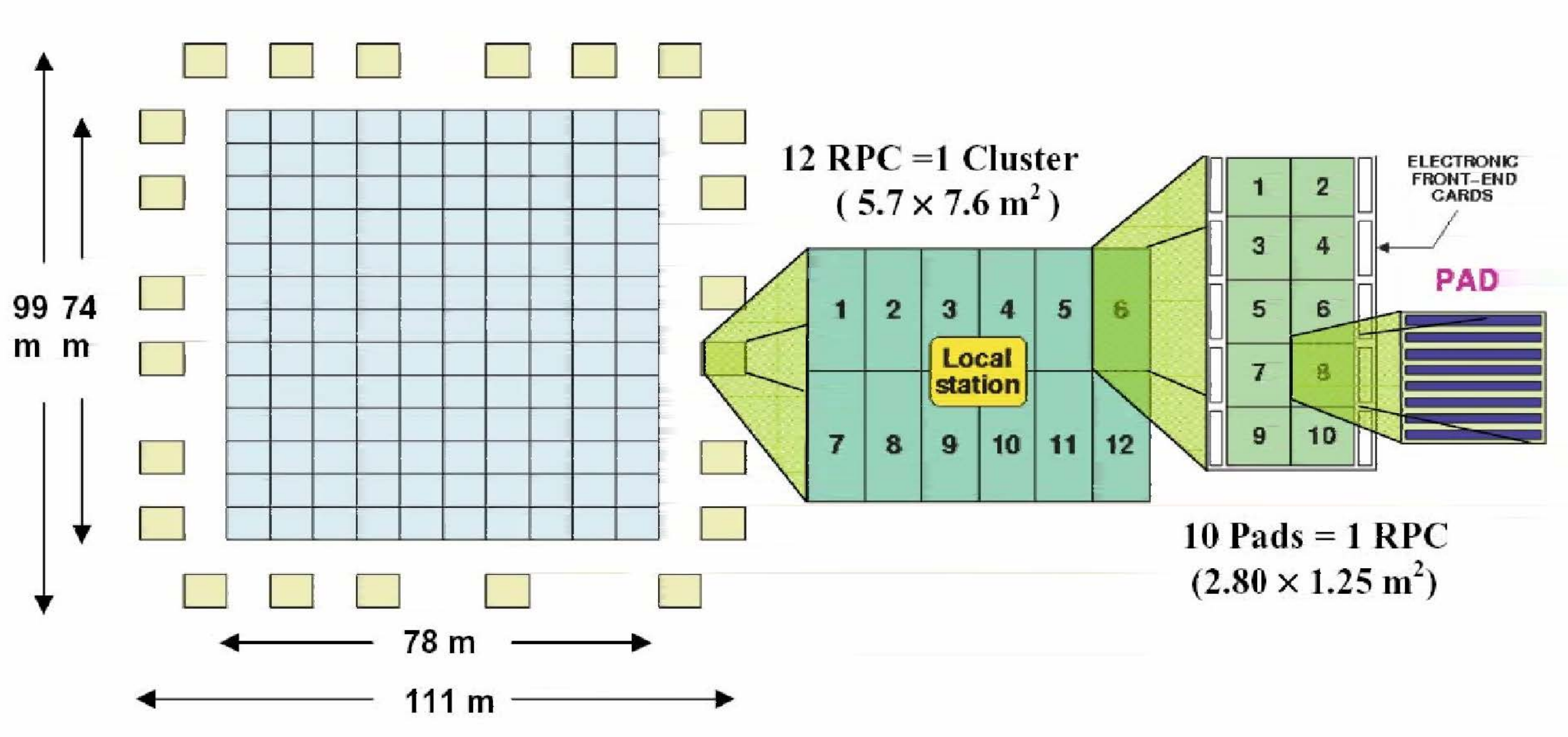}}
  \caption{Layout of the ARGO-YBJ experiment (see text for a detailed description of the detector).
  \label{fig:fig-01}}
\end{figure}

The detector, located at the Yangbajing Cosmic Ray Observatory (Tibet, PR China, 4300 m a.s.l., 606 g/cm$^2$), is constituted by a central carpet $\sim$74$\times$78 m$^2$, made of a single layer of resistive plate chambers (RPCs) with $\sim$93$\%$ of active area, enclosed by a guard ring partially instrumented ($\sim$20$\%$) up to $\sim$100$\times$110 m$^2$. The apparatus has a modular structure, the basic data acquisition element being a cluster (5.7$\times$7.6 m$^2$), made of 12 RPCs (2.85$\times$1.23 m$^2$ each). Each chamber is read by 80 external strips of 6.75$\times$61.80 cm$^2$ (the spatial pixels), logically organized in 10 independent pads of 55.6$\times$61.8 cm$^2$ which represent the time pixels of the detector \cite{aielli06}. 
The readout of 18,360 pads and 146,880 strips is the experimental output of the detector. The relation between strip and pad multiplicity has been measured and found in fine agreement with the Monte Carlo prediction \cite{aielli06}.
In addition, in order to extend the dynamical range up to PeV energies, each chamber is equipped with two large size pads (139$\times$123 cm$^2$) to collect the total charge developed by the particles hitting the detector \cite{argo-bigpad}.
The RPCs are operated in streamer mode by using a gas mixture (Ar 15\%, Isobutane 10\%, TetraFluoroEthane 75\%) for high altitude operation \cite{bacci00}. The high voltage settled at 7.2 kV ensures an overall efficiency of about 96\% \cite{aielli09a}.
The central carpet contains 130 clusters (hereafter ARGO-130) and the full detector is composed of 153 clusters for a total active surface of $\sim$6700 m$^2$ (Fig. \ref{fig:fig-01}). The total instrumented area is $\sim$11,000 m$^2$.
%
%%%%%%%%%%%%%%%%%%%%%%%%%%%%%%%%%%%%%%%%%%%%%%%%%%%%%%
%%%%%%%%%%%%%%%%%%%%%%%%%%%%%%%%%%%%%%%%%%%%%%%%%%%%%%
%
The information on strip multiplicity and the arrival times recorded by each pad are received by a local station devoted to manage the data of each cluster. A central station collects the information of all the local stations. 
The time of each fired pad in a window of 2 $\mu$s around the trigger time and its location are used to reconstruct the position of the shower core and the arrival direction of the primary particle.
In order to perform the time calibration of the 18,360 pads, a software method has been developed \cite{aielli09b}. To check the stability of the apparatus a control system monitors continuously the current of each RPC, the gas mixture composition, the high voltage distribution as well as the environment conditions (temperature, atmospheric pressure, humidity).

The detector is connected to two different data acquisition systems, working independently, and corresponding to the two
operation modes, shower and scaler. In shower mode, for each event the location and timing of every detected particle is recorded, allowing the reconstruction of the lateral distribution and the arrival direction \cite{llf,DiS07}. 
In scaler mode the total counts on each cluster are measured every 0.5 s, with limited information on both the space distribution and arrival direction of the detected particles, in order to lower the energy threshold down to $\sim$1 GeV \cite{aielli08}.

In shower mode, a simple, yet powerful, electronic logic has been implemented to build an inclusive trigger. This logic is based on a time correlation between the pad signals depending on their relative distance. In this way, all the shower events giving a number of fired pads N$_{pad}\ge$ N$_{trig}$ in the central carpet in a time window of 420 ns generate the trigger.
This trigger can work with high efficiency down to N$_{trig}$ = 20, keeping negligible the rate of random coincidences.

\begin{figure}[t!]
\begin{minipage}[t]{.47\linewidth}
  \centerline{\includegraphics[width=\textwidth]{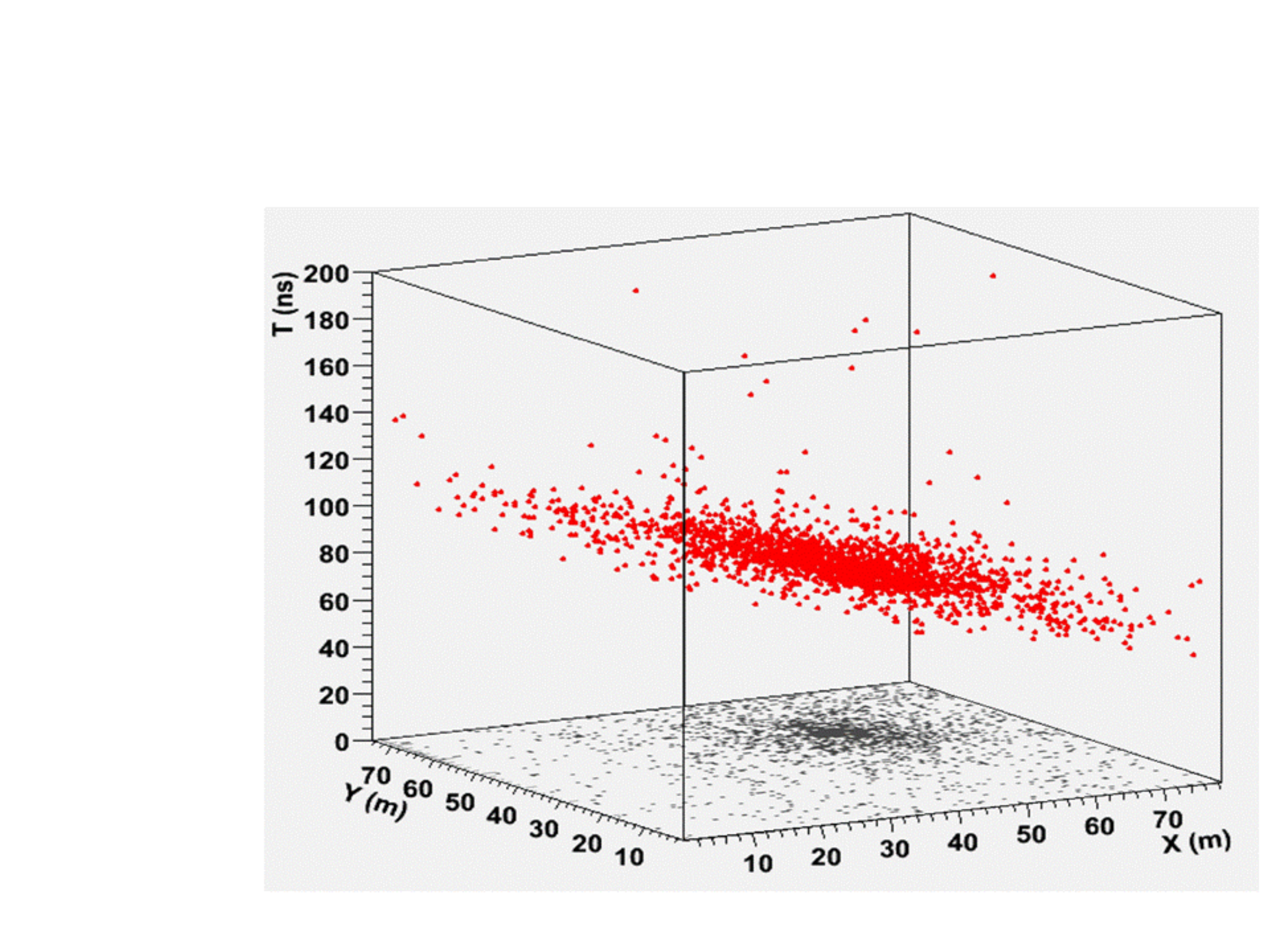} }
\end{minipage}\hfill
\begin{minipage}[t]{.47\linewidth}
  \centerline{\includegraphics[width=0.8\textwidth,angle=90]{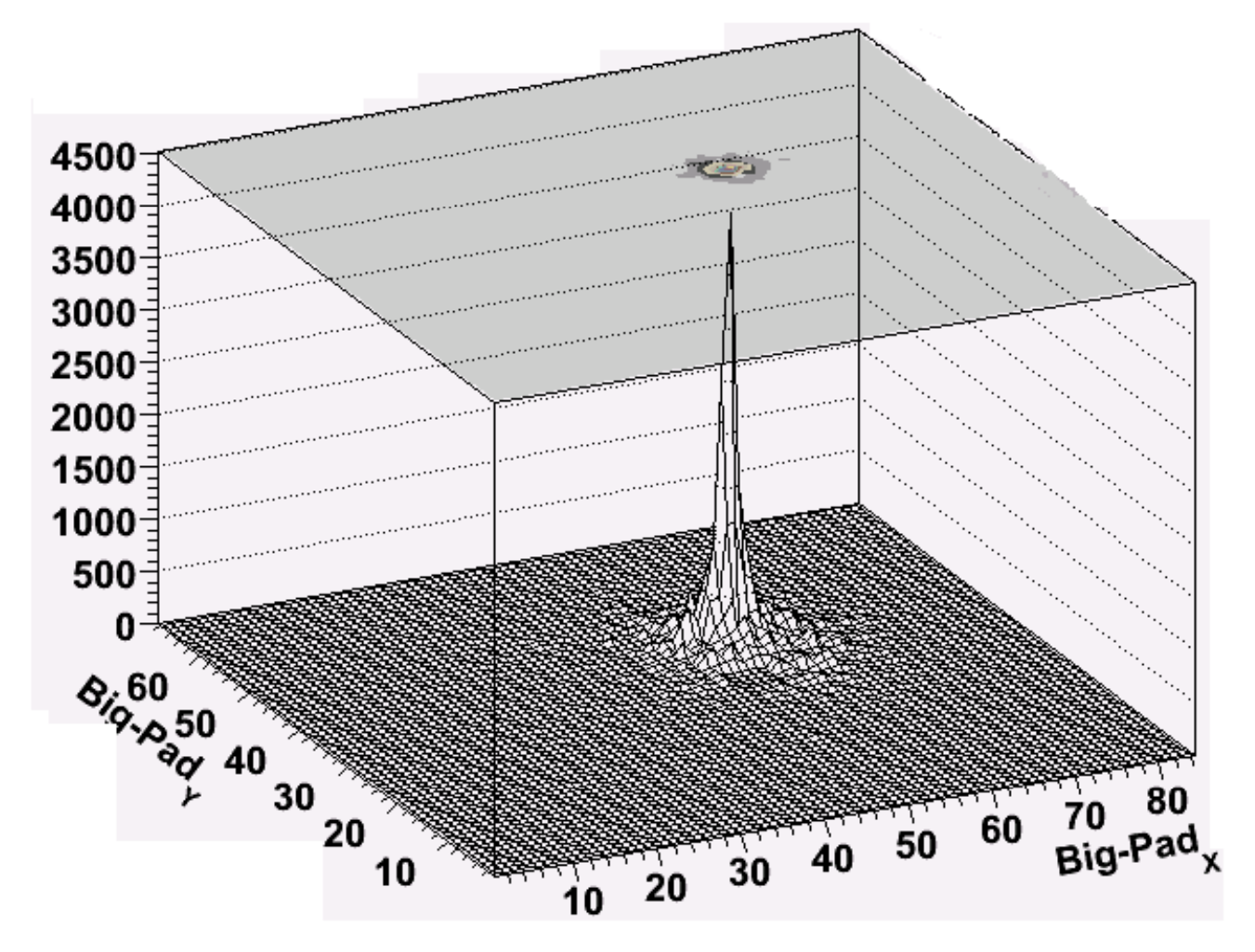} }
\end{minipage}\hfill
\caption[h]{Showers imaged by the ARGO-YBJ central carpet with the digital (left plot: the space-time structure of a low energy shower provided by the strip/pad system) and the analog (right plot: the core of a high energy shower imaged by the charge read-out) read-outs.} 
\label{fig:argo-shower}
\end{figure}

Because of the small pixel size, the detector is able to record events with a particle density exceeding 0.003 particles m$^{-2}$, keeping good linearity up to a core density of about 15 particles m$^{-2}$.
This high granularity allows a complete and detailed three-dimensional reconstruction of the front of air showers at an energy threshold of a few hundred GeV, as can be appreciated in Fig. \ref{fig:argo-shower} (left plot) where a typical shower detected by ARGO-YBJ is shown. Showers induced by high energy primaries ($>$100 TeV) are also imaged by the charge readout of the large size pads (see Fig. \ref{fig:argo-shower}, right plot), which allow the study of shower core region with unprecedented resolution \cite{argo-bigpad}.
%
%%%%%%%%%%%%%%%%%%%%%%%%%%%%%%%%%%%%%%%%%%%%%%%%%%%%%%%%%%%
%%%%%%%%%%%%%%%%%%%%%%%%%%%%%%%%%%%%%%%%%%%%%%%%%%%%%%%%%%%
%
The whole system, in smooth data taking since July 2006 with ARGO-130, has been in stable data taking with the full apparatus of 153 clusters from November 2007 to January 2013, with the trigger condition N$_{trig}$ = 20 and a duty cycle $\geq$86\%. The trigger rate is $\sim$3.5 kHz with a dead time of 4$\%$.

The reconstruction of the shower parameters is carried out through the following steps.

At first, a plane surface is analytically fitted (with weights equal to 1) to the shower front. This procedure is repeated up to 5 times, each iteration rejecting hits whose arrival time is farther than 2 standard deviations from the mean of the distribution of the time residuals from the fitted plane surface. This iterative procedure is able to reject definitively from the reconstruction the time values belonging to the non-Gaussian tails of the arrival time distribution \cite{icrc05_risang}. 
After this first step the problem is reduced to the nearly-vertical case by means of a projection which makes the fit plane overlapping the detector plane. 
Thereafter, the core position, i.e. the point where the shower axis intersects the detection plane, is obtained fitting the lateral density distribution of the secondary particles to a modified Nishimura-Kamata-Greisen (NKG) function. The fit procedure is carried out via the maximum likelihood method \cite{llf}.
Finally, the core position is assumed to be the apex of a conical surface to be fitted to the shower front. The slope of such a conical correction is fixed to $\alpha$ = 0.03 ns/m \cite{icrc05_risang}.

The capability of reconstructing the primary arrival direction  can be further enhanced by applying robust statistical methods in the analysis of the shower front, conveniently weighting the contribution of the most delayed particles. In detail, we first fit a conical surface to the shower image, by minimizing the sum of the squares of the time residuals. At this stage, all the particles hitting the detector have the same weight $w_i$=1. After computing the RMS of the time residual distribution with respect to such a conical surface, we set K = 2.5$\cdot$RMS as a `scale parameter' and perform the minimization of the square of the time residuals weighted sum, where $w_i$=1 if the particle is onward the shower front, $w_i$=$f((t_i^{exp}-t_i^{fit})/K)$ otherwise. The function $f(x)$ is a common Tukey biweight function \cite{Tukey}. The fit procedure is iterated, every time refreshing the scale parameter, until the last reconstructed direction differs from the previous one for less than 0.1$^\circ$.

\section{EAS Observables}

The main experimental EAS observables sampled at the detection level to reconstruct the properties of the primary cosmic ray at the knee energies are the following:

\vspace{0.3cm}
1) {\bf Charged particle component}

This is the basic observable measured in all EAS experiments. The charged component is dominated by electrons and positrons and is usually sampled by an array of detectors distributed over a large area. The sensitive area is very small, $< 1 \%$ of the instrumented area, therefore this poor sampling is the source of an additional instrumental fluctuation which adds to the intrinsic large fluctuations of the shower development. The majority of EAS arrays do not distinguish between the charged particles, so that $N_e \sim N_{ch}$. From the measurement of the charged particles it is possible to determine the shower core position and, via a Lateral Density Function (LDF), reconstruct the total number of electrons at the observation level (size of the shower $N_{ch}$), from which infer the energy of the primary particle.
In addition, by measuring the shower temporal profile the direction of the incident particle can be reconstructed by a time of flight technique.

The LDF of charged particles is of phenomenological nature, determined via MC simulations for the particular experimental set-up.
The most used LDF is adapted from the Nishimura-Kamata-Greisen (NKG) function \cite{nkg,greisen1}
\begin{equation}
\rho_{ch}(r) = \frac{N_{ch}}{2\pi r_M^2} \frac{\Gamma (4.5-s)}{\Gamma (s) \Gamma (4.5-2s)}\cdot
\left ( \frac{r}{r_M} \right )^{s-2}\cdot \left (1+\frac{r}{r_M} \right )^{s-4.5}
\end{equation}
being $r$ the distance from the shower core.
Here $N_{ch}$ is the total number of charged particles, the parameter $s$ describes the shape of the distribution and is related to the longitudinal development stage (\emph{"age"} parameter). The Moliere radius $r_M$ is the distance within which about $90 \%$ of the total EAS energy is contained ($r_M\sim 80$ $m$ at the sea level, increasing with altitude). The normalization constant is expressed in terms of the Gamma Function $\Gamma$.
We note that, from MC simulations, the charged particle lateral distribution in proton-induced showers results considerably flatter than in the 
photon-induced ones. Therefore the lateral containment parameter $r_M$ must be smaller (typically by a factor of 2, see for example \cite{epas}).

\vspace{0.3cm}
2) {\bf Muonic component}

The muons sampled by air shower arrays are typically in the GeV energy range. In fact, they are measured by shielded detectors (scintillators, water tanks, streamer tubes, Geiger-Muller counters) in order to absorb the electromagnetic component that reduces the muons identification efficiency because of the punch-through of hard $\gamma$-rays or electrons.
The most sophisticated detectors also exploit some tracking devices, to better identify the particles.

Since the muons do not multiply, but only slowly lose energy by ionization, the muon content of an EAS after the maximum attenuates much slower than the electromagnetic one.
This flat shower longitudinal profile reduces the fluctuations in reconstructing the primary energy from the muon size.
But the absence of multiplicative processes induced by muons implies that their number is much lower than that of electrons (in a typical shower $N_{\mu}/N_e\sim 0.1$) with a much flatter lateral distribution.
As a consequence, the muons are typically poorly sampled, i.e., the sampling fluctuations cancel the advantage of reduced showers profile fluctuations.

The LDF of muons can be described by a Greisen-like formula \cite{greisen2} with parameters depending on the particular experimental set-up.
As an example, in the EAS-TOP experiment the data ($E_{\mu} > 1$ GeV) have been described by
\begin{equation}
\rho_{\mu}(r) = C\cdot N_{\mu}\cdot \left (\frac{1}{r_0} \right )^{1.25}\cdot r^{-0.75}\cdot
\left (1+ \frac{r}{r_0} \right )^{-2.5}
\end{equation}
with $r_0=400$ $m$ \cite{eastop_mu}. To minimize the bias due to the extrapolation of the muon lateral density function to unobserved large distances from the shower core, some experiments represent the EAS muon content via the so-called \emph{``truncated muon number''}
\begin{equation}
N_{\mu}^{tr} = 2\pi \cdot \int_{r_1}^{r_2} \rho_{\mu}(r) rdr
\end{equation}
With an appropriate choice of $r_1$ and $r_2$, $N_{\mu}^{tr}$ can be used as a nearly mass-independent energy estimator \cite{kascade_tr}.

The muon component is the one of the most sensitive observable, in combination with the size $N_e$, to select showers induced by different nuclei. 
In fact, for a given observation level, $N_e$ is smaller for heavy showers with respect to proton ones with the same energy, while $N_{\mu}$ is typically larger for heavy primaries.
This can be easily understood in the framework of the superposition model. Cascades induced by heavier nuclei develop and attenuate faster than proton induced showers of the same energy, because they have less energy per nucleon. Nucleons of lower energy produce mesons which also have lower energies, and these decay more often than high energy ones, thus giving rise to more muons. On the other hand, the rapid attenuation of cascades arising from lower energy pions results in less electrons (positrons) in the lower part of the atmosphere. Muons and hadrons can be identified by tracking detectors.

Finally, since the muons are produced in hadron decays, they are almost absent in photon-induced showers. Hence, a measurement of the muon content in sampled showers allows one to improve the signal to background ratio in high energy gamma-ray astronomy (above 10 TeV).

\vspace{0.3cm}
3) {\bf Hadronic component}

The spectrum of EAS hadrons retains important information about the primary cosmic ray spectrum, which is dominated by the lightest component, i.e., the protons.
In addition, the study of the EAS hadronic cores is a useful tool to investigate the hadronic interaction models. In fact, this observable is based on
the energy retained by the incoming primary, and therefore on different features (total cross section and inelasticity) of the high-energy hadronic processes.
In addiction, the hadron rate is an observable highly sensitive to the non-diffractive inelastic cross-section and provides important information about the dissipative mechanisms which occur in the nuclear interactions.

The hadronic component sampling has been performed at different atmospheric depths using various experimental techniques, like calorimeters, emulsions chambers and magnetic spectrometers. The most interesting information is:
\begin{itemize}
\item[a)] the measurement of the total energy content in the EAS core (i.e., the energy not transferred to the e.m. and muonic components as measured at large core distances) and its absorption characteristics;
\item[b)] the measurement of the so-called \emph{``unaccompanied hadrons''}, or \emph{``quasi-unaccompanied hadrons''}, surviving primary protons or primary protons that interact only into the last interaction length above the detector, thus producing small showers events;
\item[c)] the detection of peculiar events as, for example, the multi-structured events correlated to large $p_T$ secondaries produced in hadronic interactions.
\end{itemize}

\vspace{0.3cm}
4) {\bf Underground muons}

The EAS high energy muons originate essentially from the first interactions of the cosmic rays in the atmosphere, thus carrying information about the hadronic interactions (in particular the forward region \cite{forti}) and the nature of the primary. Nevertheless, the measurement limitations due to the fluctuations in the muon multiplicity or to the finite size of the detector make difficult the way back to the primary characteristics \cite{nusex}.
The uncertainties in the reconstruction of the corresponding EAS energy can be reduced, in principle, if the measurement is correlated to a surface EAS array.
Interesting results come from the correlated EAS-TOP/MACRO experiments \cite{eastop_macro}. Similar studies are under way with the combined IceTop/IceCube esperiments \cite{ice-muons}.

\vspace{0.3cm}
5) {\bf Cherenkov light}

The EAS arrays observe the showers at a fixed atmospheric depth. The main tool for removing this limitation is provided by the detection of optical emission in atmosphere (fluorescence or Cherenkov light). Fluorescence light can be exploited at primary energies $> 10^{17}$ eV. At
the knee energies Cherenkov light provides the information about the longitudinal development of the shower, related to the rate of energy released in the atmosphere and therefore to the primary composition.

Optical photons are little absorbed in the atmosphere, therefore the Cherenkov lateral distribution is much broader than that of charged particles. In addition, their density is much higher.
As a consequence, it is possible to construct an array of wide field of view Cherenkov telescopes to perform high sensitivity measurements, with smaller area and wider spacing than necessary for equivalent measurements of charged particles.

The amount of light received from each altitude of the shower by telescopes can be reconstructed via simple geometry calculations once that the shower axis direction and the core distance are known.
The amount of Cherenkov light, strongly correlated with the electron shower size, can be used to estimate $N_e$ as a function of atmospheric depth from which the position of the shower maximum $X_{max}$ can be calculated.
This method is essentially geometrical and is independent of MC simulations except for the reconstruction of the angular distribution of Cherenkov light around the shower axis.

The Cherenkov intensity is proportional to the primary energy, while the slope of the Cherenkov lateral distribution depends on the depth of the shower longitudinal maximum $X_{max}$ and hence on the mass of the primary.
Therefore, the study of cosmic ray composition can be made sampling the lateral distribution, i.e., the Cherenkov photon density as a function of the shower core distance.

The mean $X_{max}$ for a given primary type increases logarithmically with energy at an elongation rate of about 80 g/cm$^2$ per decade, although this value depends on the hadronic model used in the calculation. The expected $X_{max}$ value is similar for two primaries of different
masses but with the same energy per nucleon.

The observations of atmospheric EAS-Cherenkov light allow:
\begin{itemize}
\item a measurement proportional to the total e.m. energy deposit of the shower above the observational level;
\item a real ``calorimetric'' energy measurement, if the shower is completely absorbed;
\item the geometrical reconstruction of the development of the e.m. cascade in the atmosphere, by means of the temporal or angular distribution of the light signal;
\item a reduction of the primary energy threshold;
\item a good angular information on the arrival direction of the primary particle. The arrival direction can be obtained by measuring the time of flight among different telescopes.
\end{itemize}

The first hybrid experiment exploiting a combined measurement between a EAS array and a wide field of view Cherenkov telescope to reconstruct the energy spectrum the primary CR light (p+He) component has been performed by ARGO-YBJ and a prototype of the future array of telescopes of the LHAASO experiment.
A multiparametric analysis combined two mass-sensitive parameters: the particle density in the shower core measured by the charge readout of ARGO-YBJ and the shape of the Cherenkov footprint measured by WFCTA \cite{hybrid15}.

Combined with TeV muon underground observations such technique can provide:
\begin{itemize}
\item information complementary to the shower size $N_e$ and to the number of high-energy muons relevant to the analysis of the primary composition and interactions;
\item a reduction of the EAS primary energy threshold of operation, with the possibility of performing measurements at primary energies very close to the surface energies of the muons reaching the underground level ($E_0\sim$ few TeV), i.e., in an energy region where only primary protons can contribute to the TeV muon flux.
\end{itemize}
Therefore the atmospheric Cherenkov light associated with the TeV muons provide an interesting tool for calibrating the simulations and checking the cross section for pion production in the forward region at energies $E_{\mu}\sim E_0/A$ with a known primary beam.

\section{Analysis Techniques: Energy Spectrum and Elemental Composition}

The difficulty of EAS observations with an array is in the interpretation of the data, i.e., in the reconstruction of the primary particle properties (energy, mass number and arrival direction) from the quantities measured in the experiments.
EAS analysis of CR data consists in the disentanglement of the threefold problem involving primary energy $E$, primary mass $A$ and hadronic interaction details. There is an intrinsic ambiguity in the interpretation of CR data governed by our poor understanding of two basic elements: (1) the shower development, and (2) the elemental composition of the primary flux.
To describe the shower development is crucial our knowledge of the inelasticity as a function of the energy, and of the inelastic cross section.

Strictly speaking, no air shower experiment measures the primary elemental composition of cosmic rays. 
To reconstruct energy and mass of the primary particle two orthogonal measurements, at least, are needed. And a third to test hadronic interaction models.
As mentioned, because of the reduced resolution in the measurement of the primary mass, the majority of shower arrays typically separated events as \emph{"light"} ("proton-like") or \emph{"heavy"} ("iron-like"), with results which critically depend on MonteCarlo predictions.
As a consequence, the results are only displayed as a function of the total energy per particle with the so-called \emph{"all-particle''} energy spectrum, i.e., as a function of the total energy per nucleus, and not per nucleon.

The most recent multi-component EAS experiments enabled the simultaneous determination of several mass-sensitive shower observables, on an event-by-event basis, with high statistics.
Sophisticated analysis techniques have been exploited to infer energy spectra and elemental composition by measuring the correlation between different components (for a review see, for example, \cite{haungs2003,kampert} and references therein).
However, despite large progresses in building new arrays and in the analysis techniques, the key questions concerning origin and propagation of the radiation are still open.
In particular, one of the most important questions to be solved concerns the maximum energy of accelerated particles in CR sources, namely the end of the proton component (the \emph{"proton knee}).
\begin{itemize}
\item \emph{How to obtain the energy spectrum in ground-based experiments?}

This is the first step in the analysis of cosmic ray data.
We measure the spectrum in one observable and make a conversion to the energy spectrum. 
The observable is the shower size because the number of electrons at shower maximum is nearly independent on the primary mass: N$_{e,\,  max}^A\approx$ N$_{e,\, max}^p$. But, the experimental situation is more complicated, because surface detectors usually do not measure the number of electrons at shower maximum. Since heavy primaries reach their shower maximum at smaller depths than the light ones, the number of electrons on ground is expected to be composition sensitive, with a larger electron number for air showers initiated by light primaries, according to a relation of the type
\be
N_e(E,A) = \alpha(A)\cdot E^{\beta(A)}
\ee
where the parameters $\alpha$ and $\beta$ depend on the primary mass $A$. This imply a degeneracy in the reconstruction procedure because to recover the primary energy we must assume a given elemental composition to be measured.

The upper panel of Fig. \ref{fig:rigidity} shows spectra of $p$, $O$ and $Fe$  as a function of energy. For each element a knee at the same
rigidity ($= pc/Ze$) is assumed. In the lower plot the spectra for $O$ and $Fe$ are shifted to the right of the proton curve since showers induced by heavy nuclei are sampled in a later stage with respect to the proton ones of the same energy. Hence, the proton primaries dominate the size spectra, as compared to the energy spectrum, because the heavier nuclei have smaller electron sizes at the observational level. The resulting sum spectrum resembles more the proton size spectrum with its sharp knee.

As already discussed, a composition insensitive method to estimate the primary energy is crucial in order to reconstruct an accurate energy spectrum. That is, if the composition changes across the knee, then the relationship between the measured electron size and inferred energy will also vary. Without an independent technique of assessing the composition, the energy spectrum can be subject to unknown systematic shifts.

%
%%%%%%%%%%%%%%%%%%%%%%%%%%%%%%%%%%%%%%%%%%%%%%%%%%%%%%%%%%%%%%%%%%%%%
\begin{figure}[h]
\vfill  \begin{minipage}[h]{.47\linewidth}
  \begin{center}
    \mbox{\epsfig{file=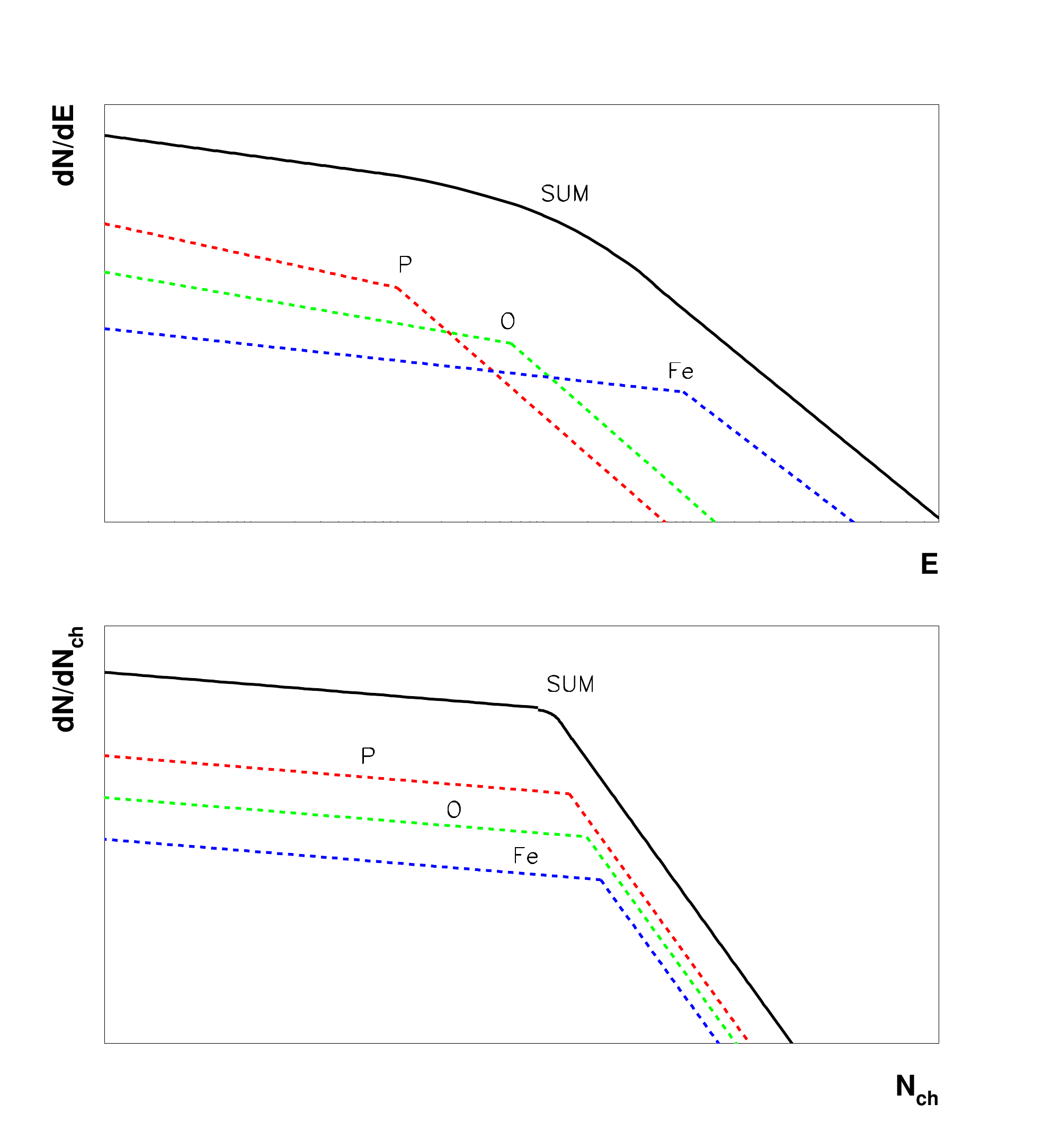,width=6.8cm}}
  \vspace{-0.5pc}
    \caption{Schematic description of the rigidity scheme for the knee. Energy (upper plot) and size (lower plot) spectra with a knee at the same rigidity ($= pc/Ze$) for each element.
    \label{fig:rigidity}}
  \end{center}
%%%\end{figure}
\end{minipage}\hfill
\hspace{-1.cm}
\begin{minipage}[h]{.47\linewidth}
%%%%\begin{figure}[t]
  \begin{center}
    \mbox{\epsfig{file=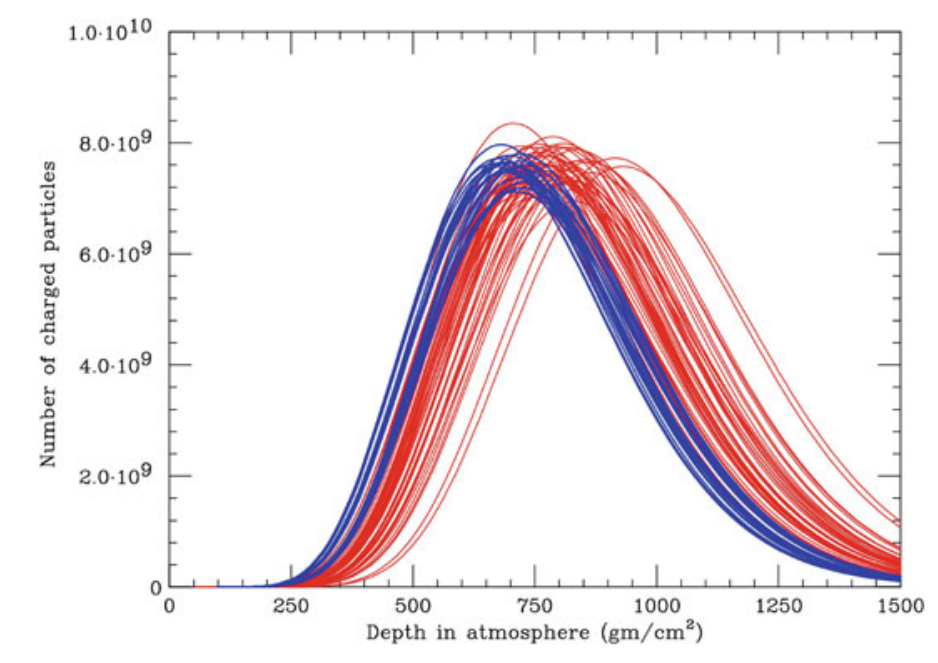,width=6.8cm}}
  \vspace{-0.5pc}
    \caption{Simulation of the longitudinal profile produced with the CORSIKA code for 50 proton-induced (red) and 50 iron-induced (blue) showers \cite{spurio}. The same total energy of 10$^{19}$ eV is assumed. Shower-to-shower fluctuations on $N_e^{max}$ and $X_{max}$ are evident.
 \label{fig:showfluct}}
  \end{center}
\end{minipage}\hfill
\end{figure}
%%%%%%%%%%%%%%%%%%%%%%%%%%%%%%%%%%%%%%%%%%%%%%%%%%%%%%%%%%%%%%%%%%%%
%
The integration of the measured particle densities by means of a phenomenological LDF, determined via MC simulations for the particular array, allows the reconstruction of the total shower size event by event and, consequently, of the size spectrum.
Assuming a given elemental composition we obtain the all-particle energy spectrum.

\item \emph{How do we measure elemental composition at ground?}\\
The different approaches to investigate the elemental composition are commonly based on the fact that inelastic cross section of the nucleus of mass $A$ is proportional to $A^{2/3}$, which leads to longer interaction mean free path (m.f.p.) of protons and short m.f.p. of nuclei. 
For example, the inelastic cross-section $\sigma^{A-Air}_{inel}$ of iron is at 1 PeV approximately six times larger than for protons of equal energy. Hence, nuclei develop higher in atmosphere (smaller $X_{max}$) than protons, producing flatter lateral distributions. 
This imply that showers by nuclei dissipate their energy faster than protons.

Then, increasing the mass $A$
\begin{itemize}
\item More secondary particles with less energy $\to$ less electrons (after maximum), more muons.
\item Surviving hadrons have less energy.
\item Larger deflection angles $\to$ flatter lateral distributions of secondary particles.
\end{itemize}

Measuring electron and muon numbers (and their fluctuations) simultaneously at ground has become the first and most commonly employed technique applied to infer the cosmic ray elemental composition from EAS data. 
As discussed, due the shorter interaction length and the smaller energy per nucleon and because of the reduced attenuation of the muon component, the electromagnetic component of a heavy-particle induced shower contain less particles at ground, but the shower carries more muons than a proton induced shower of the same energy, this is the basis of the electron-muon correlation method.

A number of additional mass-sensitive parameters are typically used by multi-component experiments: the lateral distributions of the different components in the shower core region, the distributions of the relative arrival times and angles of incidence of the muon component, characteristics of the lateral distribution of high energy muons (the so-called \emph{"muon bundles"}) measured underground, pulse shape and lateral distribution of the air Cherenkov light, depth $X_{max}$ of the EAS maximum.

Intrinsic shower to shower fluctuations limit mass resolution. In Fig. \ref{fig:showfluct} the strong fluctuations of the electron size at sea level is shown, with the consequent huge spread in the position of $X_{max}$.
\end{itemize}
Therefore, the general scheme of analysis is the following:
\begin{enumerate}
\item From the experimental data, via some phenomenological functions determined by MC simulations for the particular array, the EAS observables of the particular detector ($N_e$, $N_h$, $N_{\mu}$, $X_{max}$, ...) are reconstructed. 
\item The distributions of such observables are compared with those extracted from a detailed simulation of the EAS development in the atmosphere in which is made use of a trial cosmic rays spectrum. 
\item The input spectrum is changed in order to optimize the agreement between the reconstructed and calculated distributions of EAS observables.
\end{enumerate}
The connection between the nature of the primary particle and the actual experimental results requires a detailed understanding of particle interactions at very high energies and at very forward scattering angles. Since there is still no complete information available from accelerators on this domain, we are faced with an intrinsic ambiguity on the interpretation of cosmic ray data. The ambiguity is governed by our poor understanding of two basic elements: (a) the behaviour of the inelasticity $K$, i.e., the fraction of the primary energy carried away by the secondary particles; (b) the composition of the primary cosmic ray spectrum, i.e., the mass number $A$ of the primary particles. 
Different combinations of these elements can produce similar showers.
A \emph{"short"} shower can be produced by large cross-section, high inelasticity or large mass $A$.
On the contrary, a \emph{"long"} shower can be produced by small cross-section, low inelasticity or small mass $A$.

The typical EAS data analysis consists in finding a combination of primary spectrum, elemental composition and hadronic interaction for a consistent description of all experimental results. In case of discrepancy it is difficult to identify the origin, in case of agreement, is the parameters combination unique?

\section{Analysis Techniques: CR Anisotropy}

The CR arrival direction distribution and its anisotropy has been a long-standing problem ever since the 1930s. In fact, the measurement of the anisotropy is a powerful tool to investigate the propagation mechanisms and the spatial source distribution determining the CR world as we know it.

As CRs are mostly charged nuclei, their paths throughout the Galaxy are deflected and highly isotropized by the action of galactic magnetic field (GMF) they propagate through before reaching the Earth. 
The GMF is the superposition of a regular and a chaotic contribution. Although the strength of the non-regular component is still under debate, the local total intensity is supposed to be $B=2\div 4\textrm{ $\mu$G}$ \cite{beck01}. In such a field, the gyroradius of CRs is given by $r_{a.u.}=100\,R_{\textrm{\scriptsize{TV}}}$, where $r_{a.u.}$ is in astronomic units  and $R_{\textrm{\scriptsize{TV}}}$ is the particle rigidity in TeraVolt.

The high degree of isotropy observed in the CR arrival direction distribution suggests that the propagation in the Galaxy of CRs trapped by the magnetic field can be described in terms of diffusion, at least up to 10$^{16-17}$ eV \cite{berez90}.

The measurement of the anisotropy is complementary to the study of the CR energy spectrum and elemental composition to understand the origin and propagation of the radiation and to probe the structure of the magnetic fields through which CRs travel.
In fact, while the elemental composition of CRs observed at the Earth is a quantity averaged over all possible propagation trajectories and large time intervals, thus mainly probing the diffusive propagation mechanisms, the anisotropy, on the contrary, can give information on the structure of the magnetic field near the solar system.

In principle any anisotropy reflects a motion. 
As an example, three different effects may lead to a CR anisotropy. The first effect is related to the motion of the Earth/Solar System with respect to the isotropic CRs rest frame (the so-called Compton-Getting effect \cite{comptongetting1935}). 
The second is due to nearby and recent CR sources (pulsars or SNRs). For an isotropic propagation, a CR source distribution in the Galaxy is expected to lead to a dipole anisotropy pointing toward the average CR source, with an intensity inversely proportional to the distance to these sources \cite{erlykin2006,blasi2012,pohl2013,sveshnikova2013,battaner2015}. 
The third effect is due to the leakage from the Galaxy.

Data show that the almost perfect isotropy is broken by a dipole-like feature with an amplitude of $\sim$10$^{-4}$ - 10$^{-3}$ evolving with the energy (the so-called \emph{"Large-Scale Anisotropy"}, LSA).
The existence of two distinct broad anisotropy regions in sidereal time, one showing an excess of CRs (called \emph{``tail-in''}), distributed around 40$^{\circ}$ to 90$^{\circ}$ in Right Ascension (R.A.), the other a deficit (the \emph{``loss cone''}), distributed around 150$^{\circ}$ to 240$^{\circ}$ in R.A., has been clearly observed by many experiments with increasing sensitivity and details in both hemispheres (for a review see, for example, \cite{disciascioiuppa13,disciascio15,ahlers16,ahlers2018}). 
%
%%%%%%%%%%%%%%%%%%%%%%%%%%%%%%%%%%%%%%%%%%%%%%%%%%%%%%%%%%
\begin{figure}[t]
  \centerline{\includegraphics[width=0.7\textwidth]{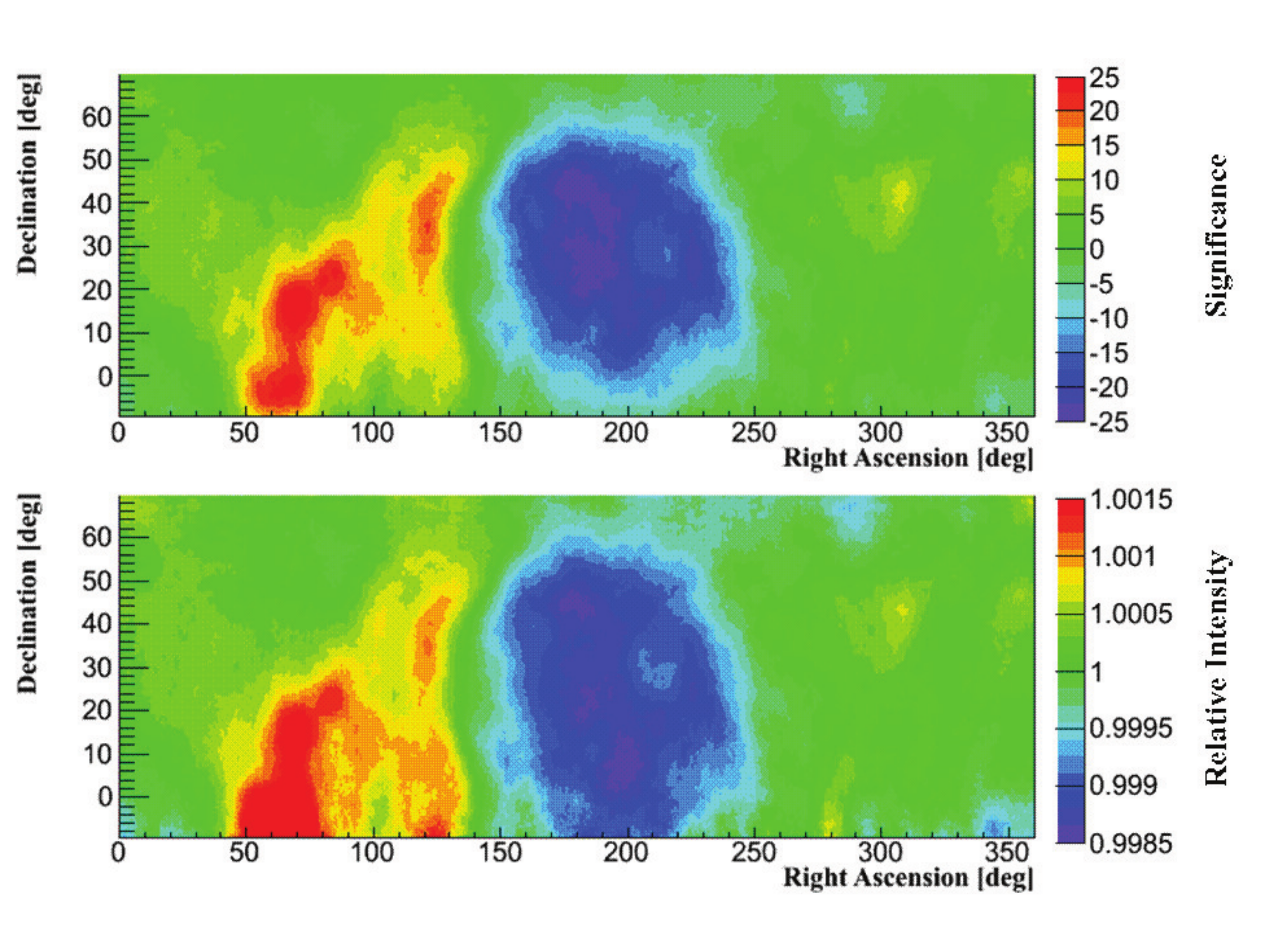}}
\caption[h]{Large scale anisotropy observed by ARGO-YBJ \cite{argo-lsa}. The upper panel shows the significance map in standard deviations (s.d.), the lower panel gives the CR relative intensity. The sky maps are in the equatorial coordinate system. The corresponding proton median energy is about 1 TeV.}
\label{fig:fig1} 
\end{figure}
%%%%%%%%%%%%%%%%%%%%%%%%%%%%%%%%%%%%%%%%%%%%%%%%%%%%%%%%%%
%
%
%%%%%%%%%%%%%%%%%%%%%%%%%%%%%%%%%%%%%%%%%%%%%%%%%%%%%%%%%%%
\begin{figure}[t!]
  \centerline{\includegraphics[width=0.7\textwidth]{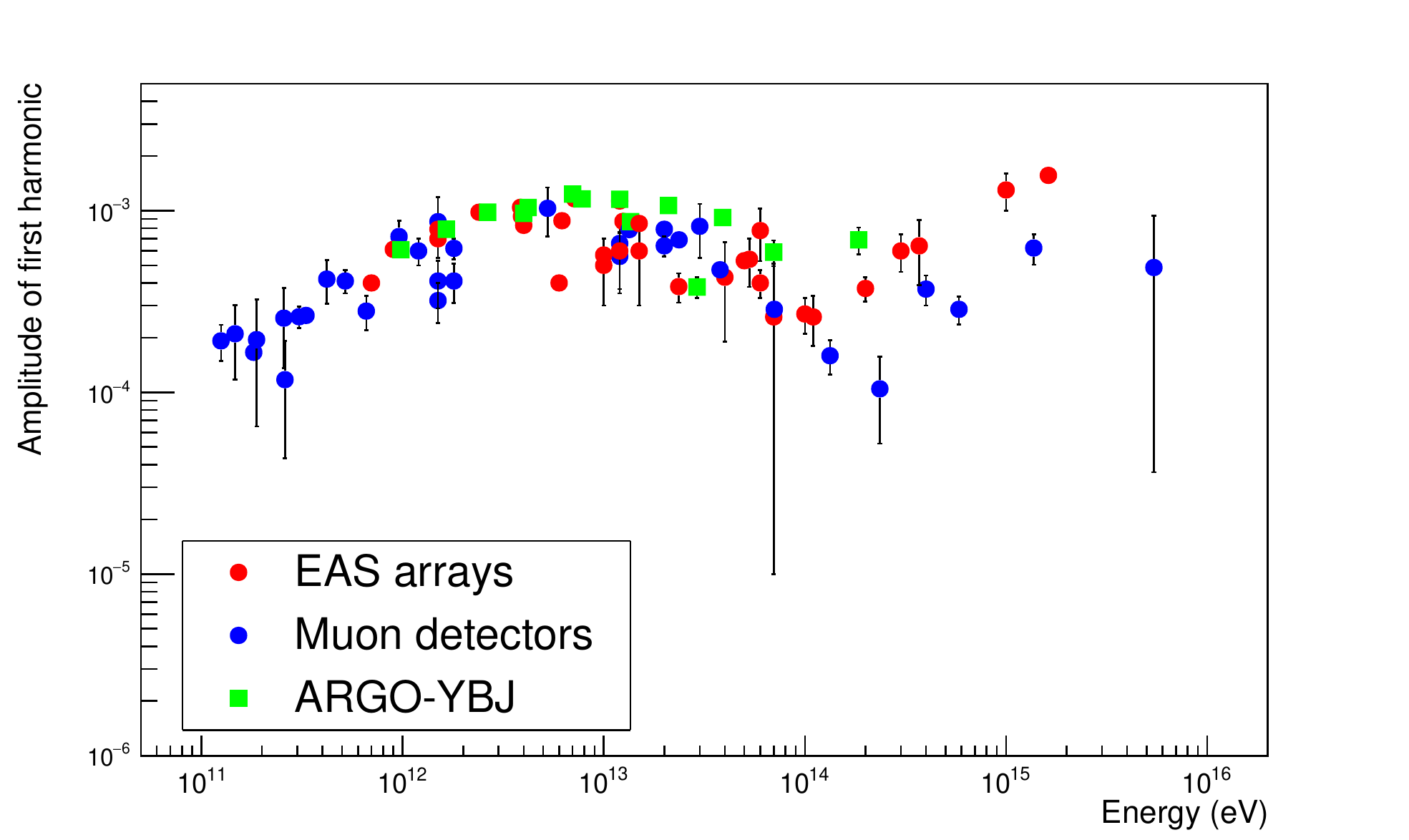}}
    \centerline{\includegraphics[width=0.7\textwidth]{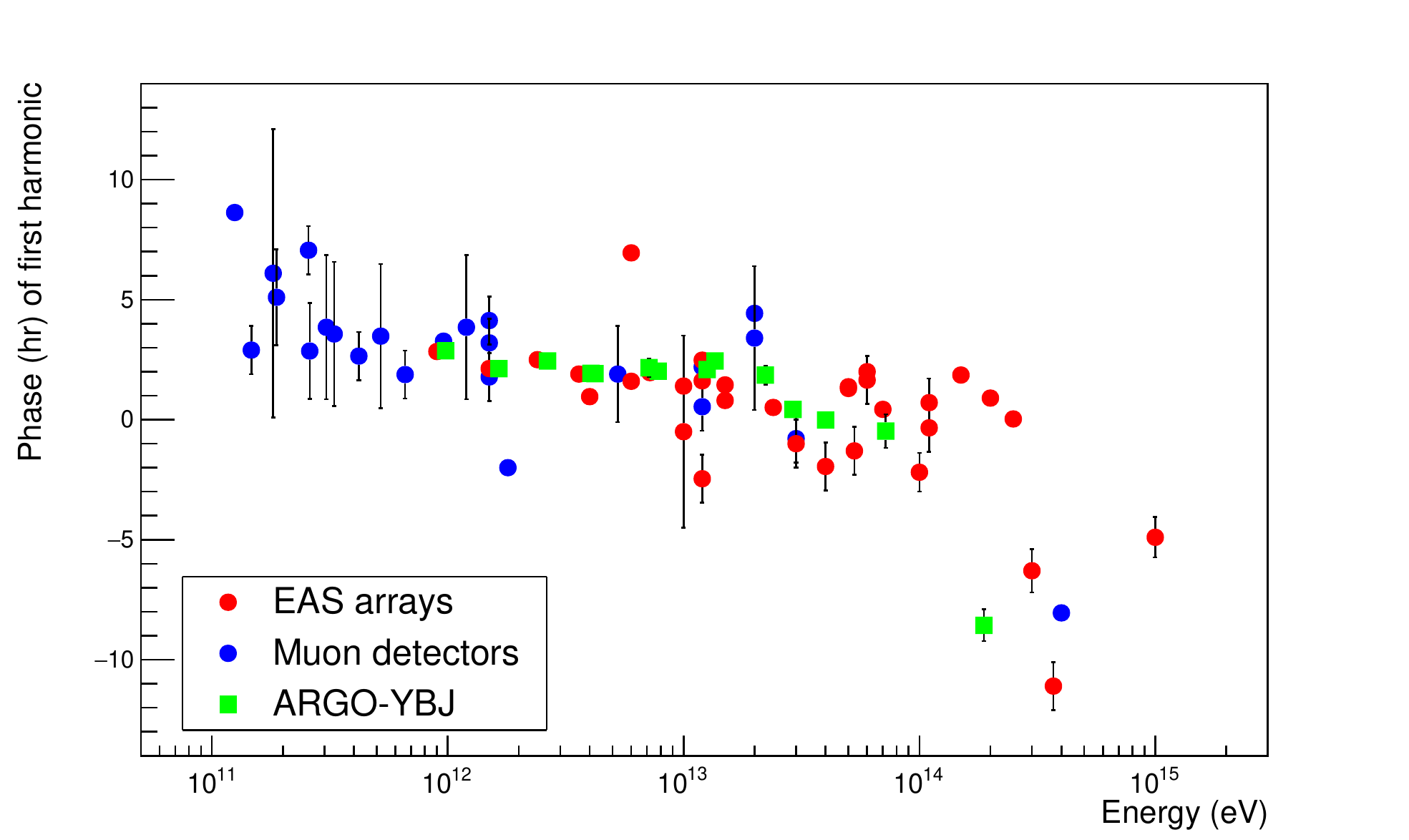}}
\caption[h]{Amplitude and phase of the first harmonic (upper and lower plots, respectively) measured by ARGO-YBJ \cite{argo-lsa,argo-lsaicrc17} compared with a compilation of data obtained by different experiments (muon detectors or EAS-arrays) as a function of the CR primary energy (for details and references see \cite{disciascioiuppa13}).} 
\label{fig:amplit-phase}
\end{figure}
%%%%%%%%%%%%%%%%%%%%%%%%%%%%%%%%%%%%%%%%%%%%%%%%%%%%%%%%%%%
%
%
%%%%%%%%%%%%%%%%%%%%%%%%%%%%%%%%%%%%%%%%%%%%%%%%%%%%%%%%%%
\begin{figure}[t]
  \centerline{\includegraphics[width=0.7\textwidth]{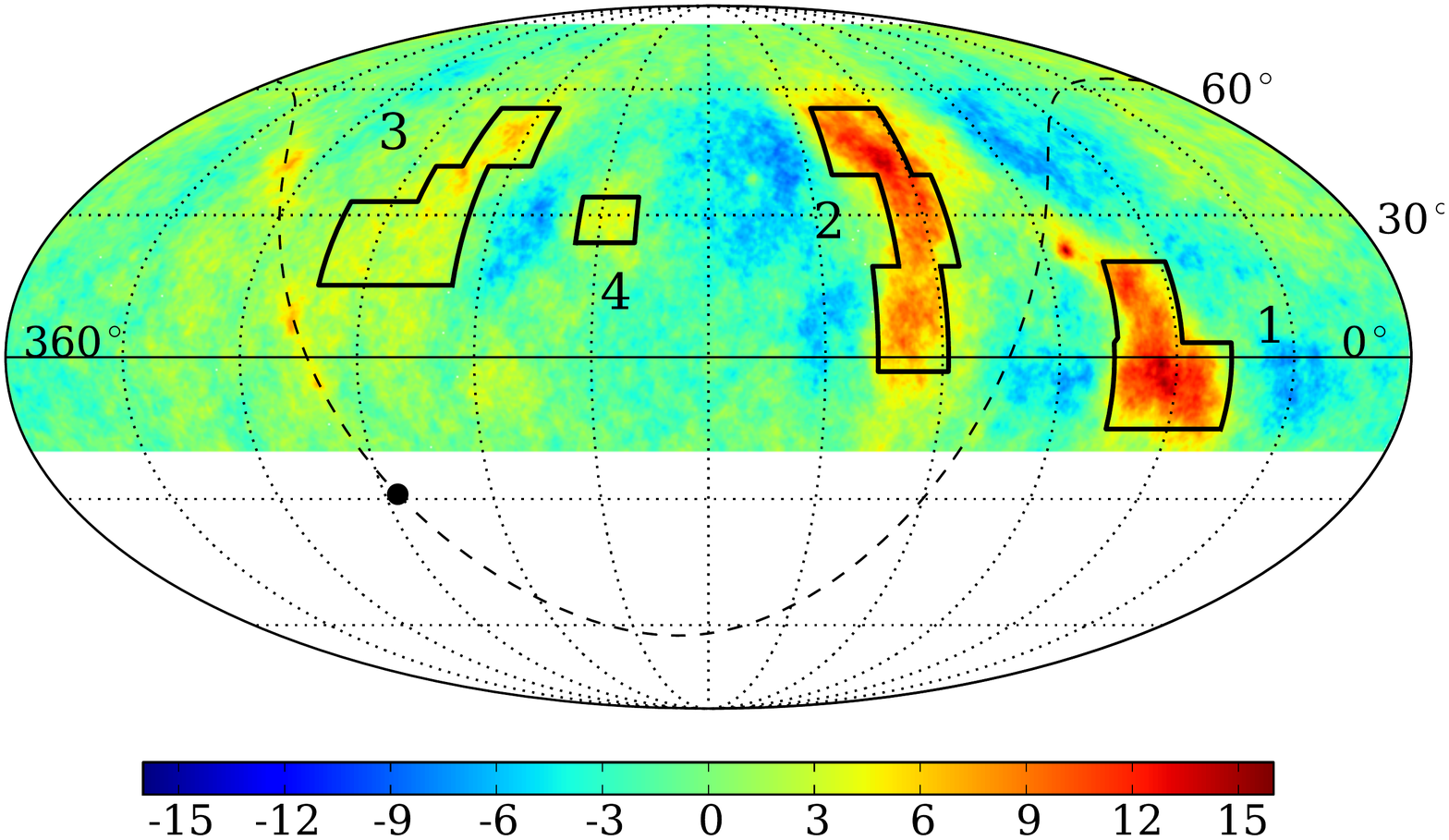}}
\caption{ARGO-YBJ sky-map in equatorial coordinates \cite{argo-msa}. The color scale shows the statistical significance of the observation in s.d.. The dashed line represents the Galactic Plane and the black point the Galactic Center.}
\label{fig:fig3}       % Give a unique label
\end{figure}
%%%%%%%%%%%%%%%%%%%%%%%%%%%%%%%%%%%%%%%%%%%%%%%%%%%%%%%%%%
%
As an example, the LSA observed by the ARGO-YBJ experiment at about 1 TeV in 2008 and 2009, during the latest minimum of the solar activity, is shown in Fig. \ref{fig:fig1} \cite{argo-lsa}.
The center of the ``tail-in'' component is close to the direction of the heliospheric tail, which is opposite to the proper motion direction of the solar system. The center of the ``loss cone'' deficit component points to the direction of the north Galactic pole.
These observations rule out the hypothesis that a Compton-Getting effect due to the motion of the heliosphere with respect to the local insterstellar medium (expected as a dipole with a maximum in the direction of the Galactic Center decl.$\simeq$49$^{\circ}$, R.A.$\simeq$315$^{\circ}$ and a larger amplitude 3.5$\times$10$^{-3}$) is a major source of the anisotropy.

In Fig. \ref{fig:amplit-phase} the amplitude and phase of the first harmonic (upper and middle plots, respectively) measured by different experiments (muon telescopes or Extensive Air Shower (EAS) arrays) are shown as a function of the primary CR energy. 
As can be seen from the plots
\begin{itemize}
\item[a)] The amplitude of the CR anisotropy is extremely small (10$^{-4}$ to 10$^{-3}$).
\item[b)] A slow increase of the amplitude to a maximum at few TeV is observed. After the maximum the anisotropy decreases to a minimum at $\sim$100 TeV. Evidence for a new increase for higher energies appears from data.
\item[c)] The phase of the first harmonic is nearly constant (slowly decreasing) around 0 hrs. A dramatic change of phase is observed around $\sim$100 TeV, suggesting a dipole opposite to the initial one. This observation clearly rules out that the Compton-Getting effect is a major source of the anisotropy.
\end{itemize}
An intriguing result by IceCube \cite{abbasi2012} is the confirmation of the EAS-TOP finding \cite{eastop2009} in the Northern hemisphere, that the anisotropy 'flip' around 100 TeV and its morphology changes. Below about 100 TeV, the global anisotropy is dominated by the dipole and quadrupole components. At higher energies the non-dipolar structure of the anisotropy challenges the current models of CR diffusion. At PeV energies the IceTop experiment showed that anisotropy persists with the same structure as at $\sim$400 TeV, but with a deeper deficit \cite{aartsen2013}.

In recent years, many experiments collected so large statistics to allow the investigation of anisotropic structures at the level of one per-mille and below, see {\it e.g.} \cite{disciascioiuppa13}. Along this line, besides a dominant dipole anisotropy expected from diffusion theory, the observation of some regions of excess down to angular scales of $\sim$10$^{\circ}$ (the so-called \emph{"Medium/Small Scale Anisotropy"}, MSA) stands out \cite{disciascioiuppa13}.

In 2007, modeling the LSA of 5 TeV CR, the Tibet-AS$\gamma$ collaboration ran into a ``skewed'' feature over-imposed to the broad structure of the "tail-in" region \cite{amenomori07,amenomori09}.
Afterwards, Milagro claimed the discovery of two localized regions of excess 10 TeV CRs on angular scales of 10$^{\circ}$ \cite{abdo2008}, observation confirmed by ARGO-YBJ in 2009 \cite{vernetto2009}.
The observation of similar small scale anisotropies has been reported also by IceCube \cite{abbasi11} in the Southern hemisphere.
The importance of this observation lies in the unexpected confinement of a large flux of low rigidity particles in such narrow beams.

The Fig. \ref{fig:fig3} shows the ARGO-YBJ sky map in equatorial coordinates as obtained with about 3.7$\cdot$10$^{11}$ events reconstructed with a zenith angle $\leq$50$^{\circ}$ (selecting the declination region $\delta\sim$ -20$^{\circ}\div$ 80$^{\circ}$) \cite{argo-msa}. According to the simulation, the median energy of the isotropic CR proton flux is E$_p^{50}\approx$1.8 TeV (mode energy $\approx$0.7 TeV). The boxes represent the parametrization of the 4 regions of interest selecting the part of signal more than 3 s.d. .

The most evident features are observed by ARGO-YBJ around the positions $\alpha\sim$ 120$^{\circ}$, $\delta\sim$ 40$^{\circ}$ and $\alpha\sim$ 60$^{\circ}$, $\delta\sim$ -5$^{\circ}$, spatially consistent with the regions detected by Milagro \cite{abdo2008}. These regions are observed with a statistical significance of about 15 s.d. .
On the left side of the sky map, several new extended features are visible, though less intense than the ones aforementioned. The area $195^{\circ}\leq R.A. \leq 290^{\circ}$ seems to be full of few-degree excesses not compatible with random fluctuations (the statistical significance is up to 7 s.d.). We note that the region 4 is located in the "loss cone" of the LSA, near the North Galactic pole.
The observation of regions 3 and 4 is reported by ARGO-YBJ for the first time. Recently HAWC confirmed the observation of the region 4 \cite{hawc-msa}.
We note that the regions over which ARGO-YBJ observes significant MSA have total extension $\sim0.8$~sr, i.e one third of the ARGO-YBJ field of view in celestial coordinates. 

The amplitude of the anisotropy is so small ($\approx 10^{-3}$ or less) that the experimental approach to detect it has to be sensitive on necessity. In order to measure such a tiny effect, large exposures are needed, i.e. large instrumented areas and long-lasting data acquisition campaigns. Until now, only ground-based detectors demonstrated to have the required sensitivity, making use of the Earth's rotation to order data in solar and sidereal time.

As the Earth rotates, the field of view of ground-based detectors points towards different directions at different times, sweeping out a cone of constant declination. 

The CR arrival direction is given after the definition of a reference frame. If Sun-related effects are looked for, a system of coordinates co-moving with the Earth has to be used. Usually, it is a spherical system centered on the Earth. Instead, if phenomena originated outside the heliosphere are considered, a reference frame co-moving with the Sun is set, usually a spherical system centered there\footnote{Or on the Earth, that is the same because of the distances under consideration.}. Whatever the spherical system is, because of the Earth rotation, the direction $\phi$ is periodically observed from ground-based detectors, that is why the $\phi$ axis is regarded as a time axis. The ``solar'' or ``universal'' time is used when the Sun is wanted to be at rest in the reference frame, whereas the ``sidereal'' time is used when the Galactic center is wanted to be at rest.

The \emph{sidereal anisotropy} is spoken about when the sidereal time is used, whereas the \emph{solar anisotropy} is measured by using the solar time. The \emph{anti-sidereal} time analysis is often performed too, as no signal is expected in this non-physical reference frame and any positive result there can be used as estimation of the systematic uncertainty on the sidereal time detection.

Once the reference frame is defined, data are collected and ordered accordingly. There are a number of ways in which the degree of anisotropy of a given distribution can be defined. Perhaps the most traditional one in CR physics is:
\begin{equation}
  \delta=\frac{I_{max}-I_{min}}{I_{max}+I_{min}}
  \label{eq:delta_anisotropy}
\end{equation}
where $I_{max}$ and $I_{min}$ are the maximum and the minimum observed intensity. This definition is the most general to be given, as no hypothesis on the form of the anisotropy is needed: either it is a peak in a smooth distribution or a di-polar modulation, the definition (\ref{eq:delta_anisotropy}) can be applied. 

In 1975 Linsley proposed to perform analyses in right ascension only, through the so-called ``harmonic analysis'' of the counting rate within a defined declination band, given by the field of view of the detector \cite{linsley75}.
In general, one calculate the first and second harmonic from data by measuring the counting rate as a function of the sidereal time (or right ascension), and fitting the result to a sine wave. The Rayleigh formalism allows to evaluate the amplitude of the different harmonics, the corresponding phase (the hour angle of the maximum intensity) and the probability for detecting a spurious amplitude due to fluctuations of a uniform distribution.

Let $\alpha_1,\alpha_2,\dots\alpha_n$  be the right-ascension of the $n$ collected events. From this data series, the experimenter has to determine the components of the two-dimensional vector amplitude (or, equivalently, the scalar amplitude $r$ and phase $\phi$).

The r.a. distribution $f(\alpha)$ can be represented with a Fourier series:
\begin{equation}
f(\alpha) =\frac{a_0}{2}+ \sum_{k=1}^{\infty} (a_k \cos k\alpha + b_k \sin k\alpha )= \frac{a_0}{2} + \sum_{k=1}^\infty r_k \sin (k\alpha + \phi_k)
\end{equation}
where $a_k=r_k \sin\phi_k$ and $b_k=r_k \cos\phi_k$ are the Fourier coefficients and 
\begin{equation}
r_k = \sqrt{a_k^2 + b_k^2}
\end{equation}
\begin{equation}
\phi_k=\tan\frac{a_k}{b_k}
\end{equation}
are the amplitude and the phase of the $k^{th}$ harmonic, respectively. It is known that the  Fourier coefficients can be computed from the $\alpha_m$ data series by means of the equations:
\begin{equation}
a_k = \frac{2}{n}\sum_{i=1}^n \cos k\alpha_i
\end{equation}
\begin{equation}
b_k = \frac{2}{n}\sum_{i=1}^n \sin k\alpha_i
\end{equation}
where $n$ is the number of data points.
This formalism can be applied up to whichever order $k$, but historically experiments never had the sensitivity to go beyond $k=3$ and anthologies are usually compiled about $k=1$ only.
If a non-zero first harmonic amplitude is found, it is important to estimate the chance probability $P$ that it is a fluctuation of an isotropic distribution.
If the sample $\alpha_1$, $\alpha_2$, ..., $\alpha_n$ is randomly distributed between 0 and 2$\pi$, then, as $n\rightarrow \infty$, the probability $P$ of obtaining an amplitude greater than or equal to $r$ is given by the well-known Rayleigh formula  
\begin{equation}
P(\geq r) = \exp(-k_0)
\end{equation}
where $k_0 = (nr^2)/4$.

In general, harmonic analysis is effective for revealing broad directional features in large samples of events measured with poor angular resolution but good stability.

The technique is rather simple: the greatest difficulty lies in the treatment of the data, that is, of the counting rate themselves. In fact, the expected amplitudes are very small with related statistical problems: long term observations and large collecting areas are required.
Spurious effects must be kept as low as possible: uniform detector performance both over instrumented area and over time are necessary as well as operational stability. In addition, CR experiments typically suffer from large variations of atmospheric parameters as temperature and pressure, which translate into changes of the effective atmospheric depth, affecting the CR arrival rate.

Therefore, the measured rate must be corrected precisely for instrumental and atmospheric effects to prevent any misinterpretation of CR flux variations.

The \emph{East-West method} was designed to avoid performing such corrections, preventing potential subsequent systematic errors introduced by data analysis.
The original idea was proposed in the early 1940s to study asymmetries in the flux of solar CRs \cite{kolhoster41,alfver43,elliot51} and was later applied by Nagashima to extensive air showers (EASs) at higher energy \cite{nagashima89}. It is a differential method, as it is based on the analysis of the difference of the counting rates in the East and West directions \cite{bonino11}. 
By considering this, the method is able to eliminate any atmospheric effect or detector bias producing a common variation in both data groups.

For primary CRs with energy lower than 10$^{12}$ - 10$^{13}$ eV, underground muon detectors have been widely used. 
The muon rate deep underground can be affected by a spurious  periodic modulation as the result of the competition between pion decay and interaction in the upper atmosphere. As the atmosphere cools at night (or during the winter), the density increases and the pions produced in CR collisions with the atmosphere are fractionally more likely to interact than decay when compared with the day (or summer) when the climate it is warmer. The daily modulation introduced by solar heating and cooling is seen when muons are binned in solar-diurnal time; yearly or seasonal modulations are seen when events are binned with a period of a (tropical) year \cite{macro97,macro02}.
With increasing energy an increasing fraction of high energy secondary pions interact in the higher atmosphere instead of decaying to muons and the muon intensities decrease too rapidly to yield sufficient statistics for high-energy primary CRs.

The detection of EASs by employing ground detector arrays with large extensions (from 10$^{4}$ to 10$^{9}$ m$^{2}$) and long operation time (up to 20 years), extended the anisotropy studies up the highest energy region. The information about the individual shower was very limited but the very high counting rate, in particular for arrays located at high altitude, made this energy range attractive for anisotropy studies.  

%
%%%%%%%%%%%%%%%%%%%%%%%%%%%%%%%%%%%%%%%%%%%%%%%%%%%%%%%%%%
\begin{figure}[t]\centering
\includegraphics[width=0.7\linewidth]{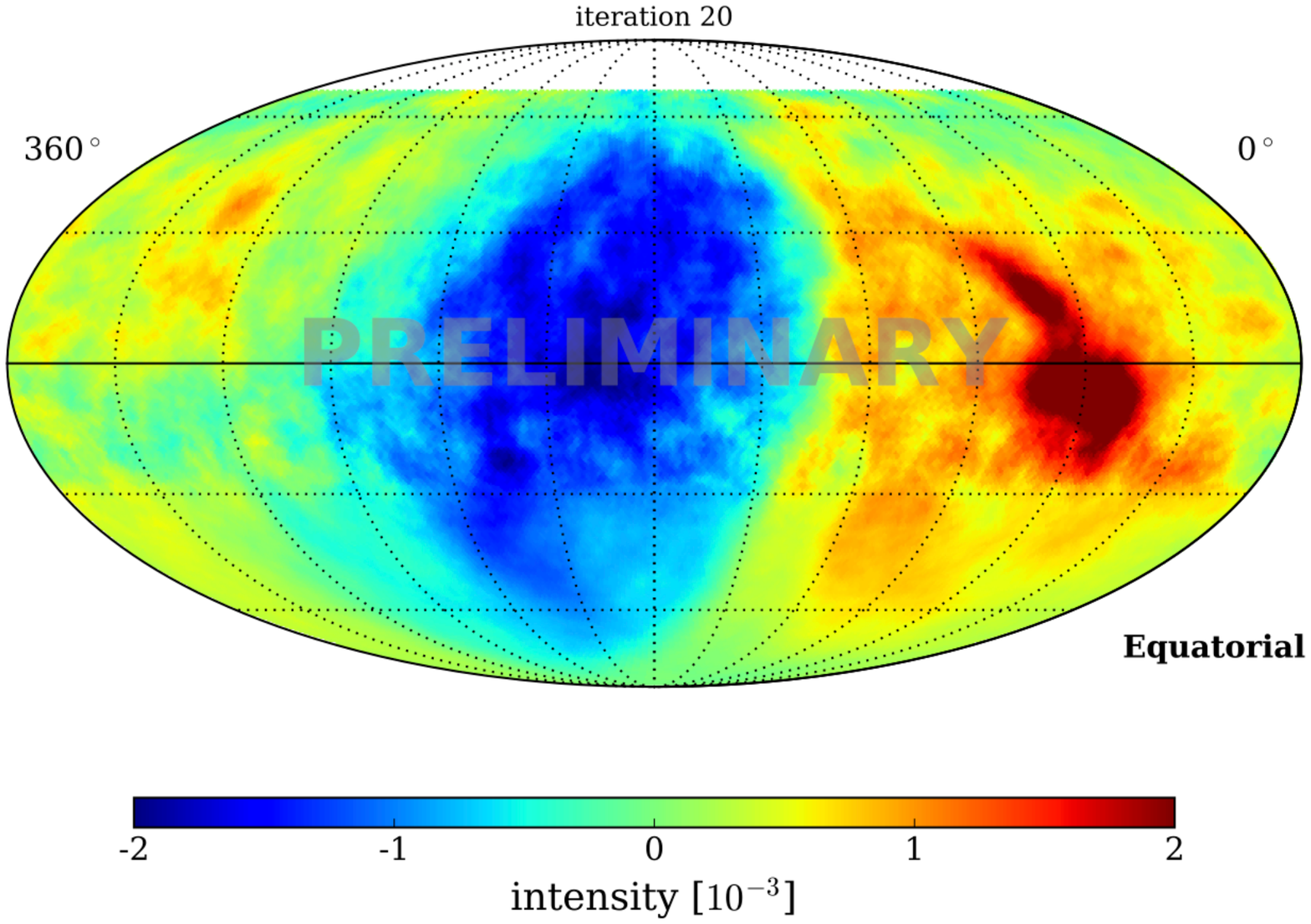}
\includegraphics[width=0.7\linewidth]{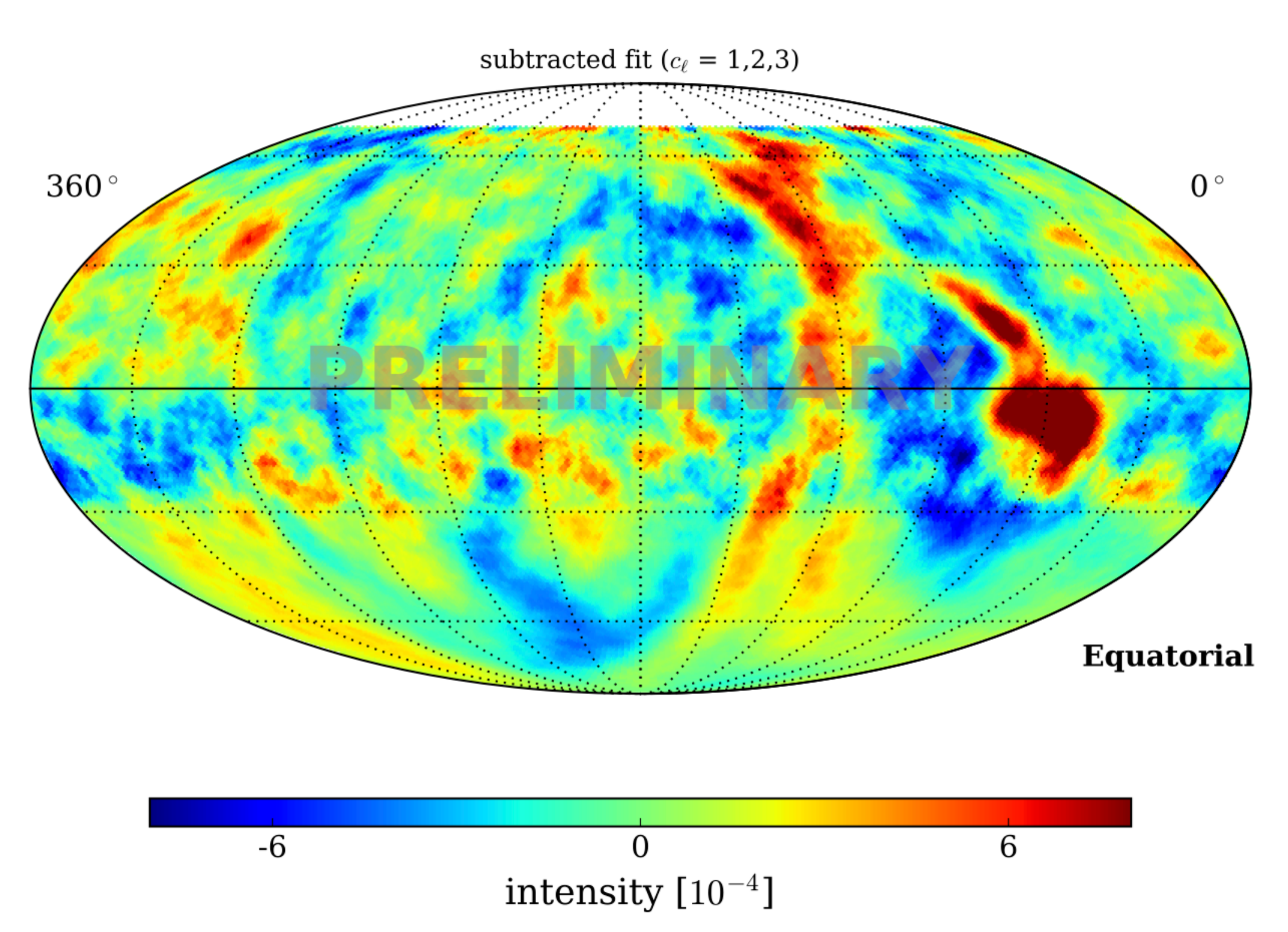}
\caption[]{Preliminary results of the 10~TeV cosmic ray anisotropy from a combined analysis of HAWC ($-30^\circ\leq\delta\leq64^\circ$) and IceCube ($-90^\circ\leq\delta\leq-20^\circ$) data (from Ref. \cite{ahlers2018}). The upper map show the full anisotropy in the equatorial coordinate system. The lower map shows the same anisotropy after removal of the dipole ($\ell=1$), quadrupole ($\ell=2$) and octopole ($\ell=3$).}
\label{fig:hawc-icecube}
\end{figure} 
%%%%%%%%%%%%%%%%%%%%%%%%%%%%%%%%%%%%%%%%%%%%%%%%%%%%%%%%%%
%

The observation of anisotropy effects at a level of 10$^{-4}$ with an EAS array is a difficult job, since it is complicated to control this kind of devices.
A wrong estimation of the exposure may affect the CR arrival rate distribution, even creating artifacts (i.e. fake excesses or deficit regions).
In fact, drifts in detector response and atmospheric effects on air shower development are quite hard to be modeled to sufficient accuracy\footnote{In fact, the temperature can modify the lateral extension of a shower trough the variation of the Moliere radius, and the pressure can influence the absorption of the electromagnetic component.}.
The envisaged solution is to use the data to estimate the detector exposure, but data contain either signal and background events, so that some distortions could be present in the results. 
The shape and the size of possible artifacts depend on the characteristic angle and time scale over which all the aspects of the data acquisition vary more than the effect to be observed \cite{iuppadisciascio13}.
Therefore, if an anisotropy of the order 10$^{-4}$ is looked for, operating conditions must be kept (or made up) stable down to this level,
all across the field of view and during all the acquisition time.
In addition, as it is well known, the partial sky coverage carried out by a given experiment will bias the observations of structure on largest scales.
Finally, the observation of a possible small angular scale anisotropy region contained inside a larger one relies on the capability for suppressing the anisotropic structures at larger scales without, simultaneously, introducing effects of the analysis on smaller scales \cite{iuppadisciascio13}.

Both multi-directional and uni-directional detectors have been widely used to investigate different declination intervals defined by the center direction of viewing of the telescopes. 

Old generation experiments were not able to reconstruct the arrival direction of the primary, but the anisotropy induces periodicity in the response from a CR detector, having a fundamental frequency corresponding to the duration of a sidereal day. Physicists exploited the rotation of the Earth to build 1D distributions of the CR relative intensity.
Due to the poor statistics, any attempt to reconstruct the all-sky structure of the anisotropy was quite unrealistic and the only chance was to check if any strong statistical indication of a genuine CR intensity variation was there or not.

In the last decade, a number of experiments have been able to reconstruct the primary arrival direction with good precision on large amount of data, thus allowing to build two-dimensional (2D) maps of the anisotropy (see, for example, \cite{disciascioiuppa13,hampel17,juancarlos17}).
When 2D sky maps are realized, excesses and deficits observed in one dimension appear to be localized and to have their own extension and morphology. Any hypothesis about the correlation of the observed features with astrophysical sources can be confirmed or denied by the exact information on the direction.

Figure \ref{fig:hawc-icecube} shows a recent (preliminary) anisotropy map from a combined analysis of HAWC and IceCube data that illustrates the complexity of the anisotropy pattern \cite{ahlers2018}.

The arrival directions of Galactic TeV-PeV CRs exhibit complex anisotropy patterns up to the level of one per-mille over various angular scales. In addition, the dipole anisotropy has a strong energy dependence with a phase-flip around 100 TeV. 
The observed dipole data is consistent with the expectation from diffusion theory, if one accounts for the combination of various effects: the anisotropic diffusion of CRs, the presence of nearby sources, the Compton-Getting effect from our relative motion and the reconstruction bias of ground-based observatories (see Ref. \cite{ahlers2018} for further details).

%%%%%%%%%%%%%%%%%%%%%%%%%%%%%%%%%%%%%%
\section {Measurement of the Cosmic Ray Energy Spectrum with ARGO-YBJ}
%%%%%%%%%%%%%%%%%%%%%%%%%%%%%%%%%%%%%%

In this Section we describe the measurement of the CR primary energy spectrum (all-particle and light nuclei component) carried out by the ARGO-YBJ experiment in the energy range from few TeV up to about 10 PeV.
To cover this wide energy interval the ARGO-YBJ Collaboration exploited different approaches:
\begin{itemize}
\item \emph{'Digital-Bayes' analysis}, based on the strip multiplicity, i.e. the picture of the EAS provided by the strip/pad system, in the few TeV -- 300 TeV energy range. The selection of light elements (i.e. p+He) is based on the particle lateral distribution. The energy is reconstructed, on a statistical basis, by using a bayesian approach  \cite{bartoli12,bartoli15}.
\item \emph{'Analog-Bayes' analysis}, based on the RPC charge readout \cite{argo-bigpad}, covers the 30 TeV -- 10 PeV energy range. The energy is reconstructed (as in the previous analysis), on a statistical basis, by using a bayesian approach \cite{argo-rm3knee}.
\item \emph{'Hybrid measurement'}, carried out by ARGO-YBJ and a wide field of view Cherenkov telescope, in the 100 TeV - 3 PeV region. The selection of (p+He)-originated showers is based on the shape of the Cherenkov image and on the particle density in the core region measured by the ARGO-YBJ central carpet \cite{argo-hybrid1,argo-hybrid2}.
\end{itemize}

In the ARGO-YBJ experiment the selection of (p+He)-originated showers is performed not by means of an unfolding procedure after the measurement of electronic and muonic sizes, but on an event-by-event basis exploiting showers topology, i.e. the lateral distribution of charged secondary particles. This approach is made possible by the full coverage of the central carpet, the high segmentation of the read-out and the high altitude location of the experiment that retains the characteristics of showers lateral distribution in the core region.

\begin{figure}
\centerline{\includegraphics[width=0.8\textwidth]{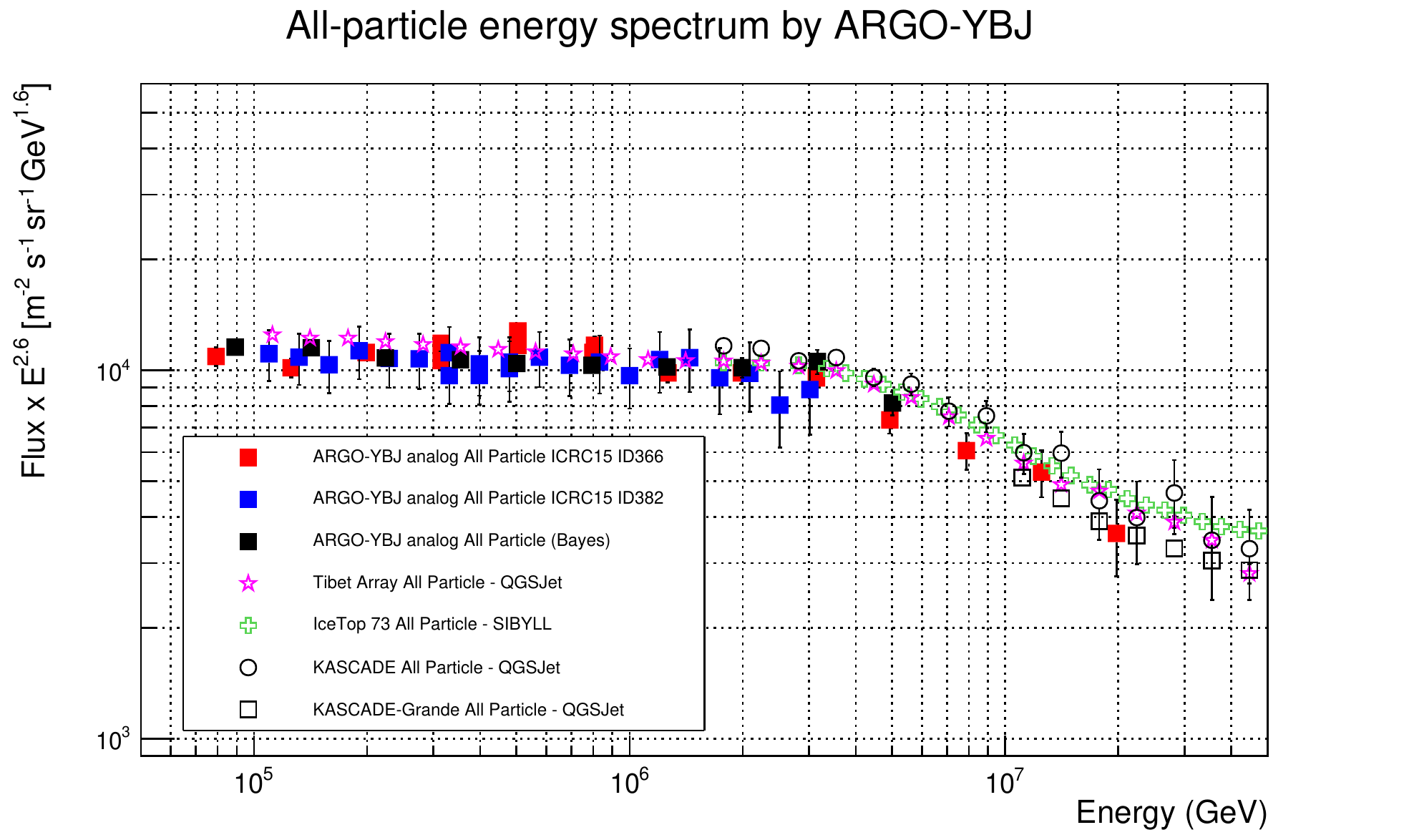} }
\caption{All-particle energy spectrum of primary CRs measured by ARGO-YBJ. Data analyzed with different techniques are compared. The statistical uncertainty is shown by the error bars. For comparison all-particle spectra measured by other experiments (Tibet AS$\gamma$  \cite{tibetIII}, IceTop 73 \cite{icetop73}, KASCADE \cite{kascade}, KASCADE-Grande \cite{kascade-grande}) are shown.}
\label{fig:argo-allpartspectra}       % Give a unique label
\end{figure}

The all-particle energy spectra measured by ARGO-YBJ by reconstructing showers with three different approaches \cite{argo-rm3knee,icrc15-id366,icrc15-id382} are shown in Fig. \ref{fig:argo-allpartspectra}.
The statistical uncertainty is shown by the error bars. A systematic uncertainty, due to hadronic interaction models, selection criteria, unfolding algorithms, aperture calculation and energy scale, of $\pm$15\% is estimated. The ARGO-YBJ all-particle spectrum clearly shows a knee-like structure at few PeVs in fair agreement with the results obtained by Tibet AS$\gamma$ \cite{tibetIII}, IceTop-73  \cite{icetop73}, KASCADE \cite{kascade} and KASCADE-Grande \cite{kascade-grande} experiments. 

%
%%%%%%%%%%%%%%%%%%%%%%%%%%%%%%%%%%%%%%%%%%%%%%%%%%%%%%%%%%
%%%%%%%%%%%%%%%%%%%%%%%%%%%%%%%%%%%%%%%%%%%%%%%%%%%%%%%%%%
%
%
%%%%%%%%%%%%%%%%%%%%%%%%%%%%%%%%%%%%%%%%%%%%%%%%%%%%%%%%%%
\begin{figure}
\centerline{\includegraphics[width=0.8\textwidth]{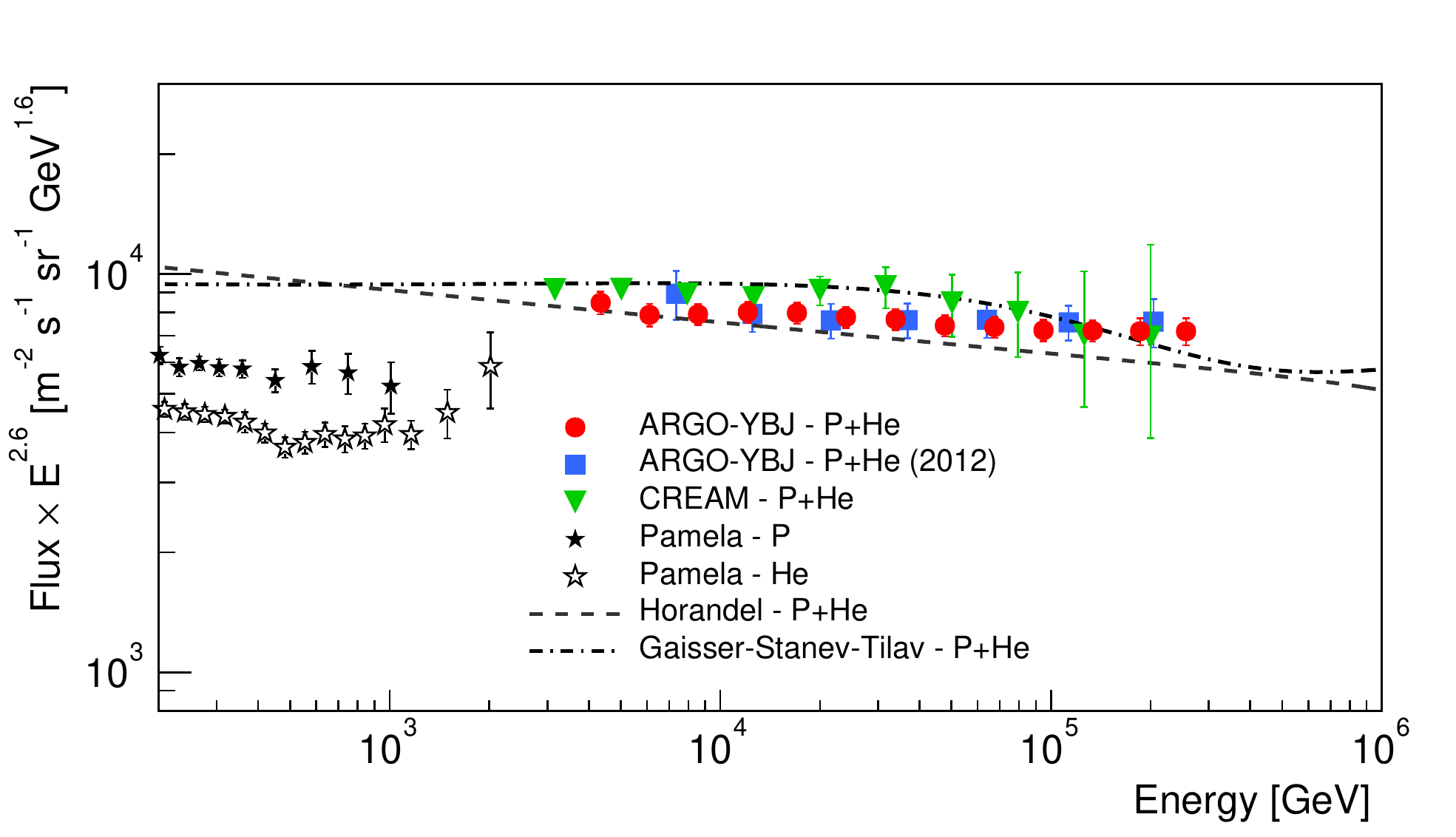} }
\caption{Light (p+He) energy spectrum of primary CRs measured by ARGO-YBJ with the digital readout (strip/pad system) using the full 2008--2012 data sample \cite{bartoli15}. 
The error bars represent the total uncertainty.  Previous measurement performed by ARGO--YBJ in a narrower energy range by analyzing a smaller data sample is also reported (blue squares) \cite{bartoli12}. The green inverted triangles represent the sum of the proton and helium spectra measured by the CREAM experiment \cite{cream11}. The proton (stars) and helium (empty stars) spectra measured by the PAMELA experiment \cite{pamelahe} are also shown. The  light component spectra according to the Gaisser-Stanev-Tilav (dashed--dotted line) \cite{gst} and H\"orandel (dashed line) \cite{horandel} models are also shown.}
\label{fig:argo-digspectra}       % Give a unique label
\end{figure}
%%%%%%%%%%%%%%%%%%%%%%%%%%%%%%%%%%%%%%%%%%%%%%%%%%%%%%%%%%
%

\subsection{Light component (p+He) energy spectrum of Cosmic Rays}

As described in \cite{bartoli12,bartoli15}, by using the read-out provided by the strip/pad system and applying a selection criterion based on the particle density to quasi-vertical showers ($\theta$ $<$ 35$^{\circ}$), a sample of events induced by p and He nuclei, with the shower core well inside the ARGO-YBJ central carpet, has been selected. The contamination by heavier nuclei is found negligible. An unfolding technique based on the bayesian approach has been applied to the strip multiplicity distribution in order to obtain the differential energy spectrum of the light component. 

The light component (p+He) energy spectrum measured by ARGO-YBJ in the few TeV -- 300 TeV energy range is shown in Fig. \ref{fig:argo-digspectra} \cite{bartoli12,bartoli15}.  
Data agree remarkably well with the values obtained by adding up the p and He fluxes measured by CREAM both concerning the total intensities and the spectral index \cite{cream11}. 
The value of the spectral index of the power-law fit to the ARGO-YBJ data is -2.64$\pm$0.01 \cite{bartoli15}.
ARGO-YBJ is the only ground-based experiment that overlaps with the direct measurements for more than two energy decades.

This measurement has been extended to higher energies exploiting an \emph{"hybrid measurement"} with a prototype of the future Wide Field of view Cherenkov Telescope Array (WFCTA) of the LHAASO project \cite{lhaaso}.
The telescope, located at the south-east corner of the ARGO-YBJ detector, about 78.9 m away from the center of the RPC array, is equipped with 16$\times$16 photomultipliers (PMTs), has a FOV of 14$^{\circ}\times$16$^{\circ}$ with a pixel size of approximately 1$^{\circ}$ \cite{wfcta-nim2011}.

The idea is to combine in a multiparametric analysis two mass-sensitive parameters: the particle density in the shower core measured by the analog readout of ARGO-YBJ and the shape of the Cherenkov footprint measured by WFCTA \cite{argo-hybrid1}.

From December 2010 to February 2012, in a total exposure time of 728,000 seconds, the ARGO-YBJ/WFCTA system collected and reconstructed 8218 events above 100 TeV according to the following selection criteria: (1) reconstructed shower core position located well inside ARGO-130, excluding an outer region 2 m large; (2) more than 1000 fired pads on ARGO-130; (3) more than 6 fired pixels in the PMT matrix; (4) a space angle between the incident direction of the shower and the telescope main axis less than 6$^{\circ}$.
This selection guarantees that the Cherenkov images are fully contained in the FOV, an angular resolution better than 0.3$^{\circ}$ and a shower core position resolution less than 2 m.

%%%%%%%%%%%%%%%%%%%%%%%%%%%%%%%%%%%%%%%%
\begin{figure}
\begin{minipage}[ht]{.47\linewidth}
  \centerline{\includegraphics[width=\textwidth]{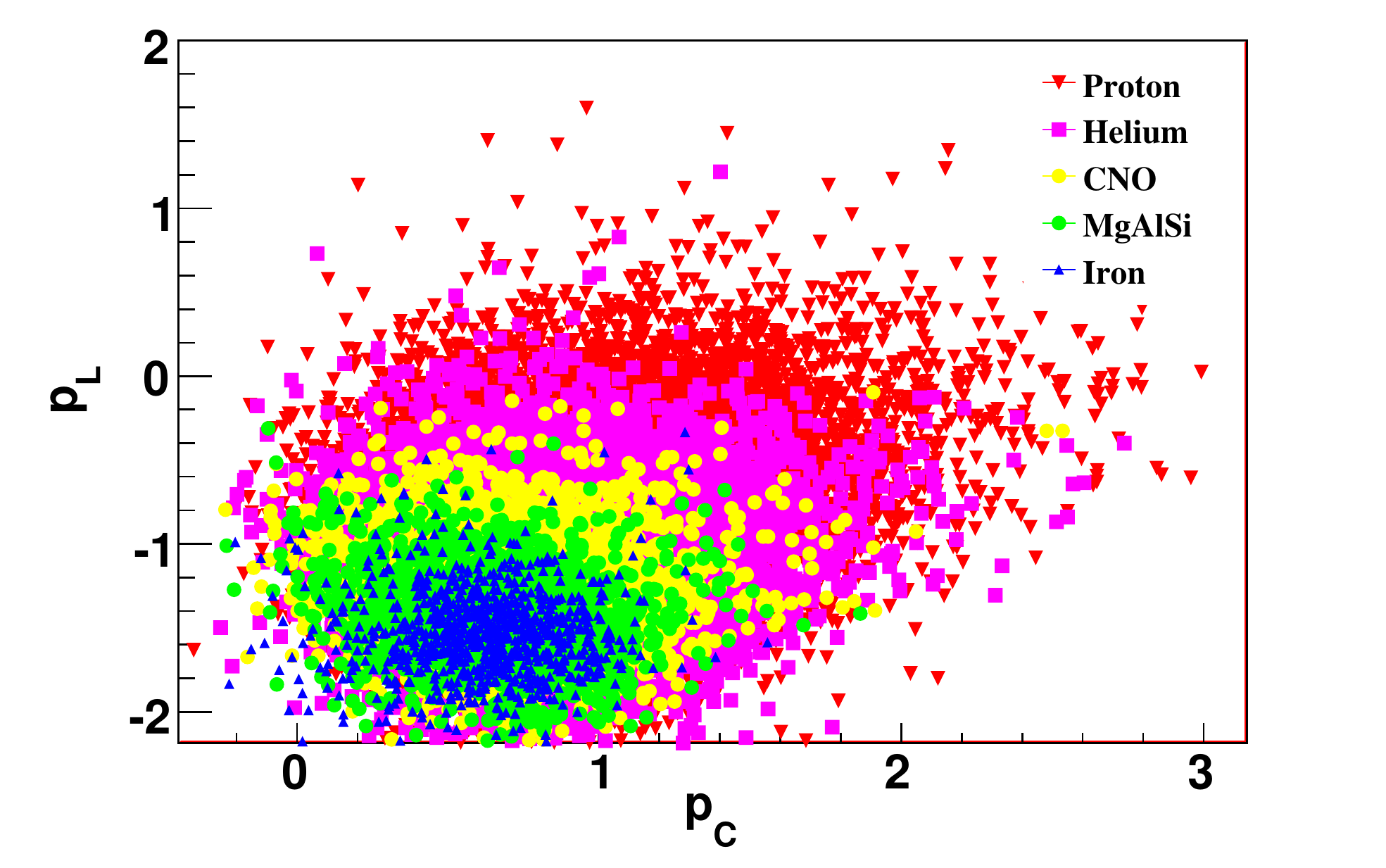}}
    \caption{Scatter plot of the parameters $p_C$ and $p_L$ for showers induced by different nuclei. The primary masses have been simulated in the same relative percentage.}
\label{fig:pl-pc}
\end{minipage}\hfill
\begin{minipage}[ht]{.47\linewidth}
  \centerline{\includegraphics[width=\textwidth]{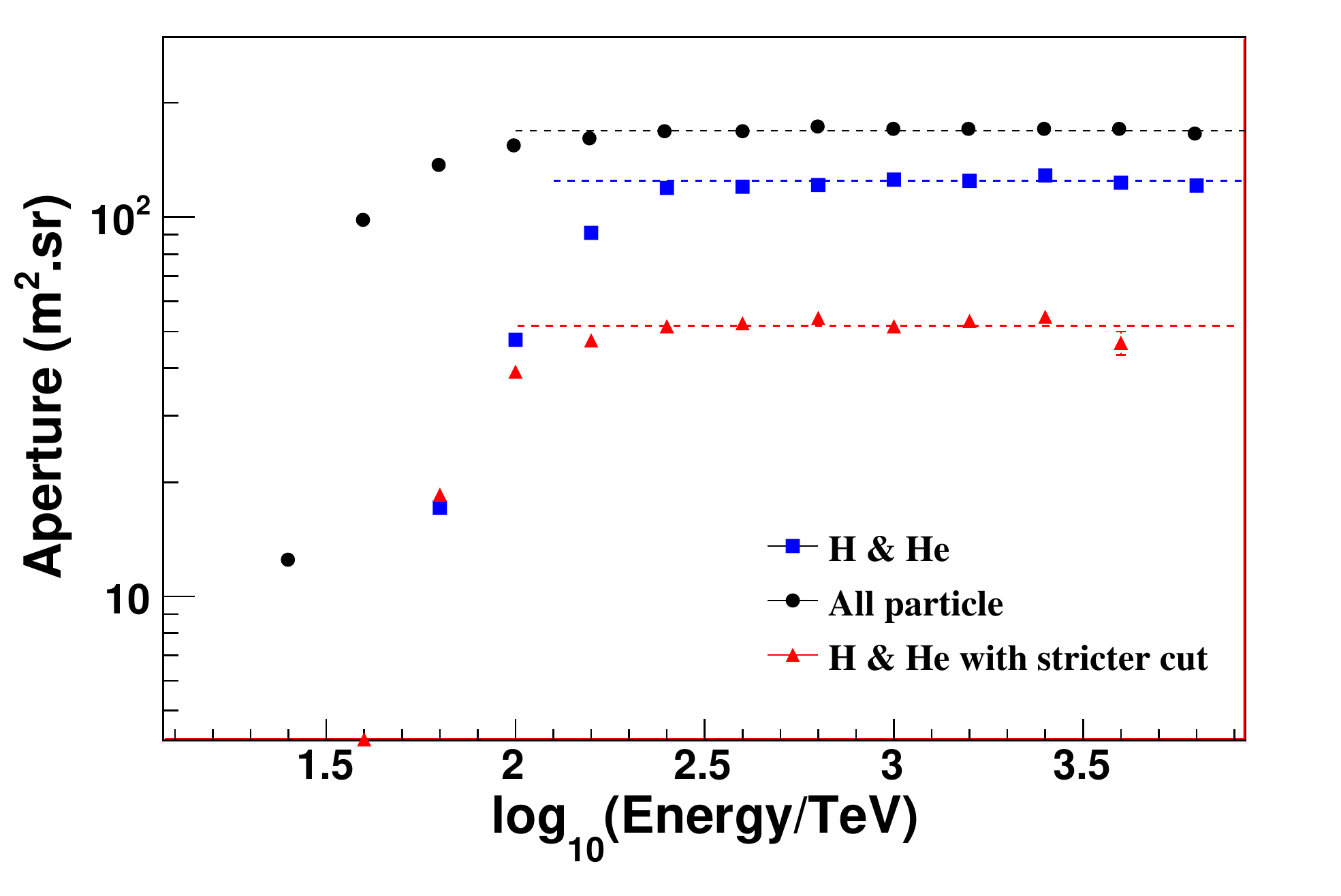} }
    \caption{The aperture of the hybrid experiment ARGO-YBJ/WFCTA. Solid circles represent the aperture for all particles,
   solid squares for the selected p+He events, triangles for the p+He events obtained with stricter cuts for calibration purposes using the low energy part of the spectrum~\cite{argo-hybrid1}.}
  \label{fig:wfcta-aperture}
        \end{minipage}\hfill
\end{figure}
%%%%%%%%%%%%%%%%%%%%%%%%%%%%%%%%%%%%%%%%

According to the MC simulations, the largest number of particles N$_{max}$ recorded by a RPC in a given shower is a useful parameter to measure the particle density in the shower core region, i.e. within 3 m from the core position.
For a given energy, in showers induced by heavy nuclei N$_{max}$ is smaller than in showers induced by light particles. Therefore, N$_{max}$ is a parameter useful to select different primary masses.
In addition, N$_{max}$ is proportional to E$_{rec}^{1.44}$, where E$_{rec}$ is the shower primary energy reconstructed using the Cherenkov telescope.
We can define a new parameter p$_L$ = $log_{10} (N_{max}) - 1.44\cdot log_{10} (E_{rec}/TeV)$ by removing the energy dependence \cite{argo-hybrid1}.

%%%%%%%%%%%%%%%%%%%%%%%%%%%%%%%%%%%%%%%%
\begin{figure}
\begin{minipage}[ht]{.47\linewidth}
  \centerline{\includegraphics[width=\textwidth]{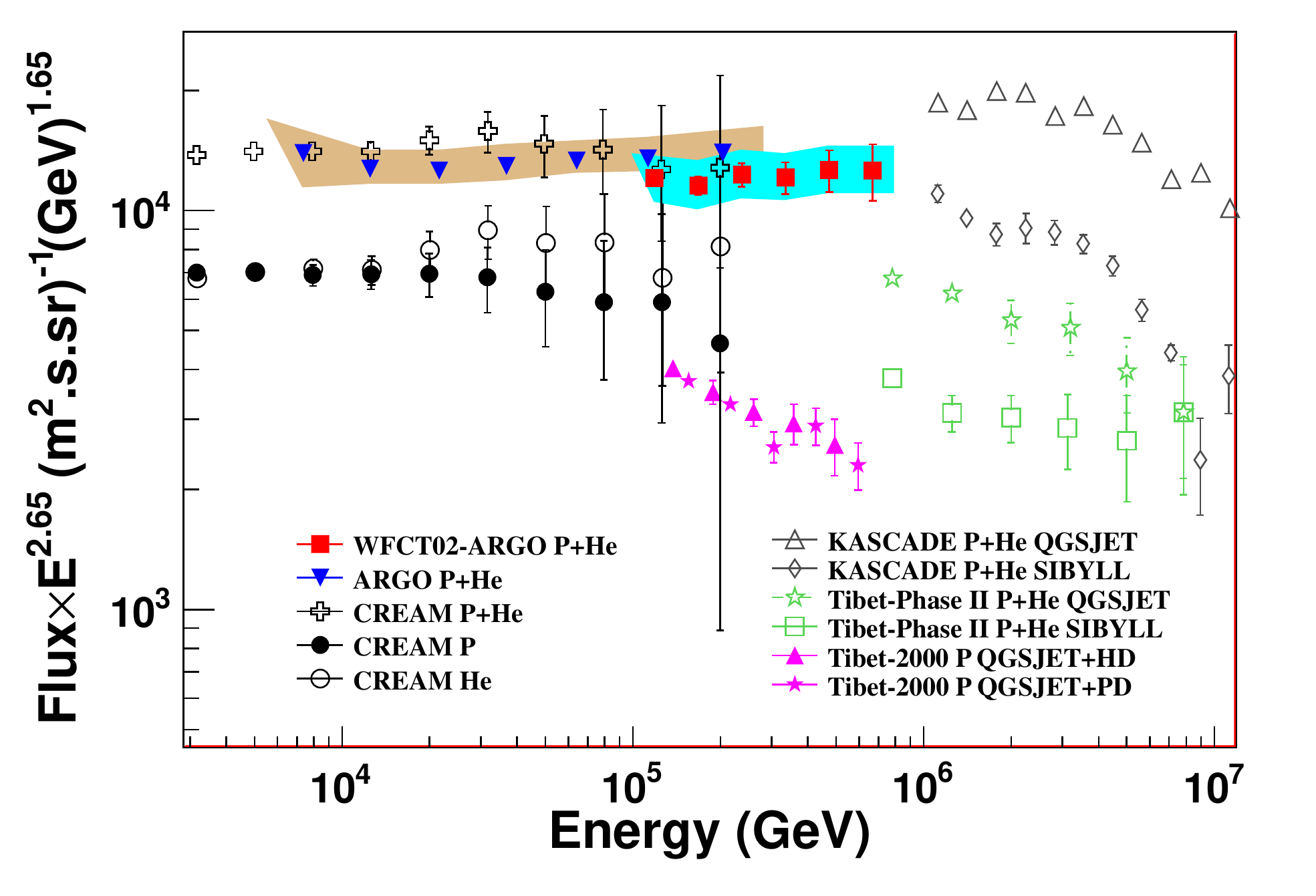}}
    \caption{Light component (p+He) energy spectrum of primary CRs measured by ARGO-YBJ/WFCTA hybrid experiment (filled red squares) in the energy range 100 -- 700 TeV, compared with other experimental results. The ARGO-YBJ data at lower energies are published in \cite{bartoli12}.}
  \label{fig:wfcta-enspt}
\end{minipage}\hfill
\begin{minipage}[ht]{.47\linewidth}
  \centerline{\includegraphics[width=\textwidth]{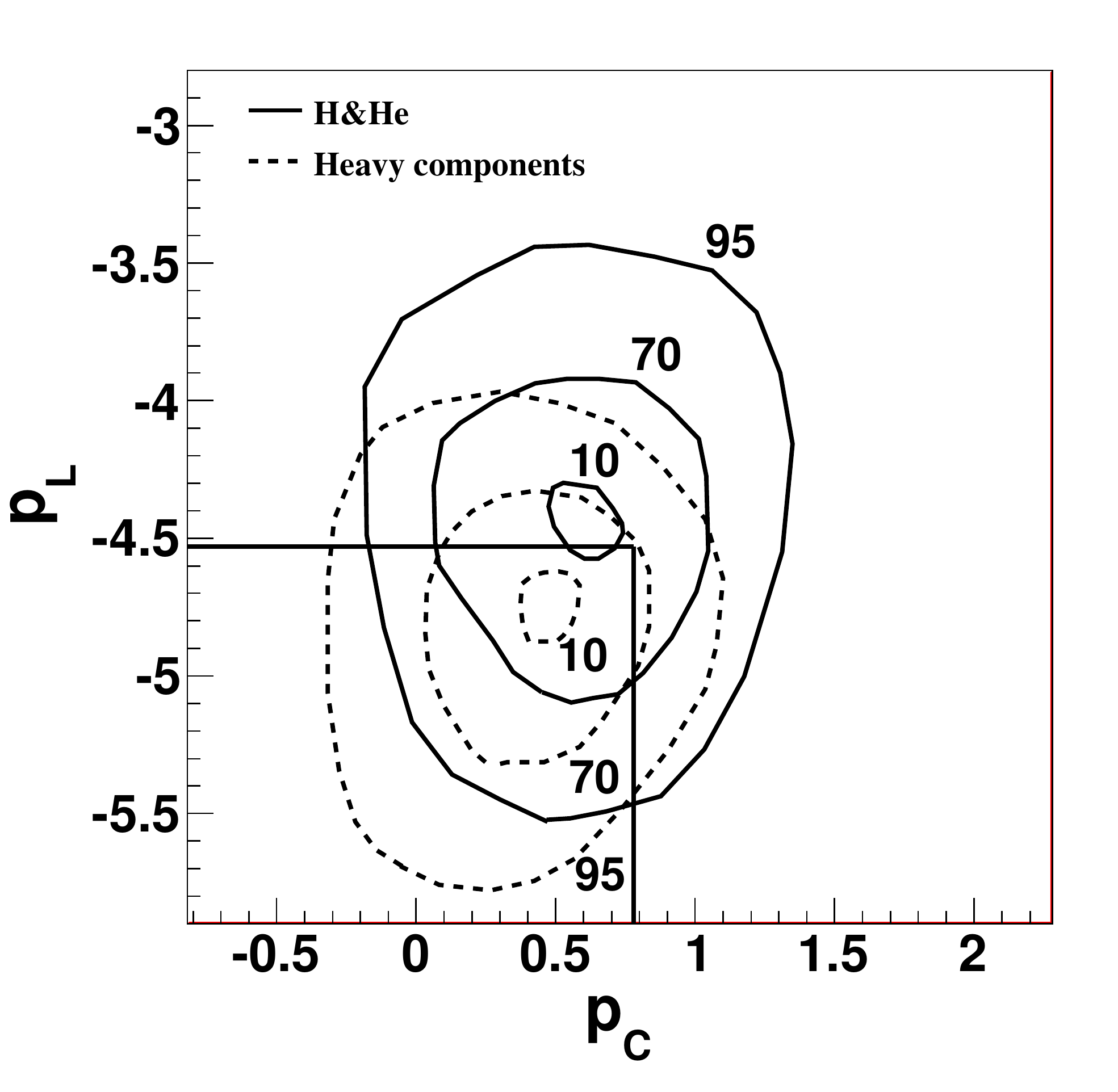}}
    \caption{Composition-sensitive parameters p$_L$ and p$_C$ for (p+He) (solid contours) and heavier masses (dashed contours) including 1:1:1 mixing of CNO, MgAlSi, and Iron. The primary energy of the plotted events is in the range 100 TeV -- 10 PeV. Numbers on the contours indicate the percentage of contained events.}
  \label{fig:pl-pc-2}  
            \end{minipage}\hfill
\end{figure}
%%%%%%%%%%%%%%%%%%%%%%%%%%%%%%%%%%%%%%%%

The Cherenkov footprint of a shower can be described by the well-known Hillas parameters \cite{hillas85}, i.e. by the width and the length of the image. 
Older showers which develop higher in the atmosphere, such as iron-induced events, have Cherenkov images more stretched, i.e. narrower and longer, with respect to younger events due to light particles which develop deeper.
Therefore, the ratio between the length and the width (L/W) of the Cherenkov image is expected to be another good estimator of the primary elemental composition.

Elongated images can be produced, not only by different nuclei, but also by showers with the core position far away from the telescope, or by energetic showers, due to the elongation of the cascade processes in the atmosphere. 
Simulations show that the ratio of L/W is nearly proportional to the shower impact parameters R$_p$, the distance between the telescope and the core position, which must be accurately measured.
An accurate determination of the shower geometry is crucial for the energy measurement. In fact, the number of photoelectrons collected in the image recorded by the Cherenkov telescope N$_{pe}$ varies dramatically with the impact parameter R$_p$, because of the rapid falling off of the lateral distribution of the Cherenkov light. 
Only an accurate measurement of the shower impact parameters R$_p$, and a good reconstruction of the primary energy allow to disentangle different effects.
A shower core position resolution better than 2 m and an angular resolution better than 0.3$^{\circ}$, due to the high-granularity of the ARGO-YBJ full coverage carpet, allow to reconstruct the shower primary energy with a resolution of 25\%, by using the total number of photoelectrons N$_{pe}$. The uncertainty in absolute energy scale is estimated about 10\%.

Therefore, in order to select the different masses we can define another new parameter p$_C$ = $L/W - 0.0091\cdot(R_p/1\>m) - 0.14\cdot log_{10}(E_{rec} /TeV)$ by removing both the effects due to the shower distance and to the energy \cite{argo-hybrid1}.

The values of these parameters for showers induced by different nuclei are shown in Fig. \ref{fig:pl-pc}. The events have been generated assuming a $-2.7$ spectral index in the energy range 10 TeV -- 10 PeV for all the five mass groups (p, He, CNO, MgSi, Fe) investigated. The primary masses have been simulated in the same relative percentage.
As can be seen from the figure, a suitable selection in the p$_L$ -- p$_C$ space allows to pick out a light composition sample with high purity. In fact, by cutting off the concentrated heavy cluster in the lower-left region in the scatter plot, i.e. p$_L\leq$ -0.91 and p$_C \leq$ 1.3, the contamination of nuclei heavier than He is less than 5\%. 
About 30\% of H and He survives the selection criteria \cite{argo-hybrid1}.

The aperture of the ARGO-YBJ/WFCTA system has been estimated using the Horandel model for the primary spectrum \cite{horandel}.
Its value, $\sim$170 m$^2$sr above 100 TeV, shrinks to $\sim$50 m$^2$sr after the selection of the (p+He) component (see Fig. \ref{fig:wfcta-aperture}).
In the sample of 8218 events recorded above 100 TeV by the hybrid system, 1392 showers are selected in the (p+He) sub-sample.

The light component energy spectrum measured by the ARGO-YBJ/WFCTA hybrid system is shown in Fig. \ref{fig:wfcta-enspt} by the filled red squares. A systematic uncertainty in the absolute flux of 15\% is shown by the shaded area. The error bars show the statistical errors only.
The spectrum can be described by a single power-law with a spectral index of $-2.63 \pm 0.06$ up to about 600 TeV. 
The absolute flux at 400 TeV is (1.79$\pm$0.16)$\times$10$^{-11}$ GeV$^{-1}$ m$^{-2}$ sr$^{-1}$ s$^{-1}$. 
This result is consistent for what concern spectral index and absolute flux with the measurements carried out by ARGO-YBJ below 200 TeV and by CREAM. The flux difference is about 10\% and can be explained with a difference in the experiments energy scale of $\pm$3.5\% \cite{argo-hybrid1}. 

This result is very important to fix the energy scale of the experiment. Below 10 TeV the absolute energy scale of ARGO-YBJ is calibrated at 10\% level exploting the westward displacement of the Moon shadow under the effect of the GMF. Above this energy the overposition with CREAM allows to fix the energy scale at few percent level.

 %%%%%%%%%%%%%%%%%%%%%%%%%%%%%%%%%%%%%%
\subsection{Observation of the knee in the (p+He) energy spectrum}
%%%%%%%%%%%%%%%%%%%%%%%%%%%%%%%%%%%%%%

The measurement of the light component energy spectrum has been extended above PeVs exploiting two different approaches.
\begin{itemize}
\item[(1)] The ARGO-YBJ/WFCTA hybrid experiment with different selection cuts in the p$_L$ -- p$_C$ space \cite{argo-hybrid2}.
\item[(2)] A Bayesian unfolding technique applied to data recorded with the RPC charge readout \cite{argo-rm3knee}.
\end{itemize}

\noindent \emph{(1) ARGO-YBJ/WFCTA hybrid experiment.}

In order to extend the measurement of the ARGO-YBJ/WFCTA hybrid experiment to the PeVs, we modified the selection cuts in the $p_L$--$p_C$ space: events for which $p_L\geq$ -4.53 and $p_C\geq$ 0.78 are rejected (see Fig. \ref{fig:pl-pc-2}) \cite{argo-hybrid2}. The aperture is a factor 2.4 larger (see Fig. \ref{fig:wfcta-aperture}). The contamination increase and the purity of the p+He sample below 700 TeV reduces to 93\% with respect to 98\% estimated with the original cuts. At 1 PeV the contamination is less than 13\% increasing to 44\% around 6 PeV.  About 72\% of p+He events survive the selection criteria.

%
%%%%%%%%%%%%%%%%%%%%%%%%%%%%%%%%%%%%%%%%%%%%%%%%%%%%%%%%%%
\begin{figure}
\centerline{\includegraphics[width=0.8\textwidth,clip]{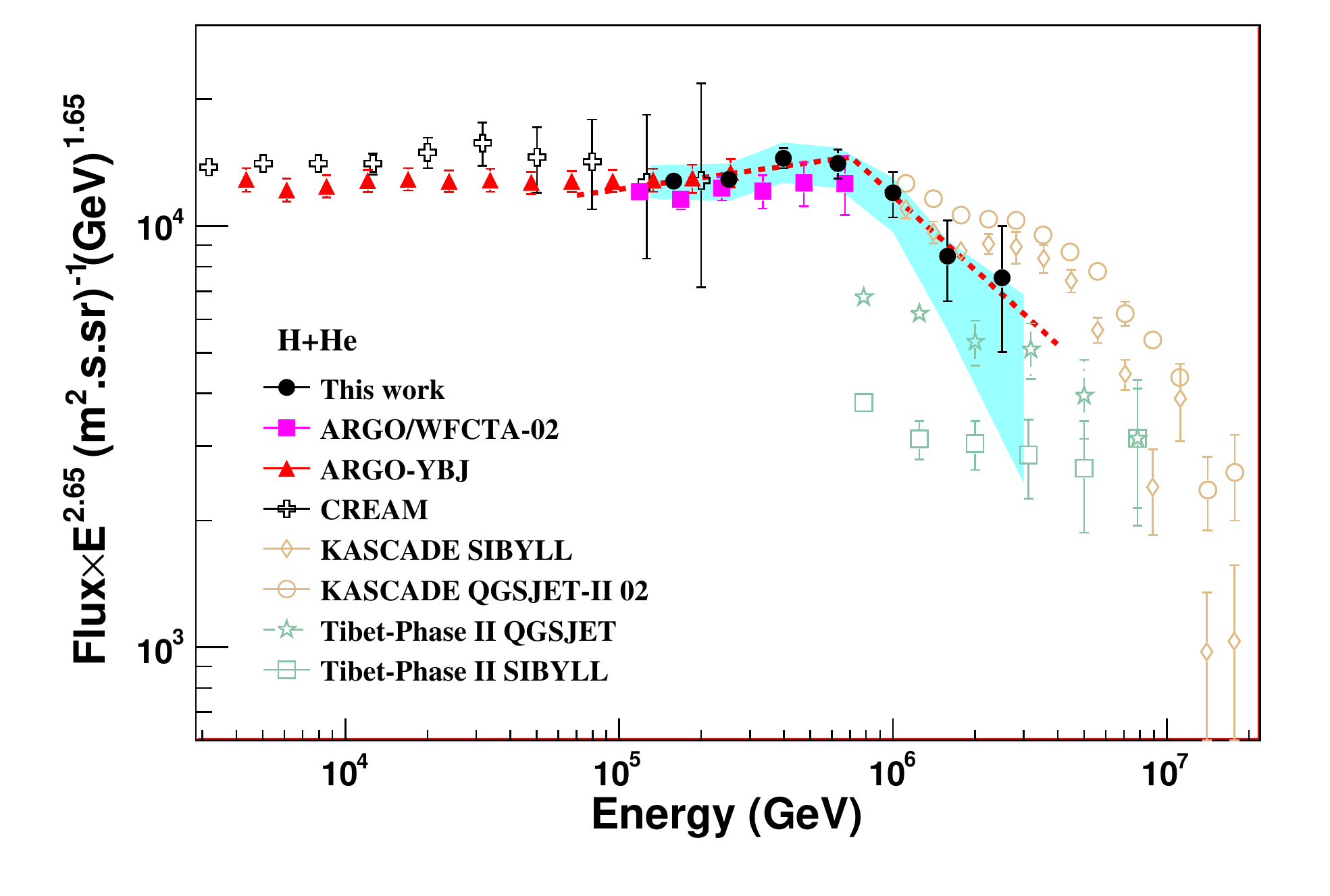} }
\caption{Light (p+He) component energy spectrum measured by the hybrid experiment ARGO-YBJ/WFCTA \cite{argo-hybrid2}. The error bar is the statistical error, and the shaded area represents the systematic uncertainty. The (p+He) spectra measured by CREAM \cite{cream11}, ARGO-YBJ \cite{bartoli15}, Tibet AS$\gamma$ \cite{tibetIII}, KASCADE \cite{kascade} and the hybrid experiment \cite{argo-hybrid1} below the knee are shown for comparison.}
  \label{fig:wfcta-spt2}
     % Give a unique label
\end{figure}
%%%%%%%%%%%%%%%%%%%%%%%%%%%%%%%%%%%%%%%%%%%%%%%%%%%%%%%%%%
%

The resulting energy spectrum is shown in Fig. \ref{fig:wfcta-spt2} and can be fitted with a broken power-law function

\begin{equation}
J(E) = \left\{
  \begin{array}{l l}
  J(E_k) \cdot (E/E_k)^{\beta_{1}}   ~~~~~~~~~~~~~      (E<E_{k})\\
  J(E_k) \cdot (E/E_{k})^{\beta_{2}}  ~~~~~~~~~~~~~ (E>E_{k})
  \end{array}\right.
\end{equation}

with E$_k$= 700 $\pm$ 230 TeV, J(E$_k$)=(4.65$\pm$0.27)$\times$10$^{-12}$ GeV$^{-1}$ m$^{-2}$ s$^{-1}$ sr$^{-1}$, $\beta_1$= -2.56$\pm$0.05 and $\beta_2$= -3.24$\pm$0.36. The relatively large error on the breaking energy E$_k$ is due to the limited statistics.
Considering a systematic uncertainty in the absolute energy scale of about 10\% (see \cite{argo-hybrid2} for a detailed discussion), the systematic uncertainty in $E_{k}$ is estimated to be $\sim$70 TeV.

The number of expected events in the three energy bins above the knee are 82, 39 and 20, respectively. The difference between the observed number of events and the expectation from a single power law spectrum corresponds to a deficit with a statistical significance of 4.2 standard deviations.

\noindent \emph{(2) Bayesian unfolding technique.}
The shower structure in the core region has been deeply investigated thanks to the peculiar characteristics of the ARGO-YBJ detector. According to MC simulations, the truncated shower size ($N_8$), defined as the number of particles within a radius of 8 m from the shower core, is a good estimator of the primary energy for a given mass and is not affected by bias effects due to the finite detector size \cite{bernardini}.

In a shower produced by heavy nuclei a substantial amount of secondary particles is spread further away from the core region. On the contrary, in a shower produced by light elements, the largest amount of particles is concentrated in a small region around the shower core. The ratio between the particle density measured at several distances from the core and the one measured very close to the core can be exploited in order to identify showers produced by light elements. According to Monte Carlo simulations, the parameters ${\beta}_5 = \rho_5/\rho_0$, where $\rho_0$ and $\rho_5$ are respectively the particle density measured in  the core  region and at 5 m from the core and ${\beta}_{10} = \rho_{10}/\rho_0$, where $\rho_{10}$ is the particle density measured  at 10 m from the core, are sensitive to primary mass. The resolution of the core position reconstruction in ARGO-YBJ determines an uncertainty of about 5\% on the measurement of the parameters $\beta_5,\beta_{10}$. Primaries have been grouped into two mass groups: light (H, He) and heavy (CNO, NeMgSi, Fe). The parameters $\beta_5$ and $\beta_{10}$ have been combined in order to estimate the probability that showers have been produced by primaries of different mass. 

The determination of the primary energy from the measured quantities is a classical unfolding problem that can be dealt by using an iterative procedure based on the Bayes' theorem  \cite{argo-rm3knee}. \\
In a probabilistic approach the probability $P(N_{8},{\beta}_{5},{\beta}_{10}|E,A)$ of measuring a shower size $N_8$ and a certain value of $\beta_5$ and $\beta_{10}$ given a primary energy $E$ and mass $A$, relates the characteristics of the primary particle to the experimental observables. 
In a discrete formulation $P(N_{8},{\beta}_{5},{\beta}_{10}|E,A)$ represents the response matrix defined by the following integrals:
\begin{equation}
%\begin{multline}
\label{eq:respmat}
P(N_{8},{\beta}_{5},{\beta}_{10}|E,A)  =   
\frac {\sum_{A}  \int_{\Delta E_o} dE_o \int_{\Delta \Omega} d\Omega \int dS \ \Phi_M(E_o) \cdot  P(N_{8},{\beta}_{5},{\beta}_{10}|E,A)}
{\sum_{A}  \int_{\Delta E_o} dE_o \int_{\Delta \Omega} d\Omega \int dS \ \Phi_M(E_o) }
\end{equation}
%\end{multline}
%
where the flux model and the conditional probability are integrated over the primary energy bin $E^i_o$, the solid angle $\Omega$ and the area S projected on a plane perpendicular to the shower direction.
The response matrix is normalised to the flux integrated over each bin. The flux model affects weighting events in each bin, but since the energy bins
result independent, the flux normalisation cancels out in the response matrix.
The quantity $P(N_{8},{\beta}_{5},{\beta}_{10}|E,A)$ can be evaluated by means of a full Monte Carlo simulation of the shower development in the atmosphere and of the detector response giving the primary energy and mass.
The number of showers  $N(N_{8,i},{\beta}_{5,j},{\beta}_{10,k})$ in the bin $(i\, j\, k)$ of the measured quantities  $(N_{8},{\beta}_{5},{\beta}_{10})$ is therefore related to the number of primaries $N(E_l,A_m)$  by the equation
\begin{equation}
%\begin{multline}
\label{eq:unfolding}
N(E_l,A_m)  =   
\sum_{i,j,k}  P(E_l,A_m|N_{8,i},{\beta}_{5,j},{\beta}_{10,k})\cdot N(N_{8,i},{\beta}_{5,j},{\beta}_{10,k}),
\end{equation}
%\end{multline}
%
where $P(E_l,A_m|N_{8,i},{\beta}_{5,j},{\beta}_{10,k})$ is the probability that a shower of size $N_{8,i}$ and ${\beta}_{5,j},{\beta}_{10,k}$ values has been generated by a nucleus $A_m$ with energy $E_l$. The probability $P(N_{8},{\beta}_{5},{\beta}_{10}|E,A)$ has to be inverted to obtain the primary energy spectrum. 

An iterative procedure based on the Bayes' theorem allows to solve the equation \ref{eq:unfolding} taking into account the bin--to--bin migration due to the fluctuations. Starting from a prior distribution $P^{(n)}_0(E_l,A_m)$ in the $n$th iteration:
\begin{equation}
%\begin{multline}
\label{eq:bayes}
 P^{(n)}(E_l,A_m|N_{8,i},{\beta}_{5,j},{\beta}_{10,k}) = 
 \frac{P^{(n)}(N_{8,i},{\beta}_{5,j},{\beta}_{10,k}|E_l,A_m)P^{(n)}_0(E_l,A_m)}{\sum_{p,q} P^{(n)}(N_{8,i},{\beta}_{5,j},{\beta}_{10,k}|E_p,A_q)P^{(n)}_0(E_p,A_q) } .
%\end{multline}
 \end{equation}
An estimate of the energy spectrum for a give mass $A_m$ is therefore obtained from the distribution of the experimental observables 
%\begin{multline}
\begin{equation}
N^{(n)}(E_l,A_m) = 
 \sum_{i,j,k} P^{(n)}(E_l,A_m|N_{8,i},{\beta}_{5,j},{\beta}_{10,k})\cdot N(N_{8,i},{\beta}_{5,j},{\beta}_{10,k}),
 \end{equation}
%\end{multline}
and is used to obtain an updated value of the prior $P^{(n)}_0(E_l,A_m)$. 
The iterative procedure ends when a variation of the measured flux in two consecutive steps is less than 1\%.

In a discrete formulation of the bayesian unfolding approach, the width of the bins has been chosen in order to better evaluate the conditional probabilities namely minimizing the statistical error, reducing  bin--to--bin migration effects and stabilising the iterative procedure. Data and Monte Carlo events have been sorted in 20 $N_8$ logarithmic bins in the range $(3 \div 6)$ and 3 $\beta_5$ bins $(0 \div 0.5)$ and 3 $\beta_{10}$ bins $(0\div 0.25)$. Monte Carlo events have been sorted in 25 logarithmic energy bins taking into account the energy resolution and 2 mass bins.\\
The fraction of selected light elements increases with energy and is about 60\% above 50 TeV, while contamination is well below 10\% over the whole energy range  \cite{argo-rm3knee}.

An overall picture of the energy range TeV - 100 PeV is shown in Fig. \ref{fig:phe-allp} where light and all-particle energy spectra measured by ARGO-YBJ are summarized. Results obtained by other experiments are also reported.
For comparison, the parametrization of the light component provided by Horandel \cite{horandel} is shown by the dashed line.
As can be seen, the different analyses show clear evidence of a knee-like structure starting from about 700 TeV. 
The energy spectrum measured by the analog read-out can be described by the broken power-law formula (1) with these parameters: 
 E$_k$= 805 $\pm$ 27 TeV, J(E$_k$)=(2.74$\pm$0.25)$\times$10$^{-12}$ GeV$^{-1}$ m$^{-2}$ s$^{-1}$ sr$^{-1}$, $\beta_1$= -2.63$\pm$0.004 and $\beta_2$= -3.76$\pm$0.05, in good agreement with the results obtained by the ARGO-YBJ/WFCTA  measurement.

These results demonstrate the possibility of exploring the CR properties in a wide energy range with a single ground-based detector without exploting the measurement of the muon size, thus reducing the uncertainties due to hadronic interaction models,
%
%%%%%%%%%%%%%%%%%%%%%%%%%%%%%%%%%%%%%%%%%%%%%%%%%%%%%%%%%%
\begin{figure}
\centerline{\includegraphics[width=\textwidth,clip]{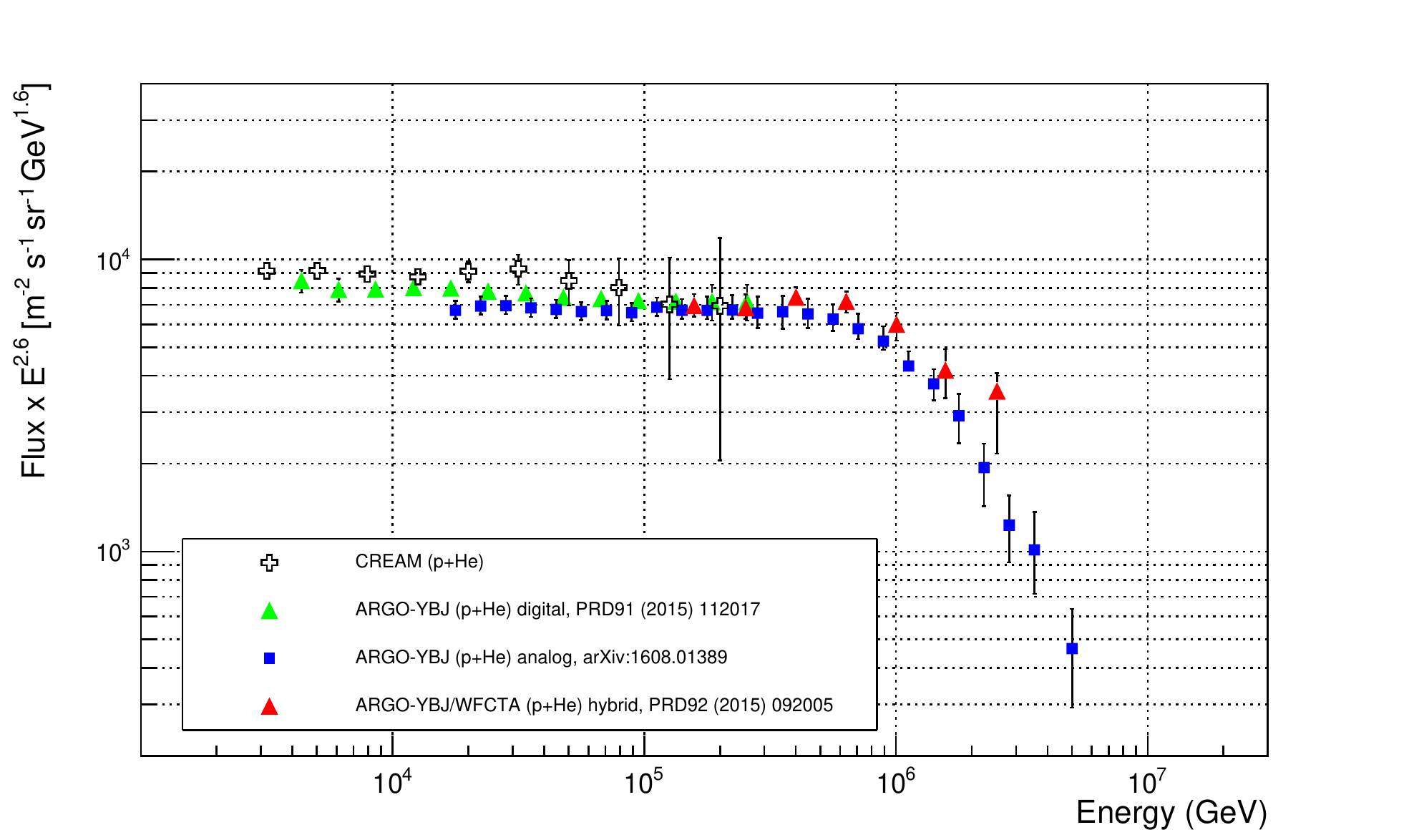} }
\caption{Light (p+He) component energy spectrum of primary CRs measured by ARGO-YBJ. The (p+He) energy spectrum measured by CREAM \cite{cream11} is shown for comparison. The statistical uncertainty is shown by the error bars.}
\label{fig:phe-knee}       % Give a unique label
\end{figure}
%%%%%%%%%%%%%%%%%%%%%%%%%%%%%%%%%%%%%%%%%%%%%%%%%%%%%%%%%%
%

\section{What's Next?}

\subsection{The LHAASO experiment}

A new project, developed starting from the experience of the high altitude experiment ARGO-YBJ, strategically built to study with unprecedented sensitivity the energy spectrum, the elemental composition and the anisotropy of CRs in the energy range between 10$^{12}$ and 10$^{17}$~eV, as well as to act simultaneously as a wide aperture ($\sim$2 sr), continuosly-operated gamma-ray telescope in the energy range between 10$^{11}$ and $10^{15}$~eV is the LHAASO experiment \cite{lhaaso1}.
The remarkable sensitivity of LHAASO in CR physics and gamma astronomy will play a key-role in the comprehensive general program to explore the \emph{``High Energy Universe''}.
%
%%%%%%%%%%%%%%%%%%%%%%%%%%%%%%%%%%%%%%%%%%%%%%%%%%%%%%%%%%
\begin{figure}
\centerline{\includegraphics[width=0.7\textwidth,clip]{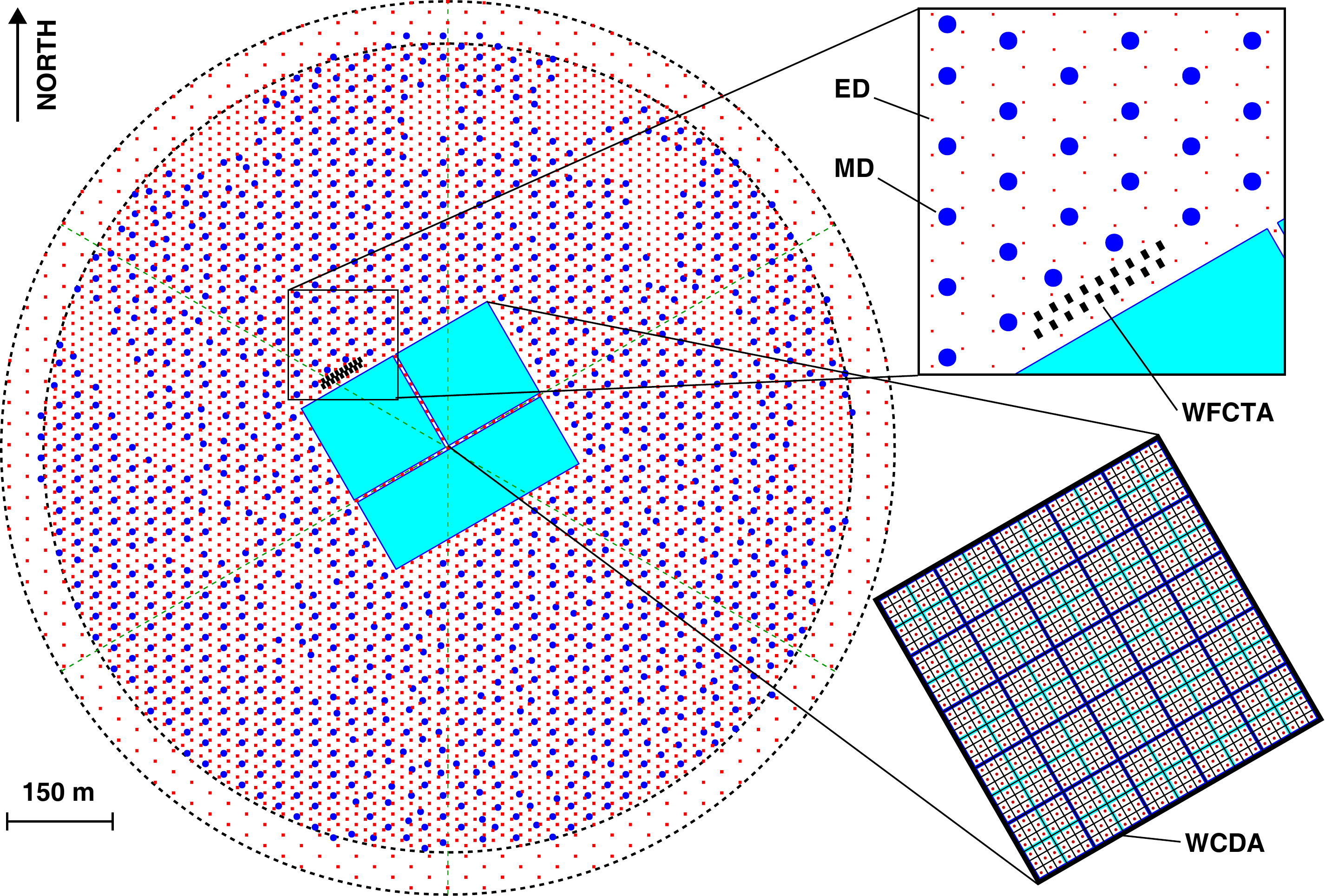} }
\caption{Layout of the LHAASO experiment. The insets show the details of one pond of the WCDA and of the KM2A array constituted by two overimposed arrays of electromagnetic particle detectors (ED) and of muon detectors (MD). The telescopes of the WFCTA, located at the edge of a pond, are also shown.} 
\label{fig:lhaaso-layout}       % Give a unique label
\end{figure}
%%%%%%%%%%%%%%%%%%%%%%%%%%%%%%%%%%%%%%%%%%%%%%%%%%%%%%%%%%
%

The first phase of LHAASO will consist of the following major components (see Fig. \ref{fig:lhaaso-layout}):
\begin{itemize}
\item 1 km$^2$ array (LHAASO-KM2A) for electromagnetic particle detectors (ED) divided into two parts: a central part including 4931 scintillator detectors 1 m$^2$ each in size (15 m spacing) to cover a circular area with a radius of 575 m and an outer guard-ring instrumented with 311 EDs (30 m spacing) up to a radius of 635 m.
\item An overlapping 1 km$^2$ array of 1146 underground water Cherenkov tanks 36 m$^2$ each in size, with 30 m spacing, for muon detection (MD, total sensitive area $\sim$42,000 m$^2$).
\item A close-packed, surface water Cherenkov detector facility with a total area of about 78,000 m$^2$ (LHAASO-WCDA).
\item 18 wide field-of-view air Cherenkov telescopes (LHAASO-WFCTA).
\end{itemize}

LHAASO is under installation at high altitude (4410 m asl, 600 g/cm$^2$, 29$^{\circ}$ 21' 31'' N, 100$^{\circ}$ 08'15'' E) in the Daochen site, Sichuan province, P.R. China. 
The commissioning of one fourth of the detector will start in 2019.
The completion of the installation is expected by the end of 2021.

In Table 1 the characteristics of the LHAASO-KM2A array are compared with other experiments. As can be seen, LHAASO will operate with a coverage of $\sim$0.5\% over a 1 km$^2$ area.
The sensitive area of muon detectors is unprecedented and about 17 times larger than CASA-MIA, with a coverage of about 5\% over 1 km$^2$.

%
%%%%%%%%%%%%%%%%%%%%%%%%%%%%%%%%%%%%%%%%%%%%%%%%%%%%%%%%%%
\begin{figure}
\centerline{\includegraphics[width=0.7\textwidth,clip]{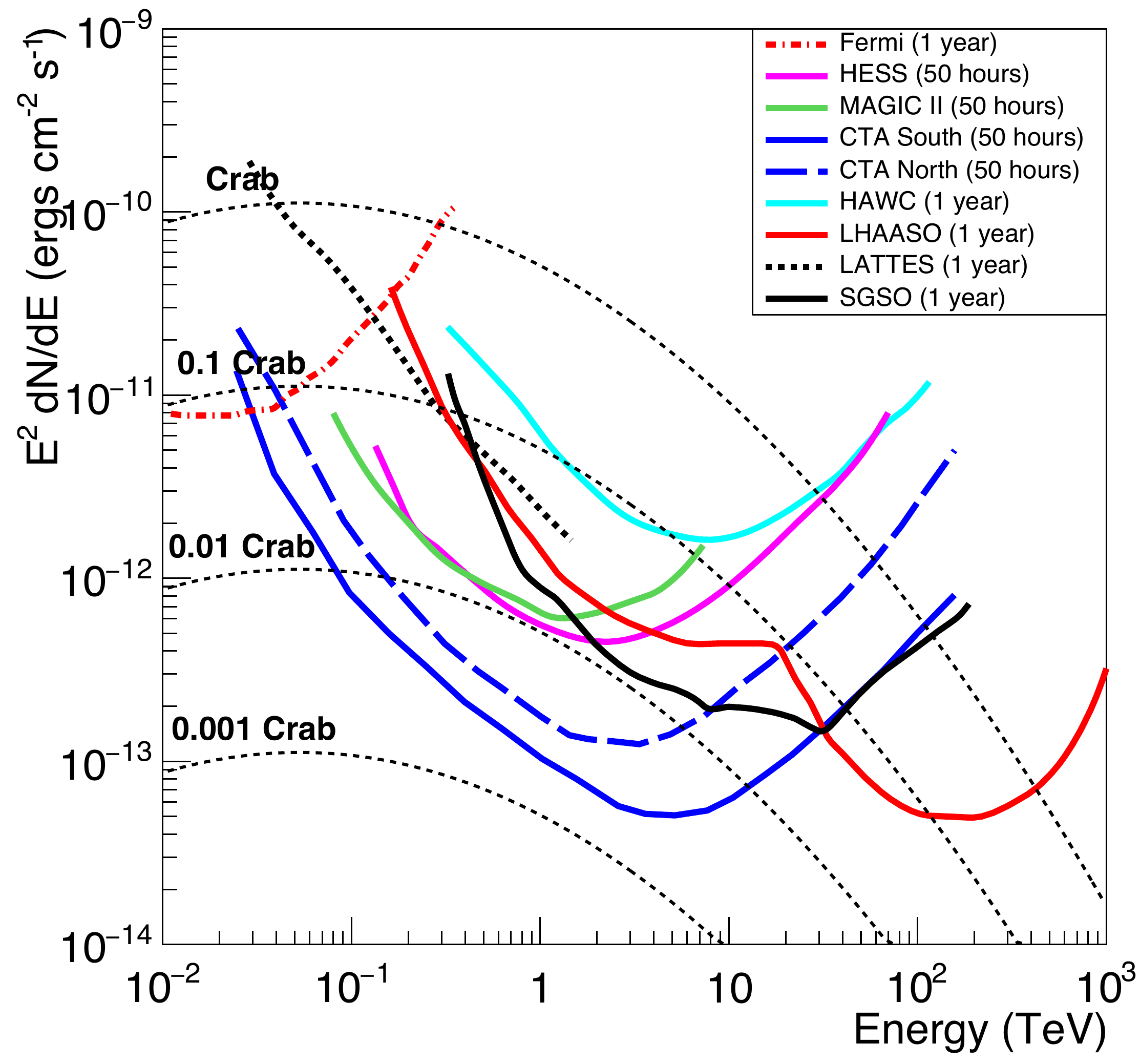} }
\caption{Differential sensitivity of LHAASO to a Crab-like gamma-ray point source compared to other experiments (multiplied by E$^2$). The Crab Nebula spectrum, extrapolated to 1 PeV, is reported as a reference together with the spectra corresponding to 10$\%$, 1$\%$ and 0.1$\%$  of the Crab flux.} 
\label{fig:gamma-diffsens}       % Give a unique label
\end{figure}
%%%%%%%%%%%%%%%%%%%%%%%%%%%%%%%%%%%%%%%%%%%%%%%%%%%%%%%%%%
%

LHAASO will enable studies in CR physics and gamma-ray astronomy that are unattainable with the current suite of instruments:
\begin{itemize}
\item[1)] LHAASO will perform an \emph{unbiased sky survey of the Northern sky} with a detection threshold of a few percent Crab units at sub-TeV/TeV energies and around 100 TeV in one year (Fig. \ref{fig:gamma-diffsens}). This sensitivity grants a high discovery potential of flat spectrum Geminga-like sources not observed at GeV energies.
This unique detector will be capable of continuously surveying the $\gamma$-ray sky for steady and transient sources from about 100 GeV to 1 PeV. \\
From its location LHAASO will observe at TeV energies and with high sensitivity about 30 of the sources catalogued by Fermi-LAT at lower energy, monitoring the variability of 15 AGNs (mainly blazars) at least.
\item[2)] The sub-TeV/TeV LHAASO sensitivity will allow to observe AGN flares that are unobservable by other instruments, including the so-called TeV orphan flares. 
\item[3)] LHAASO will study in detail the high energy tail of the spectra of most of the $\gamma$-ray sources observed at TeV energies, opening for the first time the 100--1000 TeV range to the direct observations of the high energy cosmic ray sources.
\emph{LHAASO's wide field-of-view provides a unique discovery potential.}
\item[4)] LHAASO will map the Galactic \emph{diffuse gamma-ray emission} above few hundreds GeV and thereby measure the CR flux and spectrum throughout the Galaxy with high sensitivity. 
The measurement of the space distribution of diffuse $\gamma$-rays will allow to trace the location of the CR sources and the distribution of interstellar gas.
\item[5)] The high background rejection capability in the 10 -- 100 TeV range will allow LHAASO to measure the \emph{isotropic diffuse flux of ultrahigh energy $\gamma$ radiation} expected from a variety of sources including Dark Matter and the interaction  of 10$^{20}$ eV CRs with the 2.7 K microwave background radiation. 
In addition, LHAASO will be able to achieve a limit below the level of the IceCube diffuse neutrino flux at 10 -- 100 TeV, thus constraining the origin of the IceCube astrophysical neutrinos.
\item[6)] LHAASO will allow the reconstruction of the energy spectra of different CR mass groups in the 10$^{12}$ -- 10$^{17}$ eV with unprecedented statistics and resolution, thus tracing the light and heavy components through the knee of the all-particle spectrum.
\item[7)] LHAASO will allow the measurement, for the first time, of the CR anisotropy across the knee separately for light and heavy primary masses.
\item[8)] The different observables (electronic, muonic and Cherenkov components) that will be measured in LHAASO will allow a detailed investigation of the role of the hadronic interaction models, therefore investigating if the EAS development is correctly described by the current simulation codes.
\item[9)] LHAASO will look for signatures of WIMPs as candidate particles for DM with high sensitivity for particles masses above 10 TeV. Moreover, axion-like particle searches are planned, where conversion of gamma-rays to/from axion-like particles can create distinctive features in the spectra of gamma-ray sources and/or increase transparency of the universe by reducing the Extragalactic Background Light (EBL) absorption. 
Testing of Lorentz invariance violation as well as the search for Primordial Black Holes and Q--balls will also be part of the scientific programme of the experiment.
\end{itemize} 

LHAASO is expected to be able to measure the energy spectra of different CR mass groups in the 10$^{12}$ -- 10$^{17}$ eV with unprecedented statistics and resolution.
In addition, CTA-North and LHAASO are expected to be the most sensitive instruments to study Gamma-Ray Astronomy in the Northern hemisphere from about 20 GeV up to PeV.

\subsection{TAIGA-HiSCORE experiment}

The new TAIGA-HiSCORE non-imaging Cherenkov array aims to detect air showers induced by gamma rays above 30 TeV and to study cosmic rays above 100 TeV \cite{hampf12,epimakhov15}. 
TAIGA-HiSCORE is made of integrating air Cherenkov detector stations with a wide FoV ($\sim$0.6 sr), placed at a distance of about 100 m to cover a final area of $\sim$5 km$^2$.
%
%%%%%%%%%%%%%%%%%%%%%%%%%%%%%%%%%%%%%%%%%%%%%%%%%%%%%%%%%%
\begin{figure}[ht!]
\centerline{\includegraphics[width=0.8\textwidth,clip]{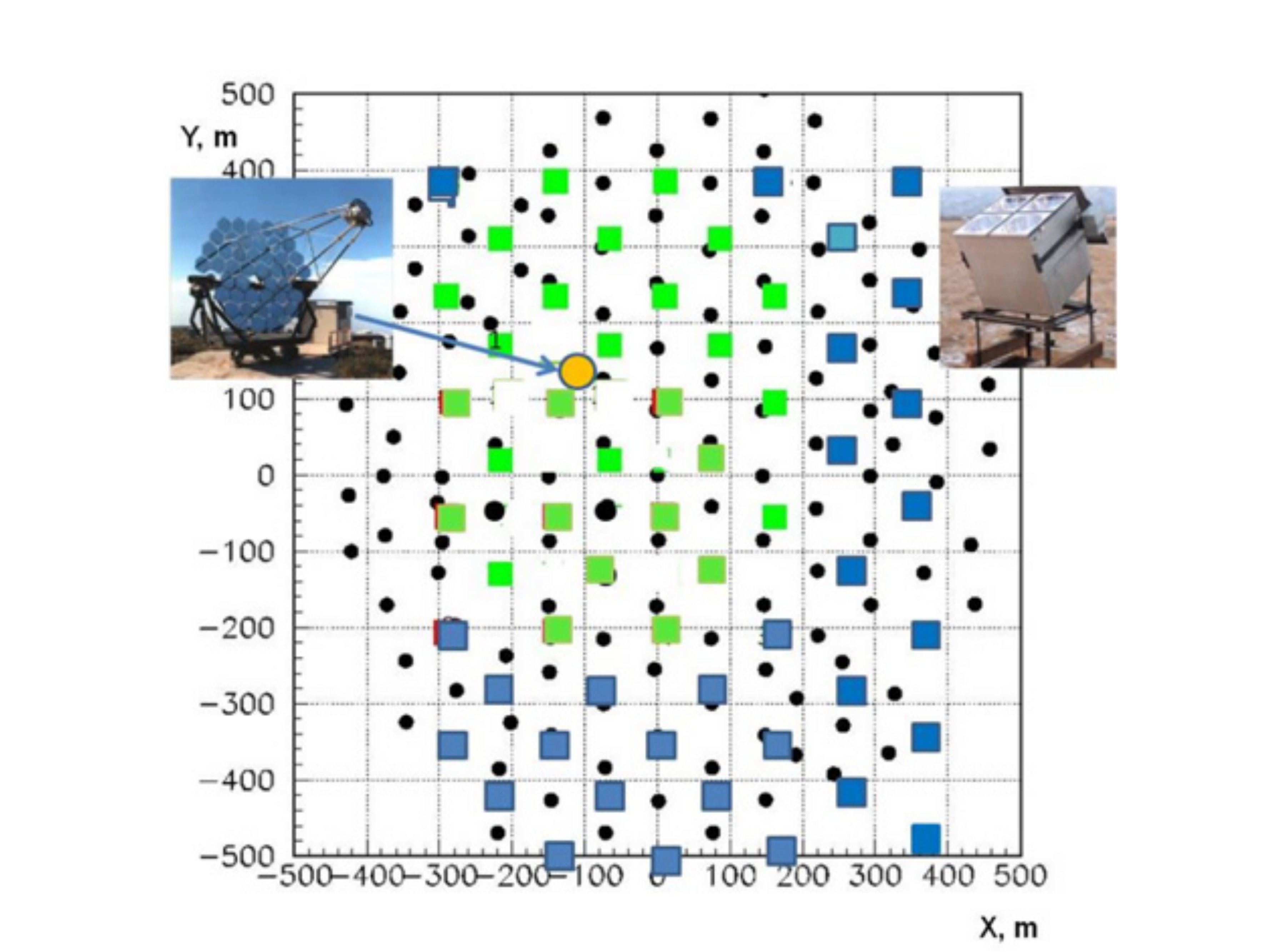} }
\caption{The TAIGA-prototype-2017. Green squares: optical stations of TAIGA-HiSCORE installed in 2014. Blue squares: optical stations to be installed in 2017. Yellow circle: position of the first TAIGA-IACT. Small black circles: optical detectors of Tunka-133.} 
\label{fig:hiscore}       % Give a unique label
\end{figure}
%%%%%%%%%%%%%%%%%%%%%%%%%%%%%%%%%%%%%%%%%%%%%%%%%%%%%%%%%%
%

The TAIGA-HiSCORE array is part of the gamma-ray observatory TAIGA (Tunka Advanced Instrument for cosmic ray physics and Gamma Astronomy). TAIGA is currently under construction in the Tunka valley, about 50 km from Lake Baikal in Siberia, Russia \cite{taiga16}. The key advantage of the TAIGA will be the hybrid detection of EAS Cherenkov radiation by the wide-angle detector stations of the TAIGA-HiSCORE array and by the Imaging Air Cherenkov Telescopes of the TAIGA-IACT array. TAIGA comprises also the Tunka-133 array and will furthermore host up a net of surface and underground stations for measuring the muon component of air showers.
The principle of the TAIGA-HiSCORE detector is the following: the detector stations measure the light amplitudes and full time development of the air shower ligth front up to distances of several hundred meters from the shower core.

Currently TAIGA-HiSCORE array is composed of more than 40 detector stations distributed in a regular grid over a surface area of $\sim$0.5 km$^2$ with an inter-station spacing of about 106 m (prototype array, see Fig. \ref{fig:hiscore}).
Each optical station contains four large area photomultipliers with 20 or 25 cm diameter. Each PMT has a light collector Winston cone with 0.4 m diameter and 30$^{\circ}$ viewing angle (FoV of $\sim$0.6 sr). Plexiglass is used on top to protect the PMTs against dust and humidity. A total station light collection area is 0.5 m$^2$ \cite{hiscore}.

Before the winter season 2017--2018 the TAIGA configuration will include 60 wide angle stations arranged over an area of 0.6 km$^2$, and one single IACT. The expected integral sensitivity for 200 hours of a source observation (about 2 seasons of operation) in the range 30--200 TeV is about 10--12 erg cm$^2$ sec$^{-1}$ \cite{taiga17}.

\section*{Acknowledgements}

I would like to thank the organizers of the \emph{ISAPP-Baikal Summer School ``Exploring the Universe through multiple messengers''} for their invitation and the warm hospitality in Bol'shie Koty.

\section*{References}

\expandafter\ifx\csname url\endcsname\relax
  \def\url#1{{\tt #1}}\fi
\expandafter\ifx\csname urlprefix\endcsname\relax\def\urlprefix{URL }\fi
\providecommand{\eprint}[2][]{\url{#2}}

\end{document}